\documentclass{aa}

\usepackage{txfonts}
\usepackage{natbib}

\usepackage{graphicx}
\usepackage[space]{grffile}
\usepackage{latexsym}
\usepackage{amsfonts,amsmath,amssymb}
\usepackage{url}
\usepackage[utf8]{inputenc}
\usepackage{fancyref}
\usepackage{hyperref}
\hypersetup{colorlinks=true, urlcolor=blue, citecolor=cyan, pdfborder={0 0 0},}

\begin{document}

\title{ASteCA - Automated Stellar Cluster Analysis}
%\subtitle{}

\author{G.~I.~Perren\inst{\ref{inst1},\ref{inst3}}
\and R.~A.~V\'azquez\inst{\ref{inst1},\ref{inst3}}
\and A.~E.~Piatti\inst{\ref{inst2},\ref{inst3}}}

\institute{
Facultad de Ciencias Astron\'omicas y Geof\'isicas (UNLP), IALP-CONICET,
La Plata, Argentina\\
\email{gabrielperren@gmail.com}\label{inst1}
\and
Observatorio Astron\'omico, Universidad Nacional de C\'ordoba,
Laprida 854, 5000, C\'ordoba, Argentina\label{inst2}
\and
Consejo Nacional de Investigaciones Cient\'{\i}ficas y T\'ecnicas,
Av. Rivadavia 1917, C1033AAJ, Buenos Aires, Argentina\label{inst3}
}

\date{Received 15 September 2014; Accepted --}

\abstract{We present \texttt{ASteCA} (\textbf{A}utomated \textbf{Ste}llar
\textbf{C}luster \textbf{A}nalysis), a suit of tools
designed to fully automatize the standard tests applied on stellar clusters
to determine their basic parameters. The set of functions included in
the code make use of positional and photometric data to obtain precise
and objective values for a given cluster's center coordinates, radius,
luminosity function and integrated color magnitude, as well as
characterizing through a statistical estimator its probability of being
a true physical cluster rather than a random overdensity of field stars.
\texttt{ASteCA} incorporates a Bayesian field star decontamination
algorithm capable of assigning membership probabilities using
photometric data alone. An isochrone fitting process based on the
generation of synthetic clusters from theoretical isochrones and
selection of the best fit through a genetic algorithm is also present,
which allows \texttt{ASteCA} to provide accurate estimates for a
cluster's metallicity, age, extinction and distance values along with
its uncertainties.\\To validate the code we applied it on a large set of
over 400 synthetic \texttt{MASSCLEAN} clusters with varying degrees of
field star contamination as well as a smaller set of 20 observed Milky
Way open clusters (Berkeley 7, Bochum 11, Czernik 26, Czernik 30,
Haffner 11, Haffner 19, NGC 133, NGC 2236, NGC 2264, NGC 2324, NGC 2421,
NGC 2627, NGC 6231, NGC 6383, NGC 6705, Ruprecht 1, Tombaugh 1, Trumpler 1,
Trumpler 5 and Trumpler 14) studied in the literature.
The results show that \texttt{ASteCA} is able to recover
cluster parameters with an acceptable precision even for those clusters
affected by substantial field star contamination.\\\texttt{ASteCA} is
written in Python and is made available as an open source code which can be
downloaded ready to be used from it's official site.
}

% A maximum of 6 key words 
% http://www.aanda.org/index.php?option=com_content&task=view&id=131&Itemid=173%E2%8C%A9=en_GB.utf8%2C+en_GB.UT#keywords

\maketitle 

\section{Introduction}
\label{sec:intro}

Stellar clusters (SCs) are valuable tools for studying the structure and
chemical/dynamical evolution of the Galaxy, in addition to provide useful
constraints for evolutionary astrophysical models. They also represent an
important step in the calibration of the distance scale because of the accurate
determination of their distances. Historically, estimating values for a 
cluster's structural characteristics along with its metallicity, age, distance 
and reddening (from here on referred as \textit{cluster parameters}), has
mostly relied on the subjective by-eye analysis of their finding charts,
density profiles, color-magnitude diagrams (CMDs),
color-color diagrams, etc.

In the last few years a number of attempts have been made in order to
partially automatize the star cluster analysis process, by developing
appropriate software.
Some studies have focused on the removal of foreground/background contaminating 
field stars from the cluster CMDs or the statistical membership probability
assignment of those stars present within the cluster region
(Sect. \ref{sec:field-star-cont}).
Others have developed isochrone-matching techniques of different degrees of
complexity, with the aim of estimating cluster fundamental parameters.
Efforts made in recent works have combined the aforementioned decontamination
procedures with theoretical isochrone based methods to provide a more
thorough analysis (Sect. \ref{sec:cl-param-deter}).

The majority of the codes available in the literature are usually
closed-source software not accessible to the community.
In this context, we have developed a new Automated Stellar Cluster Analysis
tool (\texttt{ASteCA}) that aims at being not only a
comprehensive set of functions to connect the initial determination of a
cluster's structure (center, radius) with its intrinsic (age, metallicity) and
extrinsic (reddening, distance) parameters, but also a suite of tools to
fill the current void of publicly available open source standardized tests.
Our goal consists in providing a set of clearly defined and objective rules,
thus making the final results easily reproducible and eventually
collaboratively improved, replacing the need to perform interactive by-eye
parameter estimation.
In addition, the code can be used to implement an automatic processing of
large databases (e.g., 
2MASS\footnote{\url{http://www.ipac.caltech.edu/2mass/}},
DSS/XDSS\footnote{\url{http://archive.eso.org/dss/dss}},
SDSS\footnote{\url{http://www.sdss.org/}}
and many others including the upcoming survey
Gaia-ESO\footnote{\url{http://www.gaia-eso.eu/}}),
making it applicable to generate new \textit{entirely homogeneous}
catalogs of stellar cluster parameters.

We present an exhaustive testing of the code having applied it to over 400
artificial
\texttt{MASSCLEAN}\footnote{MASSive CLuster Evolution and ANalysis Package,
\url{http://www.physics.uc.edu/~bogdan/massclean/}}
clusters \citep{Popescu_2009}, which enabled
us to determine the overall accuracy and shortcomings of the process.
Likewise we used \texttt{ASteCA} to derive cluster parameters for 20 observed
Milky Way open clusters (OCs) and compared the values obtained with values
taken from the literature.

In Sect. \ref{sec:descrip} a general introduction to the code is given along
with a detailed description of the full list of tools available. In Sect.
\ref{sec:validation} we use a large set of synthetic clusters to validate
\texttt{ASteCA}. The results of applying the code on observed OCs are presented
in Sect. \ref{sec:oc-results} followed by concluding remarks summarized in
Sect. \ref{sec:conclusions}.

\section{General description}
\label{sec:descrip}

\texttt{ASteCA} is a compilation of functions usually applied to the analysis of
observed SCs, intended to be executed as an automatic routine
requiring only minimal user intervention.
Its input parameters are managed by the user through a single configuration
file that can be easily edited to adapt the analysis to clusters observed in
different photometric systems. The code is able to run in batch mode on any 
given number of photometric files (e.g. the output of a reduction process)
and is robust enough to handle poorly
formatted data and complicated observed fields. Both a \textit{semi-automatic}
and a \textit{manual} mode are also made available in case user input is
needed for a given, more complicated, system. The former permits the user
to manually set structure parameters (center, radius, error-rejection
function) for a list of clusters, to then run the code in batch mode
automatically reading and applying those values.
The latter requires user input to set the same structure parameters and
will display plots at each step to facilitate the correct choosing of these
values.
A series of flags are raised as the code is executed, and stored in the final
output file along with the rest of the cluster parameters obtained, to warn
the user about certain results that might need more attention (e.g.: the
center assignation jumps around the frame, the density of field stars is too
close to the maximum central density of the cluster, no radius value could be
found due to a variable radial density profile, etc.)

\texttt{ASteCA} employs both spatial and photometric data to perform a complete
analysis process. Positional data is used to derive the SC structural
parameters, such as its precise center location and radius value, while an
observed magnitude and color are required for the remaining functions.
In recent years there has been a huge accumulation of photometric data
thanks to the use of large CCDs on fields and star clusters.
Although these observations are generally multi-band, those bands covering
the Balmer jump ($U$, $u$, etc) are rarely observed.
Because of this, parameter estimates for a large number of
clusters rely entirely on CMDs disregarding two-color diagrams (TCDs).
The latter is known to allow a more accurate estimation of the reddening and 
simultaneously, a substantial reduction in the number of possible solutions for
the cluster parameters.
Notwithstanding, observed bands in most databases permit primarily the
creation of CMDs rather than TCDs. We have thus developed the first version
of the code with the ability to handle photometric information from a single 
CMD, i.e.: one magnitude and one color.
We plan on lifting this limitation altogether in an immediate following
version so that an arbitrary number of observed magnitudes and colors
can be utilized in the analysis process, including of course the standard
two-color diagram.
Presently the CMDs supported by \texttt{ASteCA} include $V$ vs. $(B-V)$, $V$ vs. 
$(V-I)$ and $V$ vs. $(U-V)$ from the Johnson system,
$J$ vs $(J-H)$, $H$ vs $(J-H)$ and $K$ vs $(H-K)$ from the 2MASS system,
and $T_1$ vs. $(C - T_1)$ from the Washington system.
Any other CMD can be easily added to the list provided the theoretical
isochrones and extinction relations for its photometric system are available.

The following sub-sections introduce the entirety of functions/tools that are
implemented within \texttt{ASteCA} in the order in which they are applied to
the input cluster data; a much more thorough technical description of each one
will be provided in a complete manual.
The code is written modularly which means it is easy to replace, add or
remove a function, allowing for easy expansion and revision if a new test
is decided to be implemented or a present one modified.

In this work we make use of the \texttt{MASSCLEAN} version 2.013 (BB)
package, a tool able to create artificial stellar clusters following a King
model spatial distribution with arbitrary radius, metallicity, age, distance,
extinction and mass values, including added field star contamination.
The software was employed to generate artificial/synthetic OCs
``observed'' with Johnson's $BV$ photometric bands,
that will serve as example inputs for the plots shown throughout this section,
and as a validation set used to estimate the code's accuracy when recovering
true cluster parameters in Sect. \ref{sec:validation}.

\subsection{Center determination}
\label{sec:cent-determ}

An accurate determination of a cluster's central coordinates is of
importance given the direct impact its value will have in its
radial density profile and thus in its radius estimation (see 
Sect. \ref{sec:rad-determ}).
SCs central coordinates are frequently obtained via visual inspection of an 
observed field \citep{Piskunov_2007}, a clear example of this is the OC 
catalog by \cite{Dias_2002}
(DAML02\footnote{\url{http://www.astro.iag.usp.br/ocdb/}}) where the authors
visually check and eventually 
correct assigned central coordinates in the literature. This approach has the 
obvious drawback of being both subjective and prone to misclassifications 
of \textit{apparent} spatial overdensities as OCs or open cluster remnants 
(OCRs).

The number of algorithms for automatic center determination mentioned 
throughout the literature is quite scarce.
\cite{Bonatto_Bica_2007} start from visual estimates for the center of several 
objects from XDSS\footnote{Taken from the Canadian Astronomy Data Centre, 
\url{http://www2.cadc-ccda.hia-iha.nrc-cnrc.gc.ca/}} images and refine it 
applying a standard two-dimensional histogram based search for the maximum star 
density value.
A similar procedure is applied in \cite{Maciejewski_2007} using 
initial estimates taken from the DAML02 catalog.
In \cite{Maia_2014} the authors apply an iterative algorithm which depends on an 
initial estimate of the OC's center and radius
\citep[taken from][]{Bica_2008} and averages the stars' positions weighted
by the stellar densities around them.
Determining the center via approaches similar to the previous two has the 
disadvantage of requiring reasonable initial values for the center and the 
radius, otherwise the algorithm could converge to unexpected coordinates. 
Moreover, in the case of the latter algorithm convergence is not guaranteed.\\

\begin{figure}[t]
\begin{center}
\includegraphics[width=\columnwidth]{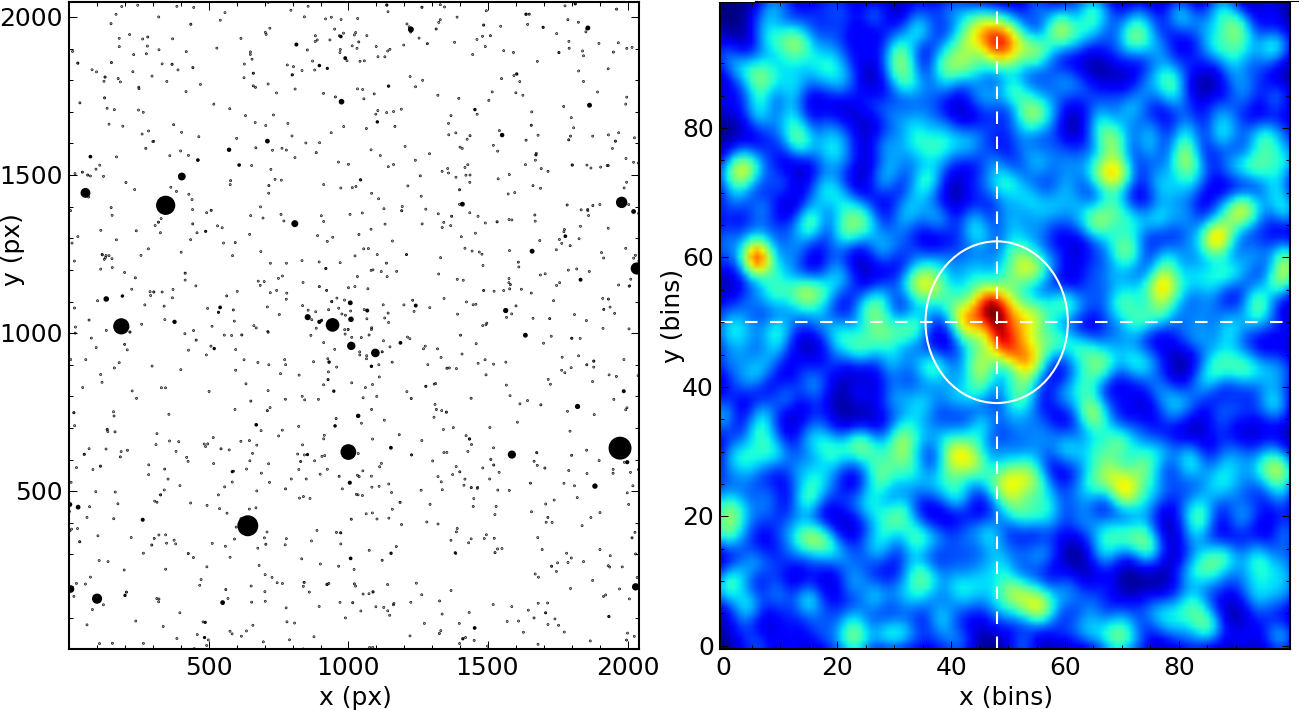}
\caption{Left: finding chart of an example artificial \texttt{MASSCLEAN} cluster
embedded in a field of stars. Right: center determination via a
two-dimensional KDE.
\label{img-center}}
\end{center}
\end{figure}

%Furthermore, assigning a unique center value to OCs with complicated 
%non-circular morphology or to poorly populated OC remnants (OCRs) often 
%carries unclear implications of what precisely that center represents,
%in terms of the cluster's own structure.
%In this work we adhere to the usual definition of OC as a spatial
%overdensity  of stars 
Unlike what usually happens with globular clusters, the center of an OC can not 
always be unambiguously identified by eye. \texttt{ASteCA} uses the standard 
approach of assigning the maximum spatial density value as the point that 
determines the central coordinates for an OC. We obtain this point searching 
for the maximum value of a two-dimensional Gaussian kernel density estimator 
(KDE) fitted on the positional diagram of the cluster, as seen in Fig. 
\ref{img-center}. The difference with the rest of the algorithms mentioned 
above is that ours does not require initial values to work (although they can 
be provided in \textit{semi-automatic} mode) and convergence is always 
guaranteed. This process eliminates the dependence on the binning of the region 
since the bandwidth of the kernel is calculated via the well-known Scott's rule 
\citep{Scott1992} and, by obtaining the maximum estimate in both spatial 
dimensions simultaneously, avoids possible deviations in the final central 
coordinates due to densely populated fields.
The process is independent of the system of coordinates used and can be equally 
applied to positional data stored in pixels or degrees.

\subsection{Radius determination}
\label{sec:rad-determ}

%"a reliable assessment of membership should use some robust estimation of
%the radius." \cite{Sanchez_2010}
A reliable radius determination is essential to the correct assignation of
membership probabilities \citep{Sanchez_2010}. There are several different
definitions of SC radius in the literature, we have chosen to assign it as
the usually employed value where the radial density profile (RDP) stabilizes
around the field density value. 
The RDP is the function that characterizes the variation of the density of
stars per unit area with the distance from the cluster's central coordinates.
The field density is the density level in \textit{stars/area} of combined
background and foreground stars that gives the approximate number of
contaminating field stars per unit of area that are expected to be spread
throughout the observed frame, including the cluster region.
Our ``cluster radius'' $r_{cl}$, is thus equivalent to the ``limiting radius''
$R_{lim}$ defined in \cite{Bonatto_2005} and \cite{Maia_2014}, the ``corona
radius'' $r_2$ defined in \cite{Kharchenko_8-2005} and the ``radial density
profile radius'', $R_{RDP}$, used in \cite{Bonatto_2009}, \cite{Pavani_2011}
and \cite{Alves_2012}.

\subsubsection{Radial density profile}
\label{sec:rdp}

The RDP is usually obtained generating concentric circular rings of increasing
radius values around the assigned cluster center, counting the number of stars
that fall within each ring and dividing it by its area.
The strategy we developed is similar but uses concentric \textit{square} rings
instead of circular ones, generated via an underlying 2D histogram/grid in the
positional space of the observed frame.
The bin width of this positional histogram is obtained as $1\%$
of whichever spatial dimension spans the smallest range in the observed frame
(i.e.: $min(\Delta x,\,\Delta y)/100$). This (heuristic) value is small enough
to provide a reasonable amount of detail but not too large as to hide
important features in the spatial distribution (e.g.: a sudden drop in
density).\footnote{As with all important input parameters, this bin width can
be manually adjusted via the input parameters file.}

The first RDP point is calculated by counting the number of stars in the central
cell (or bin) of the grid (thought of as the first square ring, with a radius of
half the bin width) divided by the area of the cell.
Following that, we move to the 8 adjacent cells (up, down, left, right; i.e.:
the second square ring with a radius of 1.5 bin widths) and repeat the calculus
to obtain the second RDP point by dividing the stars in those 8 cells by their
combined area. The process is repeated for the next 16 cells (third square
ring, radius of 2.5 bin widths), then the next 24 (fourth square ring, radius
of 3.5 bin widths) and so forth, stopping when $\sim75\%$ of the length of the
frame is reached.
This algorithm has the advantage of working with clusters located near a
frame's edge or corner with no complicated algebra needed to estimate the
area of a severed circular ring, since cells/bins that fall outside the 
frame's boundaries are easily recognized and thus not accounted for in the
calculation of the RDP point's total area. Eventually the RDP function could
be extended to accept a bad pixel mask to correctly avoid empty regions in
frames either vignetted or with complicated geometries, or even zones with
bad photometry.

\subsubsection{Field density}
\label{sec:field-dens}

The field density value is used by the radius finding function as the stable
condition where the RDP reaches the level of the assumed homogeneous field
star contamination present. This last requirement is an important one and
observed frames with a highly variable field star density should be treated
with caution, eventually providing a manual estimation for the cluster radius.

The field density is usually obtained
by manually selecting one or more field regions nearby, but not overlapping,
that of the cluster and calculating the number of stars divided by the total
area of the region(s).
This approach requires either an initial estimate of the cluster's size or a
large enough observed frame, such that the field region(s) can be selected
far enough from the cluster's center to avoid including possible members in
the count.
We developed a simple method for the determination of this parameter which
allows us to fully automatize its estimation, no matter the shape or
extension of the observed frame, using the RDP points through an iterative
process. It begins with the complete set of points and
obtains its median and standard deviation ($1\sigma$) values, to then reject
the point located the farthest outside the $1\sigma$ range around the median.
This step is repeated, with one RDP point less each time in the set, until no
points are left outside the $1\sigma$ level. The final mean density value will
have converged to the expected field density.

\subsubsection{Cluster radius}
\label{sec:clust-rad}

The cluster's radius defined above, $r_{cl}$, is obtained combining the
information from the RDP and the field star density, $d_{field}$.
The algorithm searches the RDP for the point where it ``stabilizes'' around the
$d_{field}$ value, using several tolerance thresholds to define when the
``stable'' condition is met.
This technique has proven to be very robust, assigning reasonable
radius estimates even for scarcely populated or highly contaminated SCs without
the need for user intervention at any part of the process.

\subsubsection{King profile}
\label{sec:kg-prof}
%"Few if any open clusters have enough members to be characterized in terms
%of dynamically meaningful
%quantities, such as the core radius or tidal radius." \cite{Janes_2001}
%"Tidal radii are derived from, e.g. the three-parameter King-pro{fi}le {fi}t
%to RDPs (see below), which requires large surrounding {fi}elds and adequate
%Poisson errors" \cite{Bonatto_2009}

Fitting a three-parameter (3P) King's profile \citep{King_1962} is not always
possible due to low star counts in the cluster region and high field star
contamination \citep{Janes_2001}. A two-parameter (2P) function where the
tidal radius $r_{tidal}$ is left out and only the maximum central density and
core radius $r_{core}$ are fitted, is much easier to obtain. Some authors have
recurred to somewhat elaborated schemes to achieve convergence of the fitting
process with all three parameters \citep{Piskunov_2007}. Our approach is to
first attempt a 3P fit to the RDP and if it's not possible, due either to no
convergence or an unrealistic $r_{tidal}$, fall back to a 2P fit. The 3P fit is
discarded if the tidal radius converges to a value greater than 100 times 
the core radius, which would imply a concentration parameter $c{>}2$ comparable
to that of globular clusters \citep{Hillenbrand_1998}.
The formulas for the 3P and 2P functions are the standard ones, presented for
example in \cite{Alves_2012}.\\

\begin{figure}[t]
\begin{center}
\includegraphics[width=\columnwidth]{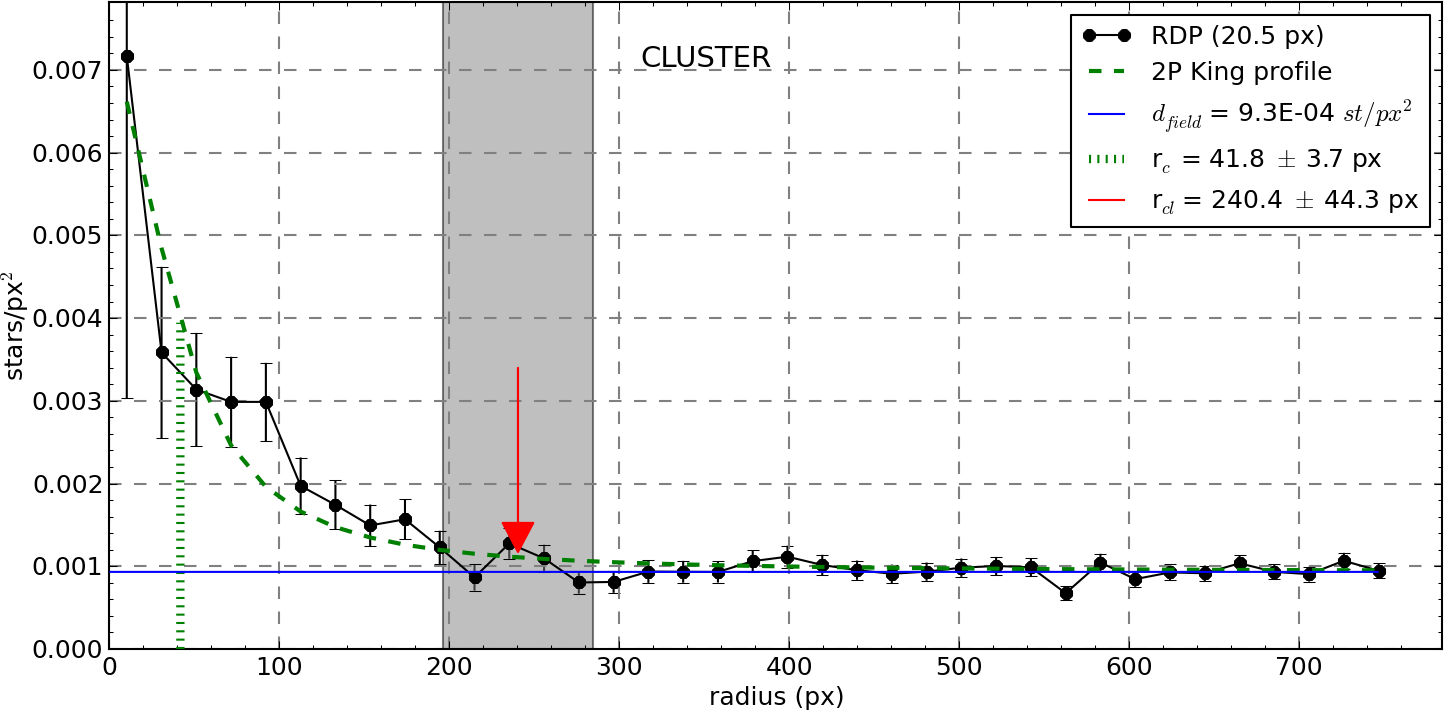}
\caption{Radial density profile of cluster region. Black dots are the
$stars/px^2$ RDP points taking the cluster center as the origin; the horizontal
blue line is the field density value $d_{field}$ as indicated in Sect.
\ref{sec:field-dens}; the red arrow marks the assigned radius with
the uncertainty region marked as a gray shaded area;
a 2P King profile fit is indicated with the green broken curve
with the $r_{core}$ (core radius) value shows as a vertical green line.
\label{img-rdp}}
\end{center}
\end{figure}

Fig \ref{img-rdp} shows as black dots (each one with its poissonian error bar)
the RDP points obtained for a cluster manually generated with a tidal radius
of $r_{tidal}{=}250\,px$. The field density value, the 2P King fit
with the obtained core radius $r_{core}$ and the value for $r_{cl}$ along with it's
uncertainty are also shown.

\subsection{Members and contamination estimation}

\subsubsection{Members estimation}

We estimate the total number of probable cluster members following two
approaches. The first one is based on the 3P King profile fitting and utilizes
the integral of the RDP from zero to $r_{tidal}$ above the estimated star field
density \citep[see Eq. 3 in][]{Froebrich_2007}.
This method will only work if the 3P
fit converged and if it did so to a reasonable tidal radius, otherwise the
result can be quite overestimated. The second approach is based on a simple
star count: multiplying $d_{field}$ by the cluster's area $A_{cl}$ (given by
the $r_{cl}$ radius), we get $n_{fl}$ which is the approximate number of
field stars inside the cluster region.
The final estimated number of cluster members, $n_{cl}$, is obtained 
subtracting this value to the actual number of stars
within the $r_{cl}$ boundary $n_{cl+fl}$:

\begin{equation}
n_{cl} = n_{cl+fl} - d_{field}\,A_{cl}
\end{equation}

\noindent Both methods give the approximate number of members down to the
faintest magnitude observed, which means that they are dependent on the
completeness level.

\subsubsection{Contamination index}
\label{sec:cont-ind}

The contamination index parameter (CI) is a measure of the field star
contamination present in the region of the SC. It is obtained as the ratio of
field stars density $d_{field}$ defined previously, over the density of stars
in the cluster region. This last value is calculated as the ratio of the
number of stars in the cluster region $n_{cl+fl}$, counting both field stars
and probable members, to the cluster's total area $A_{cl}$:

\begin{equation}
CI = \frac{d_{field}}{n_{cl+fl}/A_{cl}} = \frac{n_{fl}}{n_{fl} + n_{cl}}
\label{eq:cont-index}
\end{equation}

\noindent A CI close to zero points to a low field star contamination affecting
the cluster ($n_{cl}{\gg}n_{fl}$).
If the CI takes a value of 0.5 it means that an equal number of
field stars and cluster members are expected in the cluster region
($n_{cl}{\simeq}n_{fl}$), while a larger value means that there are on average
more field stars than cluster members expected within the limit defined by
$r_{cl}$ ($n_{cl}{<}n_{fl}$).
A large CI does not necessarily imply a high density of field stars in
general, but when compared to the density of cluster members in
the cluster region.
As we will see in Sect. \ref{sec:validation}, this parameter proves to
be a very reasonable estimator for the internal accuracy associated to the
cluster parameters derived by \texttt{ASteCA}.

\subsection{Error based rejection}
\label{sec:err-reject}

\begin{figure}[t]
\begin{center}
\includegraphics[width=\columnwidth]{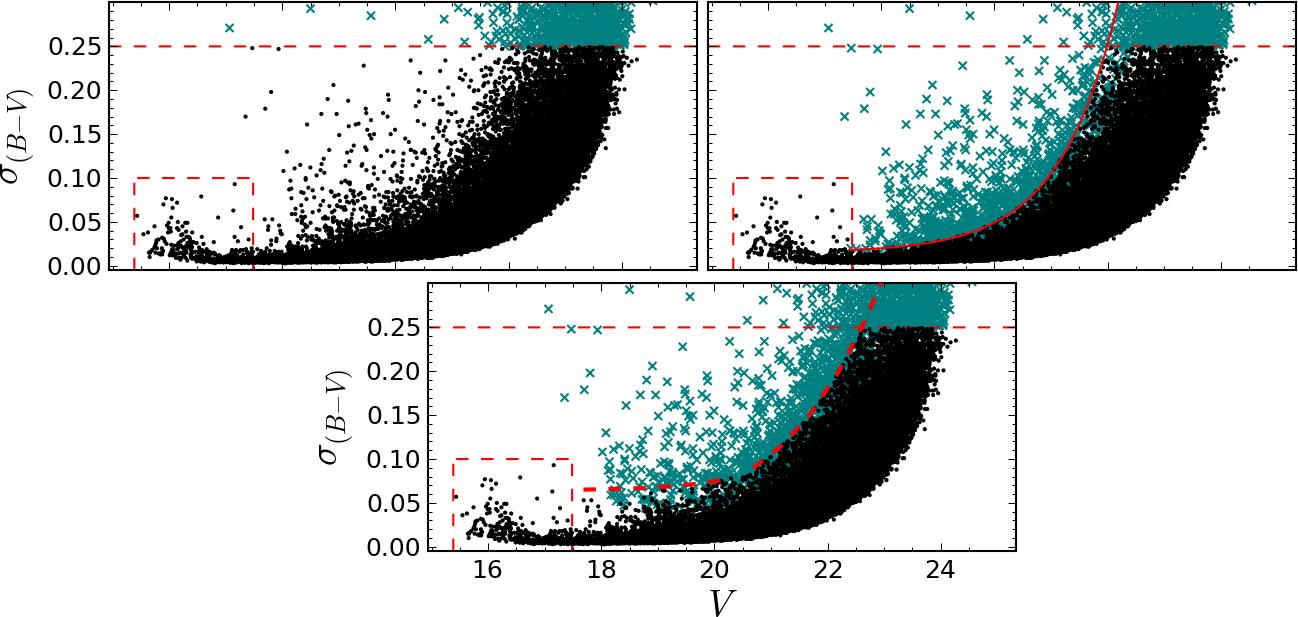}
\caption{\textit{Top}: maximum error rejection method (left) and exponential
curve method (right). \textit{Bottom}: ``eye fit'' method. Rejected stars
are shown as green crosses, horizontal dashed red line is the maximum error
value accepted. See text for details.
\label{img-errors}}
\end{center}
\end{figure}

Measured stars have photometric errors that tend to increase as they
move toward fainter magnitudes. It is necessary to
perform a filtering prior to the cluster analysis so that only stars
with error values reasonably small are taken into consideration and artifacts
left over from the photometry process are removed.
To this end \texttt{ASteCA} includes three routines to reject stars/objects with
photometric errors beyond a certain limit; the results of each can be seen in
Fig. \ref{img-errors} for a $V$ vs $(B-V)$ CMD, which we will be using in
all the example images that follow in this section.

The first routine in the figure (top left) is a simple maximum error
based algorithm that rejects any star beyond a a given limit.
A second method (top right) incorporates an exponential function to
limit the region of accepted stars.
The third method (bottom), referred as ``eye fit'' since it attempts to imitate
how one would trace an upper error envelope by eye, is similar to the previous
one but uses a combination of an exponential function and a third degree
polynomial to separate accepted and rejected stars.
Notice that stars with errors beyond the limits in \textit{either magnitude or
color} will be rejected. This explains why in the bottom diagram of
Fig. \ref{img-errors} some rejected stars can be seen lying \textit{below} the
curve for the $(B-V)$ color: it means these stars had photometric errors
\textit{above} the curve in the $V$ magnitude error diagram (not shown).
As can be seen in Fig. \ref{img-errors}, brighter stars can be treated
separately to prevent the method from rejecting early type stars with error
values above the average for the brightest region. Alternatively no rejection
method can be selected, in which case all stars are considered by the code.
The parameters of these methods can be adjusted via the input data file used
by \texttt{ASteCA}.
Those stars rejected by this function are not taken into account in any of the
processes that follow.

\subsection{Cluster \& field stars regions}
\label{sec:cl-fl-regs}

\begin{figure}[t]
\begin{center}
\includegraphics[width=\columnwidth]{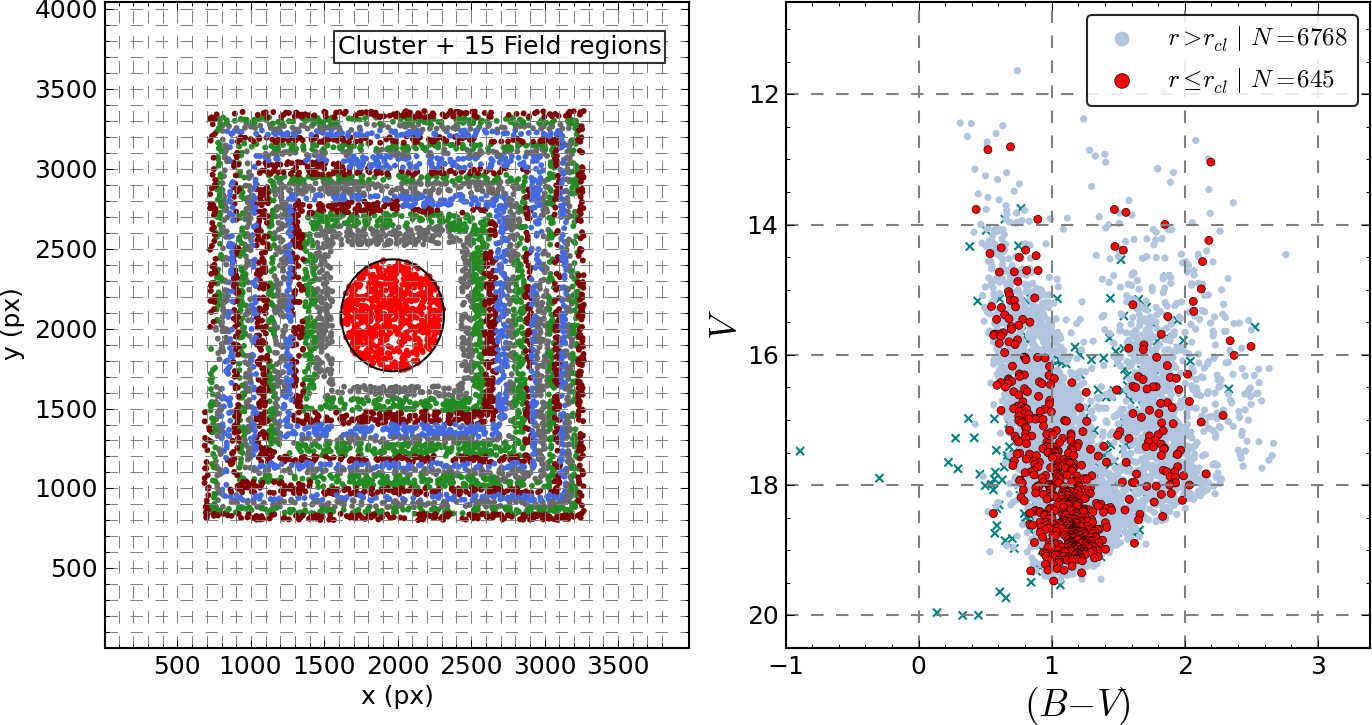}
\caption{\textit{Left}: cluster region centered in the frame and 15 field
regions of equal area defined around it. Adjacent field stars with the same
color belong to the same field region (notice that only 4 colors are used and
they start repeating themselves in a cycle). \textit{Right}: CMD showing the
cluster region stars as red points, stars in the combined surrounding field
regions in gray and rejected stars by the error rejection function as green
crosses.
\label{img-field-regs}}
\end{center}
\end{figure}

\texttt{ASteCA} delimitates several \textit{field regions} around the cluster
region, each one having the same area as that of the cluster. These field
regions are used by those functions that require removing the contribution
from field stars (see Sect. \ref{sec:lf-integ-col}) and those that compare
the cluster region's CMD with CMDs generated from field stars (see Sect.
\ref{sec:pvalue} \& \ref{sec:field-star-cont}). Each region is obtained in
a spiral-like fashion 
to maximize the available space in the observed frame. The left diagram in
Fig. \ref{img-field-regs} shows how this assignment is done with stars
belonging to the same field region plotted with the same color. To the right
the CMD of both the cluster region and the combined field regions is shown
with their stars plotted in red and gray, respectively.

\subsection{Luminosity function \& integrated color}
\label{sec:lf-integ-col}

The luminosity function (LF) of a SC gives the number of stars per magnitude
interval and may be thought of as a projection of its CMD on the
magnitude axis. This results in a simplified version of the CMD that allows
in some cases a quick estimation of certain features, for example the main
sequence turn off (TO).
Integrated colors are often used as indicators of age, especially for very
distant unresolved star clusters, and can also give insights on a SC's mass
and metallicity \citep{Fouesneau_2010,Popescu_2012}.
\texttt{ASteCA} provides both the LF and the integrated color of the SC,
cleaned from field stars contribution whenever possible, i.e.: depending
on the availability of field stars in the observed frame.

\begin{figure}[t]
\begin{center}
\includegraphics[width=\columnwidth]{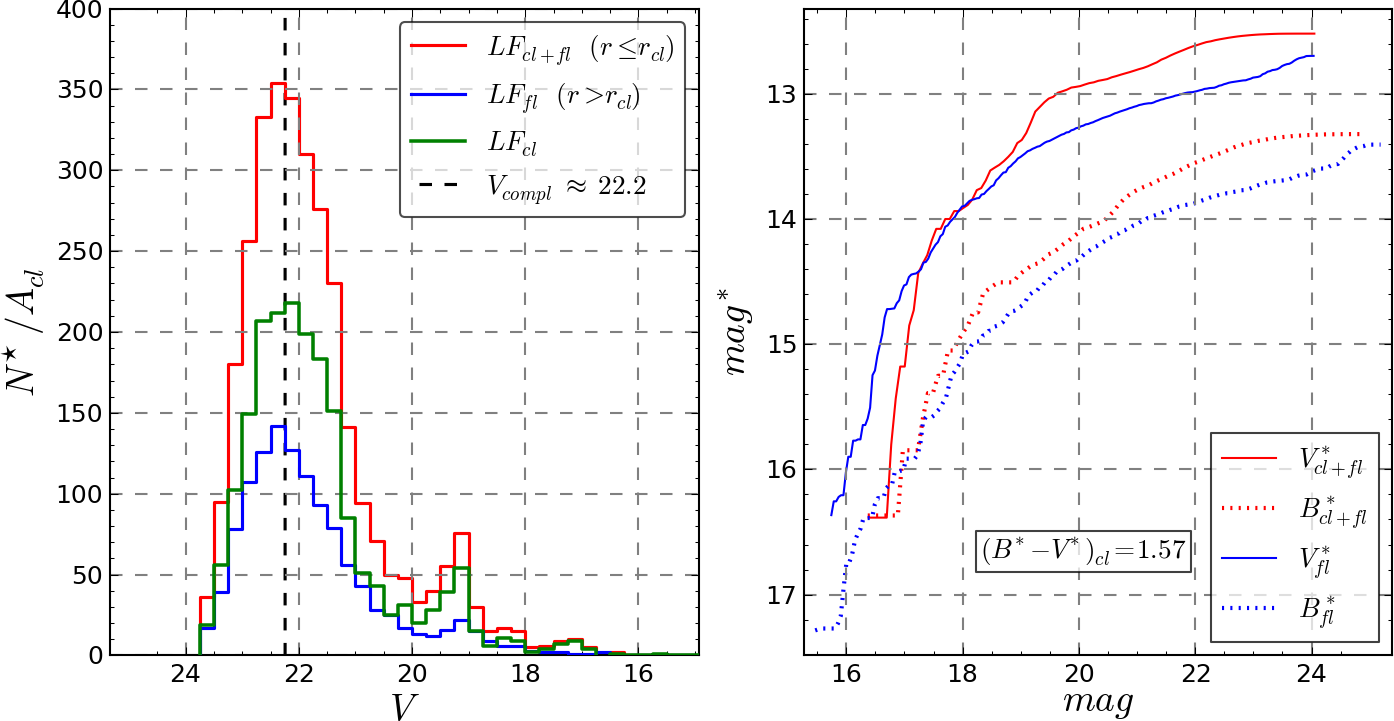}
\caption{\textit{Left}: LF curves for the cluster region plus field star
contamination (red), averaged field regions scaled to the cluster's area (blue)
and clean cluster region (green); completeness limit shown as a dashed black
line. \textit{Right}: integrated magnitudes versus magnitude values for the
cluster region (plus field star contamination) in red and average field regions
in blue.
\label{fig:lf-integ-col}}
\end{center}
\end{figure}

The LF curve for the cluster region with field stars contamination, averaged
field regions scaled to the cluster's area $A_{cl}$, and resulting clean
cluster region can be seen to the left of Fig. \ref{fig:lf-integ-col} in red,
blue and green respectively. The clean region is obtained by subtracting the
field regions LF from the cluster plus field regions LF bin by bin. The
completeness magnitude limit is also provided, estimated as the value were
the total star count begins to drop.

An integrated magnitude curve for each observed magnitude is obtained via
the standard relation \citep{Gray_1965}:

\begin{equation}
\label{eq:int-col}
m^{\star} = -2.5 \, \log \sum_{i}^N 10^{-0.4*m_i}
\end{equation}

\noindent where $m_i$ is the apparent magnitude of a single star and the sum is
performed over all the $N$ relevant stars depending on the region being
analyzed. Eq. \ref{eq:int-col} is applied to both magnitudes that make up the
cluster's CMD (when available) in the cluster region contaminated by field
stars and in the field regions defined around it as seen in
Sect. \ref{sec:cl-fl-regs}. The resulting curves are shown in the right
diagram of Fig. \ref{fig:lf-integ-col} in red and blue respectively, where the
curves for the field regions (blue) are obtained interpolating among all
the field regions to generate a single average estimate. The final integrated
magnitude value is the minimum value attained by each curve after which
Pogson's relation is used to clean each cluster region magnitude from the
field regions contribution. Combining both cleaned integrated magnitudes gives
the cleaned cluster region integrated color, as shown in the diagram mentioned
above.

\subsection{Real cluster probability}
\label{sec:pvalue}

\begin{figure}[t!]
\begin{center}
\includegraphics[width=\columnwidth]{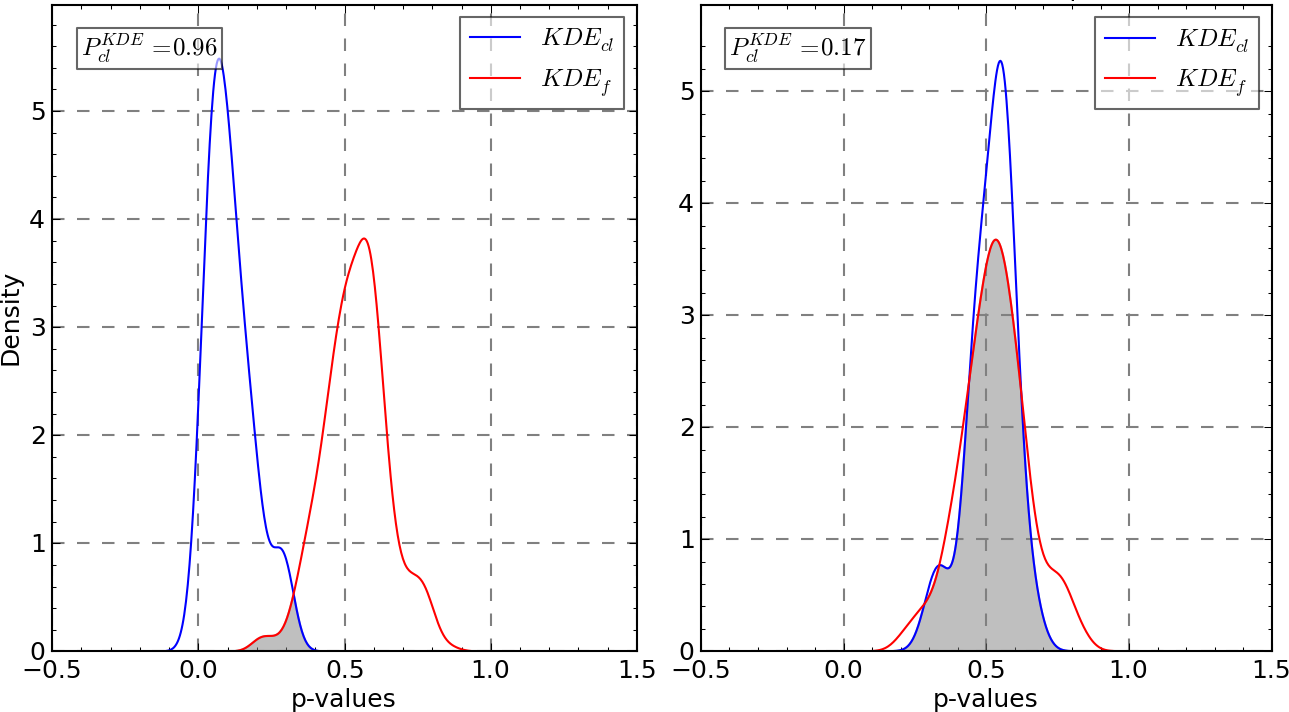}
\caption{\textit{Left}: Function applied on a synthetic cluster, the
curves are clearly separated with the blue one (cluster vs field regions CMD
analysis) showing much lower values; the final probability value obtained is
close to 1 (or 100\%). \textit{Right}: same analysis performed on a random
field region; the curves are now quite similar resulting in a very low
probability of the region containing a true stellar cluster.
\label{fig:kde-pval}}
\end{center}
\end{figure}

% ks manual: http://cran.r-project.org/web/packages/ks/ks.pdf
% kde.test descrp: http://www.inside-r.org/packages/cran/ks/docs/Hpi.kfe
Assigning a probability to a detected spatial overdensity of being a true
stellar cluster rather than a random field stars overdensity, is particularly
useful in the study of open cluster remnants \citep{Pavani_2007} and in general
for OCs poorly populated not easily distinguishable from their
surroundings.
This probability can be evaluated with the \texttt{kde.test} statistical
function provided by the \texttt{ks} package\footnote{Written for the
\texttt{R} software (\url{http://www.r-project.org/})} \citep{Duong_2007}.
The function applies a two-dimensional kernel density estimator (KDE) based
algorithm, able to broadly asses the similarity between the arrangement of
stars in two different CMDs (i.e.: a two-dimensional photometric space), where
the result is quantified by a p-value.\footnote{\textit{The p-value of a
hypothesis test is the probability, assuming the null hypothesis is true,
of observing a result at least as extreme as the value of the test statistic}
\citep{Feigelson_Babu_2012}.} A strict mathematical derivation of the method can
be found in \cite{Duong_2012}.
The null hypothesis, $H_0$, is that both CMDs were drawn from the same
underlying distribution, with a lower p-value being indicative of a lower
probability that $H_0$ is true. This function is applied to the cluster
region's CMD compared with every defined field region's CMD, which results
in a set of ``cluster vs field region'' p-values. Each field region is also
compared with the remaining field regions thus generating a second set now
of ``field vs field region'' p-values, representing the behavior of those
CMDs we expect should have a similar arrangement. The entire process is repeated
a maximum of 100 times, in each case applying a random shift in the position
of stars in the CMDs to account for photometric errors.
The final sets of p-values are smoothed by a one-dimensional KDE to obtain the
curves shown in Fig. \ref{fig:kde-pval}. The blue curve ($KDE_{cl}$) represents
the cluster vs field region CMD analysis while the red one ($KDE_{fl}$) is
the field vs field region curve. For a true cluster we would expect the blue
curve to show lower p-values than the red curve, meaning that the cluster
region CMD has a quite distinctive arrangement of stars when compared when
surrounding field regions CMDs.
Since both curves represent probability density functions, their total area
is unity; furthermore their domains are restricted between [0,1] (a small drift
beyond these limits is due to the 1D KDE processing). This means that the total
area that these two curves overlap (shown in gray in the figure) is a good
estimate of their similarity and thus proportional to the probability that the
cluster region holds a true cluster. An overlap area of 1 implies that the
curves are exactly equal, which points to a very low probability of the
overdensity being a true cluster and the opposite is true for lower overlap
values. We assign then the probability of the overdensity being a real cluster
as 1 minus the overlap between the curves and call it $P_{cl}^{KDE}$.
To the left of Fig. \ref{fig:kde-pval} we show the analysis applied to a true
synthetic cluster and as expected the $KDE_{cl}$ curve shows much lower values
than the $KDE_{fl}$ curve, with a final value of $P_{cl}^{KDE}$ close to 1.
To the right, a field region where no cluster is present is analyzed; this
time the curves are almost identical and the obtained probability very low.

\subsection{Field star contamination}
\label{sec:field-star-cont}

The task of disentangling cluster members from contaminating
foreground/background field stars is an important issue, particularly when SCs
are projected toward crowded fields and/or with an apparent variable stellar
density \citep[][hereafter KMM14]{Krone_Martins_2014}.
Most observed SCs suffer from field star contamination to some extent and only 
in those systems close and massive enough can this effect be dismissed while at 
the same time ensuring a reasonably accurate study of their properties.

Numerous \textit{decontamination algorithms} (DA) can be found throughout the 
literature, all aimed at objectively grouping observed stars into one of two 
classes: field stars or true cluster members.
One of the simplest approach consists in removing stars placed within a given
limiting distance from the cluster's Main Sequence (MS), defined by some process
\citep{Claria_1986,Tadross_2001,An_2007,Roberts_2010}.
Proper motions are known to be a good discriminant between these classes and 
techniques making use of them go back to the Vasilevskis-Sanders (V-S) method 
\citep{Vasilevskis_1958,Sanders_1971} which modeled cluster and field stars as 
a bi-variate Gaussian distributions in the vector point diagram (VPD)
solving iteratively the resulting likelihood equation.
The original method has been largely improved since and is present in numerous
works \citep{Stetson_1980,Zhao_1990,Kozhurina_Platais_1995,Wu_2002,
Balaguer_Nunez_2004,Javakhishvili_2006,Frinchaboy_2008,Krone_Martins_2010,
Sarro_2014}.
%
% \cite{Stetson_1980}
% improved in \cite{Zhao_1990} for data presenting different accuracy.
% \cite{Kozhurina_Platais_1995} combined PM with the relative positions of stars 
% from the cluster's center to improve the membership probability estimates.
% \cite{Wu_2002} employed a 9-parametric method also applied in 
% \cite{Balaguer_Nunez_2004} (BN04) along with a non-parametric approach which 
% uses a circular kernel density estimator to determine the probability density 
% function of the VPD.
% \cite{Javakhishvili_2006} developed their non-parametric ``accumulation'' method 
% based on comparing either the positions or PM of stars in a frame observed in 
% two epochs.
% Kernel density estimators are again used in \cite{Frinchaboy_2008} this time 
% also applied to radial velocities.
% \cite{Krone_Martins_2010} compared four different PM probability density 
% function parameterizations, including one where positions are considered 
% \citep[taken from][]{Zhao_2006}, and arrived at the conclusion that the one 
% presented in BN04 is the most reliable one.
% Recently \cite{Sarro_2014} expanded the original V-S method to incorporate 
% photometric data in addition to PM along with the measurement uncertainties for 
% both quantities, and account for incomplete data sets.

Although PM-based methods tend to be more accurate in determining 
membership probabilities (MP), the requirement of
precise PMs, usually only available for relatively bright stars (KMM14),
severely restricts their applicability.
Photometric multiband data on the other hand is abundant as it is much
easier to obtain, which is why many photometric-based star field
DAs have long been proposed.
\cite{Ozsvath_1960} developed one of the earliest by assigning MPs to stars
located inside the cluster region according to the difference in
stellar-density found in adjacent field regions of comparable size.
Similar algorithms can be found applied with small variations in a great number 
of articles \citep[][etc.]{Baade_1983,Mateo_1986,Chiosi_1989,
Mighell_1996,Bonatto_Bica_2007,Maia_2010,Pavani_2011,Bukowiecki_2011}. 
Some authors have attempted to refine this method by utilizing regions of 
variable sizes instead of boxes of fixed sizes, to compare the CMDs of field 
stars and stars within the cluster region \citep{Froebrich_2010,Piatti_2012}.
The recently developed UPMASK\footnote{Unsupervised Photometric Membership 
Assignment in Stellar Clusters.} algorithm presented in KMM14 is a more 
sophisticated statistical technique for field star decontamination which 
combines photometric and positional data and has the advantage of not 
requiring the presence of an observed reference field region to be able to
assign membership probabilities.

We created our own Bayesian DA, described below, broadly based on the method
detailed in \cite{Cabrera1990} (non-parametric PMs-based scheme that follows
an iterative procedure within a Bayesian framework).

\subsubsection{Method}
\label{sec:decont-algor-method}

We begin by generating three regions from the observed frame: the cluster 
region $C$ containing a mix of field stars and cluster members (i.e.: those stars 
within the cluster radius $r_{cl}$), a field region $B$ containing only field 
stars (with the same area as that of the cluster), and a \textit{hypothetical} 
clean cluster region $A$ containing only true cluster members (which we don't 
have). We can interpret this through the relation $A+B=C$ meaning that a clean 
cluster region $A$ plus a region of field stars $B$, results in the observed 
contaminated cluster region $C$. The membership problem can be reduced to this: 
if we take a random star from $C$, what is the probability that this star is a 
true cluster member ($\in A$) or just a field star ($\in B$)? In other words, 
we want to estimate its MP.

The hypotheses involved can therefore be expressed as:
\begin{itemize}
\item $H_1$: the star is a true cluster member ($\in A$).
\item $H_2$: the star is a field star ($\in B$).
\end{itemize}

These hypotheses are exclusive and exhaustive which means that either one of
them must be true; we are interested in particular in $H_1$ to derive MPs for
every star in the cluster region.

From Bayes' theorem\footnote{
Bayes' theorem can be summarized by the well-known formula:

\begin{equation}
P(H|D) = \frac{P(H)P(D|H)}{P(D)}
\end{equation}

\noindent where $P(H|D)$ is the probability of the hypothesis $H$ given the
observed data $D$, $P(H)$ is the probability of $H$ or prior, $P(D|H)$ is the
probability of the data under the hypothesis or likelihood and $P(D)$ is a
normalizing constant.} we can obtain $P(H_{1}|D)_j$ or the probability, given
the data $D$ (i.e.: the photometry for all stars), that $H_1$ is true for a
given star $j$ picked at random from the observed cluster region $C$; this
probability is thus equivalent to the MP of star $j$, $MP_j$.
The priors for $H_1$ and $H_2$ are respectively $P(H_1)=(N_A/N_C)$ and 
$P(H_2)=(N_B/N_C)$ where $N_A$, $N_B$ and $N_C$ are the number of stars in 
regions $A$, $B$ and $C$. Combining this with the likelihoods 
$L_{A,j}=P(D|H_{1})_j$ and $L_{B,j}=P(D|H_{2})_j$ for star $j$, the final form 
of the probability can be written as:

\begin{equation}
MP_j = P(H_{1}|D)_j = \frac{L_{B,j}}{(N_A/N_B) L_{A,j} + L_{B,j}}
\label{eq:memb-prob}
\end{equation}

\noindent where the formula for the likelihood of star $j$ is:

\begin{equation}
\begin{split}
L_{X,j} = \frac{1}{N_X} &\sum_{i=1}^{N_X} \frac{1}{\sigma_m(i,j) \sigma_c(i,j)} \\
&\quad exp\left[\frac{-(m_i - m_j)^2}{2 \sigma_m^2(i,j)}\right] exp\left[\frac{-(c_i - c_j)^2}{2 \sigma_c^2(i,j)}\right]
\label{eq:likel}
\end{split}
\end{equation}

\noindent with $X\in\{A, B\}$, $(m,\,c)$ are magnitude and color and
$(\sigma_m\, , \sigma_c)$ their respective photometric uncertainties
of the form:

\begin{equation}
\sigma_m^2(i,j) = \sigma_m^2(i) + \sigma_m^2(j)\; , \;
\sigma_c^2(i,j)  = \sigma_c^2(i) + \sigma_c^2(j)
\label{eq:likel-errs}
\end{equation}

Ideally we will have more than one field region defined surrounding the cluster
region\footnote{A minimum of one field region is required for the method to
be applicable, since it is based on comparing the cluster region with a nearby
field region of equal area.} and we assume $K$ such field regions:
$\{B_1, B_2, B_3, ...B_K\}$. Each one of these $K$ regions allows us to obtain
a MP value for every star $j$ in $C$: $MP_{i,j};\,i=1..K\,;\,j=1..N_C$.

The missing piece of information in this method so far is the clean cluster
region $A$. To approximate it we take the observed cluster region $C$ and
randomly remove $N_{B_i}\,(i=1..K)$ stars from it which results in a broad
estimate of $A$, under the assumption that we are removing mainly field stars.
This assumption doesn't hold for heavily contaminated SCs and, as we will see
in Sect. \ref{sec:validation}, leads to the DA behaving poorly for SCs with
high CI values.
In order to ensure that the $A$ region is a fair statistical representation of a
non-contaminated cluster region, the entire process is iterated $Q$
times\footnote{$Q=1000$ is the default value, it can be altered by the user
via \texttt{ASteCA}'s input data file.}, each time removing from $C$ a new set
of $N_{B_i}$ random stars. This means that for every $j$ star in $C$ a total
of $K*Q$ values for its MP are obtained, which can be finally combined into a
single probability taking the arithmetic mean:

\begin{equation}
\overline{MP_j} = \frac{1}{K*Q} \sum_{i=1}^{K*Q} MP_{i,j} \; ; \; j=1..N_C
\end{equation}

\begin{figure}[t!]
\begin{center}
\includegraphics[width=\columnwidth]{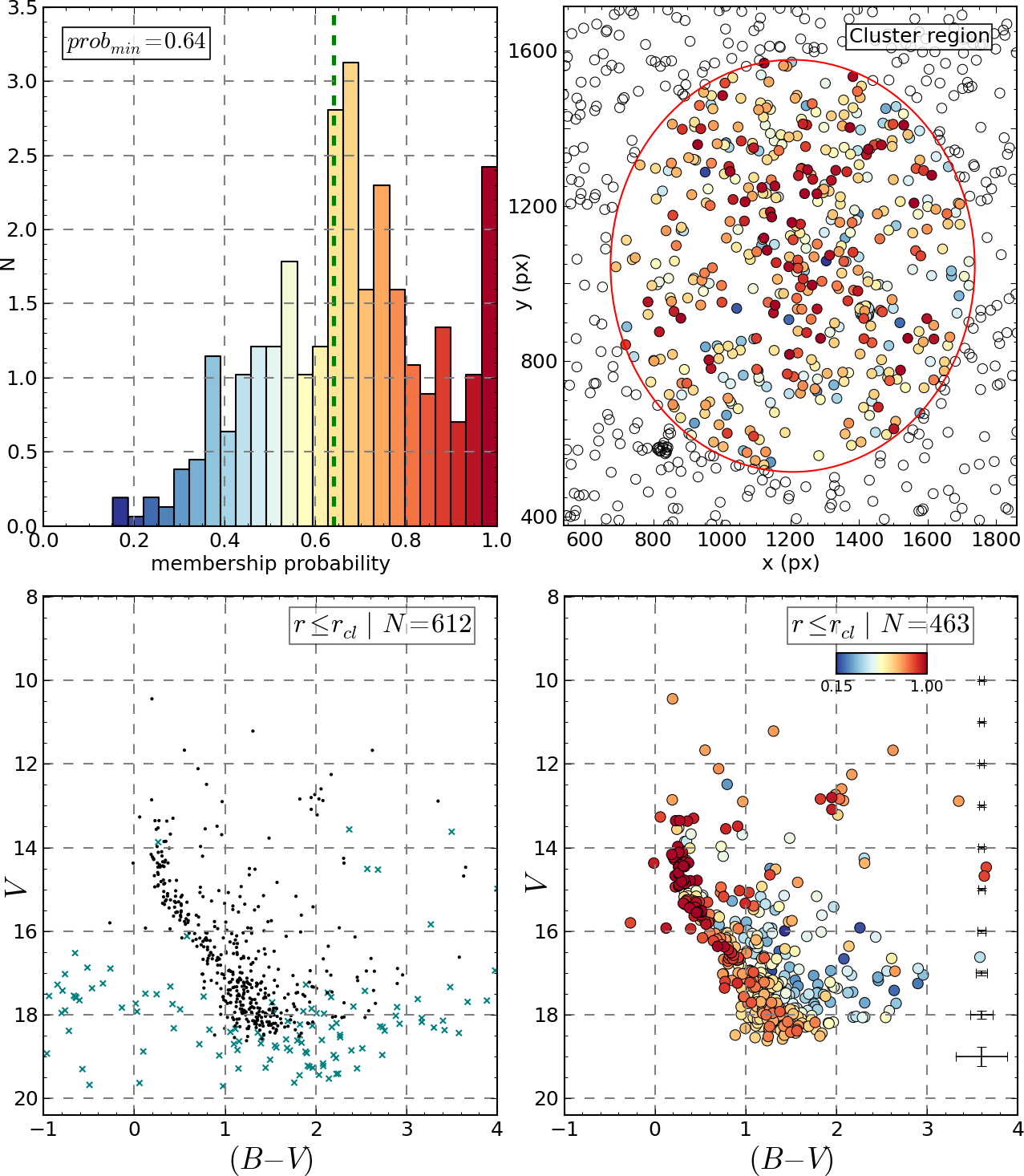}
\caption{\textit{Top}: distribution of MPs (left), $prob_{min}$ is the value
that separates the upper half of stars with the highest MPs from the lower half
and spatial chart (right) with stars in the cluster region colored by their
MPs. \textit{Bottom}: CMD of the observed cluster region (left) with rejected
stars marks as green crosses and the same CMD minus the rejected stars (right)
with coloring according to each stars MPs.
\label{fig:decont-algor}}
\end{center}
\end{figure}

\noindent Fig. \ref{fig:decont-algor} shows the result of applying the
algorithm on an example OC, see caption for details.
The DA can accept an input file with membership probabilities manually assigned 
to individual stars, this allows fixing high probabilities to known members 
obtained via a secondary method (spectroscopy).
The main strength of the method resides in its ability to eventually include 
extra information, such as other observed magnitudes, colors, radial
velocities or proper motions, by simply extending Eq. \ref{eq:likel} adding
an extra exponential term accounting for it.

\subsection{Cluster parameters determination}
\label{sec:cl-param-deter}

The most common method for obtaining the parameters of an SC is still the 
simple by-eye isochrone match on a CMD, examples of this visual approach to 
estimate theoretical isochrone vs. observed cluster best fits are abundant 
\citep[e.g.:][etc.]{Bonatto_2009,Maia_2010,Majaess_2012,Kharchenko_2013,
Carraro_2014}.
The drawbacks of this method include the obvious subjectivity involved in the 
matching process and the inability to attach an uncertainty to the values 
obtained, along with the unavoidable inefficiency when attempting to apply it 
to sets of several hundreds or even thousands of SCs, as done for
example in \cite{Buckner_2014} where the authors manually fitted over
2300 isochrones on near-infrared CMDs.

A summary of methods that make use of certain geometrical evolutionary indicators
\citep[e.g.: the $\delta$ magnitude and color indices,][]{Phelps_1994} can be
seen in \cite{Hernandez_2008} (HVG08), along with approaches to the
estimation of star cluster parameters based on full CMD analysis.
%
% \cite{Flannery_1982} developed the near point estimator $\Psi^2$ which
% derives cluster parameters by taking into account the minimum distance of
% each observed star to a given theoretical isochrone.
% \cite{Wilson_2003}, analysis of the density of points, or areal 
% density, distribution in a Hertzprung-Russel diagram;
% \cite{Naylor_2006}, application of the $\tau^2$ parameter, 
% a generalization of the $\chi^2$ statistic, to a Hess diagram \citep[further
% developed in][]{Naylor_2009};
%
This same article introduces a statistical technique based on using the
density of stars along an isochrone to lift the geometric age-metallicity
degeneracy when attempting a match.

In the pioneering work of \cite{Romeo_1989} the authors applied the
standard technique of generating synthetic populations and comparing them
with an observed simple stellar population CMD to study its properties.
Since then, the ``synthetic CMD method'' has been widely used on
simple stellar populations
\citep[][etc.]{Sandrelli_1999,Carraro_2002,Subramaniam_2005,Singh_2006,
Kerber_2007,Girardi_2009,Cignoni_2011,Donati_2014}.
An expansion of the  method can be found applied with little adjustments to
the recovery of SFHs of nearby galaxies 
\citep{Ferraro_1989,Tosi_1991,Tolstoy_1996, Hernandez_1999, 
Dolphin_2002, Frayn_2002, Aparicio_2009, Small_2013}.\footnote{We refer the 
reader to \cite{Gallart_2005} for a somewhat outdated but thorough review on 
the study of SFHs via the interpretation of composite stellar populations' 
CMDs.}

Decontamination algorithm and cluster parameters estimation processes have been 
coupled in various recent works. This can be seen for example in the series of 
articles by Kerber et al. \citep{Kerber_2002,Kerber_2005} where an estimate of 
the density of field stars in the cluster region is used to implement a field 
star removal process together with a cluster CMD modeling strategy that 
selects the best observed-artificial fit via a statistical tool; the 
white-dwarf based Bayesian CMD inversion technique developed in 
\cite{von_Hippel_2006} expanded and coupled with a basic field-star cleaning 
process in \cite{van_Dyk_2009}; the synthetic cluster fitting method introduced 
in \cite{Monteiro_2010} \citep[MDC10, further developed in the 
articles][]{Dias_2012,Oliveira_2013} which includes a likelihood-based 
decontamination algorithm; the work by \cite{Pavani_2011} where CMD 
density-based membership probabilities are given to stars within the cluster 
region to later apply a very basic isochrone fitting process that makes use of 
stars close to a given isochrone in CMD space and the articles by 
\cite{Alves_2012} and \cite{Dias_2014} who employ the same membership 
probability assignment method used in \cite{Pavani_2011} coupled with a 
slightly improved isochrone fitting algorithm based on the one developed by 
MDC10 but applied to a Hess diagram of the CMD instead of the full CMD.
In \cite{Buckner_2013} the membership assignment method presented
in \cite{Froebrich_2010} (a variation of the \citealp{Bonatto_Bica_2007}
algorithm) is used in conjunction with the Besan\c{c}on model of the
galaxy\footnote{\url{http://model.obs-besancon.fr/}}
to derive distances to OCs, based on foreground stars density estimations.

In a series of papers by Kharchenko et al. where the
COCD\footnote{Catalogue of Open Cluster Data, available at CDS 
via \url{http://cdsarc.u-strasbg.fr/viz-bin/Cat?J/A+A/438/1163}}
and MWSC\footnote{Milky Way Star Clusters, available at CDS 
via \url{http://cdsarc.u-strasbg.fr/viz-bin/qcat?J/A+A/543/A156}}
catalogs are developed
\citep[respectively:][and references therein]{Kharchenko_9-2005,Schmeja_2014},
the authors develop a pipeline to analyze OCs with available PMs,
capable of determining a handful of properties: center, radius,
number of members, distance, extinction and age. The method is neither
entirely objective nor 
automatic since the user is  still forced to intervene manually adjusting
certain variables in order to generate reasonable estimates for the
cluster parameters.

The general synthetic CMD method applied by \texttt{ASteCA} has been outlined
previously in \cite{Tolstoy_1996} and \cite{Hernandez_1999} in the context of
star formation history recovery.
We take the procedures adapted to single stellar populations
described in HVG08 and MDC10 and broadly combine them to obtain the optimal set
of cluster parameters associated with the observed SC.
The theoretical isochrones employed are taken from the \texttt{CMD} v2.5
service\footnote{\url{http://stev.oapd.inaf.it/cgi-bin/cmd}} 
\citep{Girardi_2000} but eventually any set of isochrones could be used with
minimal changes needed to the code.

\subsubsection{Method}
\label{sec:bf-method}

Given a set $A=\{a_1, a_2, ..., a_N\}$ of $N$ observed stars in a cluster 
region we want to find the model $B_i$ out of a set of $M$ models $B=\{B_1, 
B_2, ..., B_M\}$ with the highest probability of resulting in the observed set 
$A$, this is $P(B_i|A)$ or the probability of $B_i$ given $A$.
Each $B_i$ model represents a theoretical isochrone of fixed metallicity ($z$) 
and age ($a$), moved by certain distance ($d$) and extinction values ($e$); 
meaning the models in $B$ are fully determined as points in the 4-dimensional 
space of cluster parameters: $B_i=B_i(z, a, d, e)$.
Finding the highest $P(B_i|A)$ can be reduced to maximizing the probability of 
obtaining $A$ given $B_i$, $P(A|B_i)$, i.e.: the probability that the observed
SC $A$ arose from a $B_i$ model.\footnote{We will not repeat here the full
Bayesian formalism from where this is deduced, the reader is referred to
the aforementioned works in Sect. \ref{sec:cl-param-deter} for more details.}
The first step in determining $P(A|B_i)$ is to define how are the $B_i$ models 
generated. The well-known age-metallicity degeneracy is, as stated in HVG08 and 
\cite{Cervino_2011}, geometrical in nature\footnote{\textit{The effects of
increasing metal abundance on stellar isochrones are remarkably similar to
those of increasing age} \citep{Frayn_2003}.} and the result of considering
only the \textit{shapes} of the isochrones when fitting an observed SC,
instead of also taking into account the \textit{density} of stars along them.
There are two ways of accounting for the star-density in a given isochrone:
through a mass-density parameter as done in HVG08 or, as done in MDC10,
generating correctly populated $B_i$ models as synthetic clusters; we've 
chosen to apply the latter.

\begin{figure}[t!]
\begin{center}
% One-column
% \includegraphics[width=0.8\columnwidth]{ga-synth_cl.png}
% Two-columns
\includegraphics[width=\columnwidth]{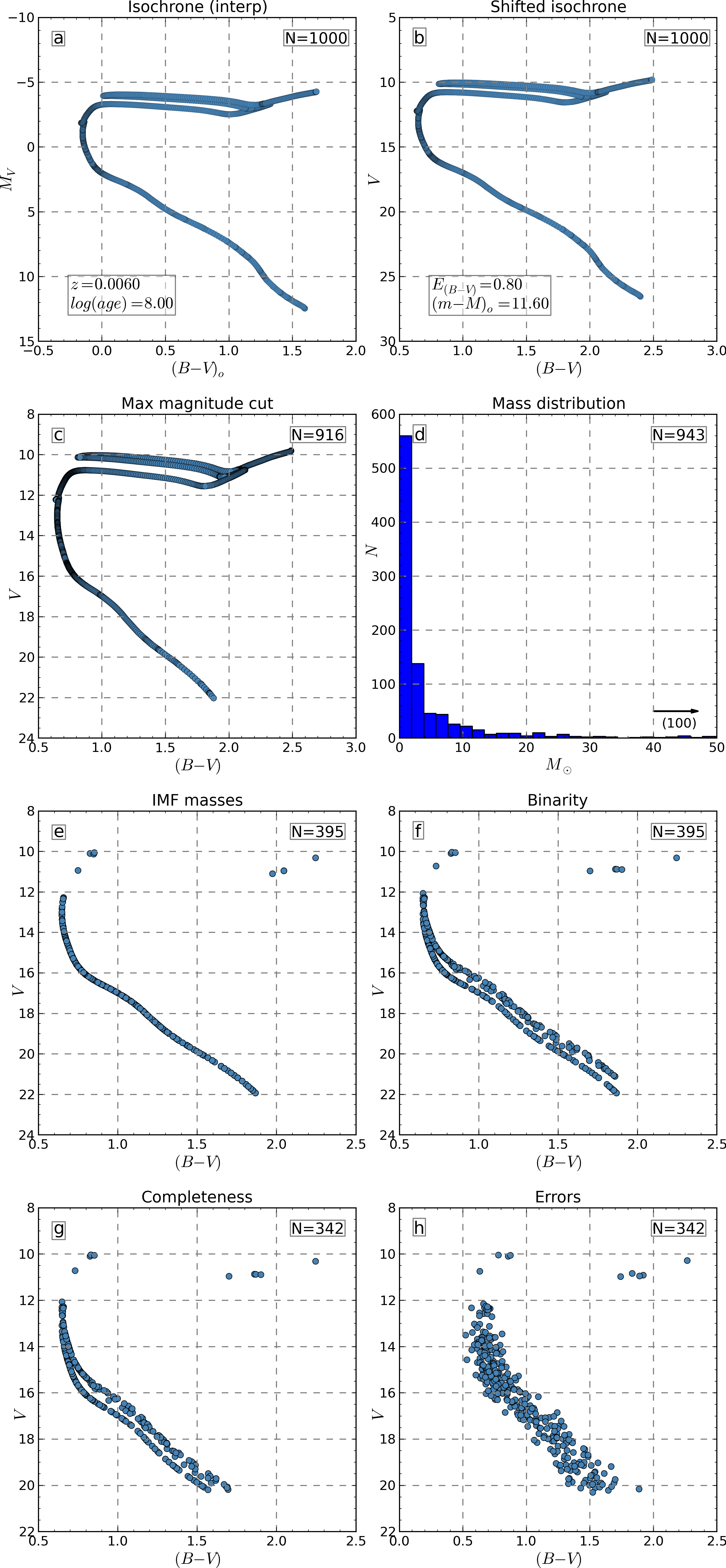}
\caption{Generation of a synthetic cluster starting from a theoretical
isochrone of fixed parameters (metallicity, age, distance and extinction)
populated using the exponential IMF of \cite{Chabrier_2001}.
See text for a description of each panel.
\label{fig:synth-cl}}
\end{center}
\end{figure}

The process by which a synthetic cluster of given $z$, $a$, $d$ and $e$ 
parameters, or $B_i(z, a, d, e)$ model, is generated is shown in Fig. 
\ref{fig:synth-cl}. Panel \textit{a} shows a random theoretical isochrone 
picked with certain metallicity and age values, densely interpolated to contain 
1000 points throughout its entire length; notice even the evolved parts are 
taken into account. In panel \textit{b} the isochrone is shifted by some 
extinction and distance modulus values, to emulate the effects these extrinsic 
parameters have over the isochrone in a CMD. At this stage the synthetic cluster 
can be objectively identified as a unique point in the 4-dimensional space of 
parameters. Panel \textit{c} shows the maximum magnitude cut performed 
according to the maximum magnitude attained by the observed SC being analyzed, 
we see that the total number of synthetic stars drops.
An initial mass function (IMF) is sampled as shown in panel \textit{d} in
the mass range $[{\sim}0.01{-}100]\,M_{\odot}$ up
to a total mass value $M_{total}$ provided via the input data file,
set to $M_{total}=5000\,M_{\odot}$ by default.\footnote{The total mass value
$M_{total}$ is fixed for all synthetic clusters generated by the method,
so we set it to a number high enough to ensure the evolved parts of the
isochrone are also sampled. We plan on removing this restriction in a future 
version of the code so that this can be left as an extra free parameter to be
fitted by the method.}
Currently \texttt{ASteCA} lets the user choose 
between three IMFs \citep{Kroupa_1993,Chabrier_2001,Kroupa_2002} but there is 
no limit to the number of distinct IMFs that could be added. The 
distribution of masses is then used to obtain a correctly populated synthetic 
cluster, as shown in panel \textit{e}, by keeping one star in the interpolated 
isochrone for each mass value in the distribution. The drop seen for the total 
number of stars is due to the limits imposed by the mass range of the 
post-magnitude cut isochrone. 
A random fraction of stars are assumed to be binaries, by default the value is 
set to $50\%$ \citep[typical for OCs,][]{von_Hippel_2005} with secondary masses 
drawn from a uniform distribution between the mass of the primary star and a 
fraction of it given by a mass ratio parameter set by default to $0.7$; both 
figures can be modified in the input data file. Panel \textit{f} shows the 
effect of binarity on the position of stars in the CMD.
Each synthetic cluster is finally perturbed by a magnitude completeness removal 
function and an exponential error function, where the parameters for both are 
taken from fits done on the observed SC and are thus representative of it. 
Panels \textit{g} and \textit{h} show these two processes with the resulting 
$B_i(z, a, d, e)$ synthetic cluster shown in the former.

With the $B$ set of synthetic clusters generated, the next step is to maximize 
$P(A|B_i)$ to find the best fit cluster parameters for the observed SC. The 
probability that an observed star $a_j$ from $A$ is a star $k$ from a given 
synthetic cluster $B_i$ can be written as (HVG08):

\begin{equation}
\begin{split}
P(a_j|B_{i,k}) = \frac{1}{\sigma_m(j,k) \sigma_c(j,k)} \\
exp\left[\frac{-(m_j - m_k)^2}{2 \sigma_m^2(j,k)}\right] &\quad
exp\left[\frac{-(c_j - c_k)^2}{2 \sigma_c^2(j,k)}\right]
\end{split}
\end{equation}

\noindent where the same notation as the one used in Eqs.
\ref{eq:likel} and \ref{eq:likel-errs} applies.

Summing over the $M_i$ stars in $B_i$ gives the probability that
$a_j$ came from the distribution of stars in the synthetic cluster:

\begin{equation}
P(a_j|B_i) = \frac{1}{M_i} \sum_{k=1}^{M_i} P(a_j|B_{i,k})
\end{equation}

\noindent where the $1/M_i$ normalization factor prevents models with
more stars having artificially higher probabilities.
The final probability for the entire observed SC, $A$, to have arisen from the
model $B_i$, also called likelihood, is obtained combining the probabilities
for each observed star as:

\begin{equation}
L_i(z, a, d, e) = P(A|B_i) = \prod_{j=1}^{N} P(a_j|B_i) \times MP_j
\label{eq:likelihood}
\end{equation}

\noindent Following MDC10 we include the MPs obtained by the DA for every star
in the cluster region, $MP_j$, as a weighting factor. The problem is then
reduced to finding the $B_i(z, a, d, e)$ model that produces the maximum
$L_i$ value for a fixed $A$ set or observed cluster region.

\subsubsection{Genetic algorithm}
\label{sec:gen-algor}

Obtaining the best fit between $A$, the observed SC, and the set of $M$ 
models/synthetic clusters $B$, is equivalent to searching for the maximum value 
in the 4-dimensional $L_i(z,a,d,e)$ surface and can be thought of as a global 
maximum/minimum optimization problem. The number $M$ depends on the resolution 
defined by the user for the grid of cluster parameters; to calculate it we 
multiply the total number of values each parameter can attain (range divided by 
a step) for all the parameters, four in our case. For a not too dense grid this 
number will grow to be very high\footnote{For example, given 32 possible values 
for each of our 4 parameters we'd have: $M = 32^4 > 1e06$ total models.} which 
makes the exhaustive search for the best solution in the entire parameter space 
not possible in a reasonable time frame.

Following HVG08 we've chosen to build a genetic algorithm \citep[GA; see:][for 
an in-depth description of the algorithm and references in 
HVG08]{Whitley_1994,Charbonneau_1995} function into \texttt{ASteCA} to solve 
this problem.
A GA is a method used to solve search and optimization problems based on a 
heuristic technique derived from the biological concept of natural evolution.
We prefer it over similar approaches like the cross entropy method described in 
MDC10 due mainly
to its flexibility, which makes it easily adaptable to different optimization 
scenarios.

Instead of finding the best fit for $A$ as the maximum likelihood value, we 
make the GA search for the minimum of its negative logarithm which is a 
computationally more efficient variant:

\begin{equation}
\mathbf{L}_A(z^{\star}, a^{\star}, d^{\star}, e^{\star}) = min\{-\log[L_i(z, a, d, e)] \,;\,i=1..M\}
\end{equation}

\noindent where the $\star$ upper-script indicates the final best fit cluster
parameters assigned to the observed SC.

The implementation of the GA can be divided into the usual operators:
initial random population, selection of models to reproduce, crossover,
mutation and evaluation of new models cycled a given number of ``generations''.
At the end of each generation the model that presents the best fit is selected
and passed unchanged into the new generation to ensure that the GA moves always
towards a better solution.\footnote{More details about the algorithm
can be found in the code's documentation, see:
\url{http://asteca.rtfd.org}}

Uncertainties for each parameter are obtained via a bootstrap process that 
consists in running the GA $N_{btst}$ times each time resampling the stars in 
the observed SC with replacement (i.e.: a given star can be selected more than 
once) to generate a new variation of the dataset $A$. After all these runs, the 
standard deviation of the values obtained for each parameter is assigned as the 
uncertainty for that parameter.
Ideally the bootstrap process would require $N[\ln(N)]^2$ runs to sample
the entire bootstrap distribution \citep{Feigelson_Babu_2012}, where $N$ is
the number of stars within the cluster region.
This is unfortunately not feasible even for a small OC\footnote{A cluster
region with as little as 20 stars would require $\sim180$ \textit{complete runs}
of the GA.} so we settle for setting $N_{btst}=10$ by default, which the user
can modify at will.

\begin{figure}[t!]
\begin{center}
\includegraphics[width=\columnwidth]{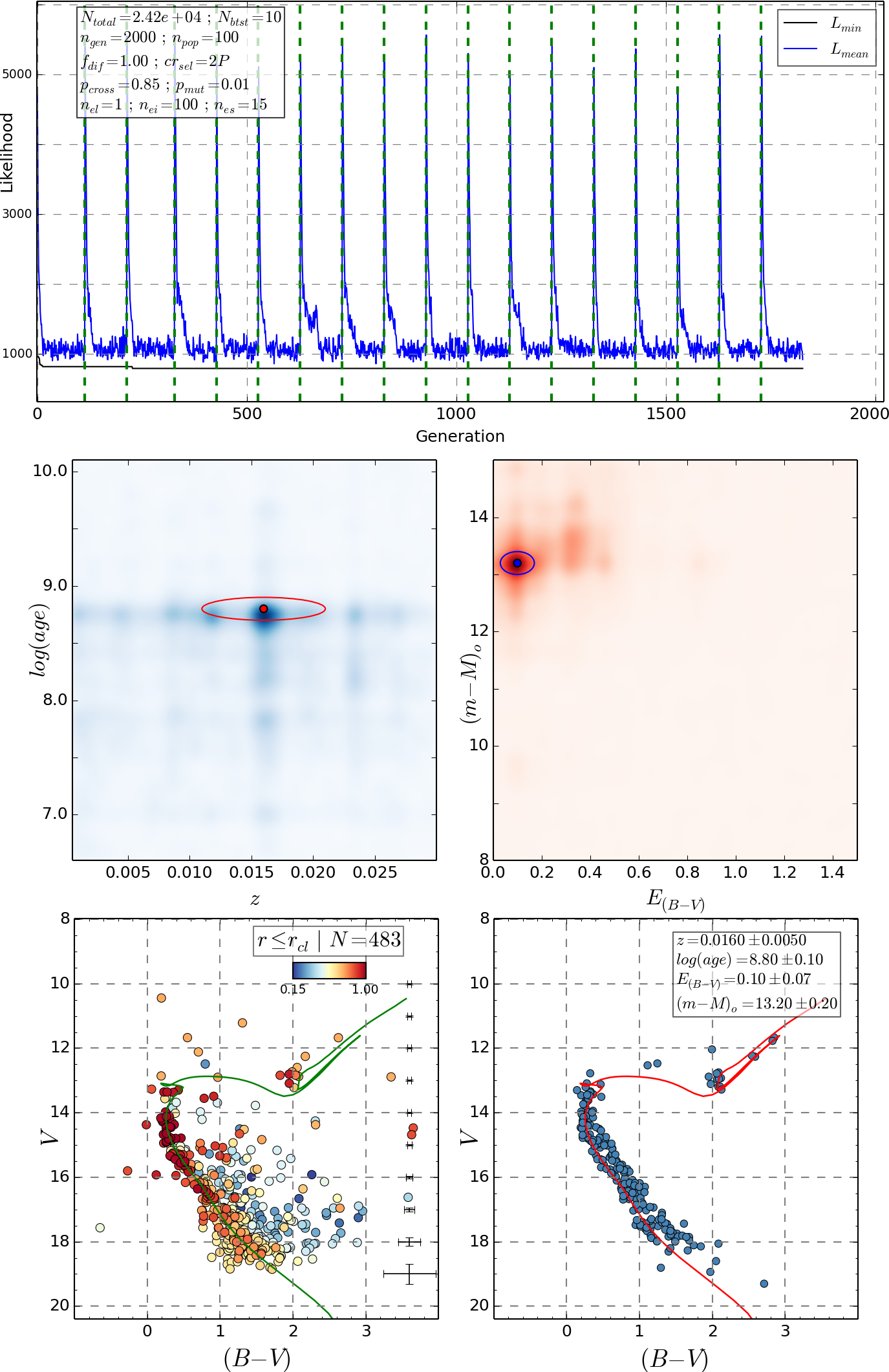}
\caption{Results of the GA applied over an example OC. See text for more details.
\label{fig:ga-results}}
\end{center}
\end{figure}

An example of the results returned by the GA is shown in Fig. 
\ref{fig:ga-results}. The top row shows the evolution of the minimal likelihood 
($L_{min} \equiv \mathbf{L}_{A}$) as generations are iterated with a black 
line, where it can be seen how the GA zooms in the optimal solution early on in 
the process; this is a know feature of the method being a very aggressive 
optimizer. The blue line is the mean of the likelihoods for all the 
chromosomes/models in a generation where each spike marked with a green dotted 
line denotes an application of the extinction/immigration operator.
% The GA halted around the 1800th generation when $N_{gen}$ was set to 2000,
% indicating the exit switch was triggered.
The middle row shows a density map of the solutions/models explored by the GA 
separated in two 2-dimensional spaces for visibility: to the left metallicity 
and age (intrinsic parameters) are shown and to the right distance modulus and 
extinction (extrinsic parameters). The position of the optimal solution is 
marked as a dot in each plot with the ellipses showing the uncertainties 
associated to it.
The bottom row shows to the left the CMD of the observed cluster region ($A$) 
colored according to the MPs obtained with the DA, and the best fit synthetic 
cluster found to the right. The isochrone from which the synthetic cluster 
originated is drawn in both panels.

\subsubsection{Brute force}
\label{sec:bf-algor}

A brute force algorithm (BFA) function is also provided in case the parameter
space can be defined small enough to allow searching throughout the entire grid
of values. Unlike the GA which has no clear stopping point, the BFA is
guaranteed to return the best global solution always, after all the points in
the grid have been analyzed. The BFA therefore doesn't need to apply a
bootstrap process to assign uncertainties to the obtained cluster parameters,
instead its accuracy depends only on the resolution of the grid. If the
required precision in the final estimation of the cluster parameters is
sufficiently small, the BFA can be preferable to the GA.

\section{Validation of ASteCA}
\label{sec:validation}

In order to validate \texttt{ASteCA} we used a sample of 432 synthetic open
clusters (SOCs) generated with the \texttt{MASSCLEAN}
package.\footnote{The complete set of SOCs is available upon
request. The scripts used to generate the set can be obtained via:
\url{https://github.com/Gabriel-p/massclean_cl}} Table \ref{tab:massclean}
lists the grid of parameters taken into account.
Each distance was assigned a fixed visual absorption value in order to cover
a wide range of extinction scenarios; all distances are from the Sun.

\begin{table}[tb]
\centering
\caption{\texttt{MASSCLEAN} parameters used for the generation of 432 SOCs.
Each $A_v$ value was associated with only one distance value.}
\label{tab:massclean}
\begin{tabular}{l c}
\hline
\hline\\[-1.85ex]
 Parameter & Values \\
\hline\\[-1.85ex]
Initial mass ($10^3 M_{\odot}$) & 0.05, 0.1, 0.2, 0.3, 0.4, 0.5, 0.6, 0.8, 1 \\
Metallicity (z) & 0.002, 0.008, 0.019, 0.03 \\
$\log(age)$ & 7.0, 8.0, 9.0   \\
Distance (kpc) & 0.5, 1.0, 3.0, 5.0 \\
\verb1 1 $A_v$ (mag) & 0.1, 0.5, 1.0, 3.0 \\
\hline
\end{tabular}
\end{table}

We built $V$ vs. $(B-V)$ CMDs and spatial stellar distributions for each SOC
with a limiting magnitude of $V=22\,mag$. The SOCs were generated with a tidal
radius $r_{tidal}= 250\,px$ and placed at the center of a star field region
$2048\times2048\,px$ wide. Since \texttt{MASSCLEAN} returns SOCs with no
photometric errors, the stars were randomly shifted according to the following
error distribution:

\begin{equation}
e_X = a e^{(b V)} + c
\end{equation}

\noindent where $X$ stands for either the $V$ magnitude or the $(B-V)$ color,
and the parameters $a, b, c$ are fixed to the values
$2{\times}10^{-5}, 0.4, 0.015$, respectively. We also randomly removed
stars beyond $V=19.5\,mag$ in order to
mimic incompleteness effects by using the expression: 

\begin{equation}
c = 1 / (1 + e^{(i - a)})
\end{equation}

\noindent where $c$ represents the percentage of stars left after the removal
process in the $V$ magnitude bin $i$, and $a$ is a random value in the range
$[2, 4)$.

\begin{figure}[t!]
\begin{center}
\includegraphics[width=\columnwidth]{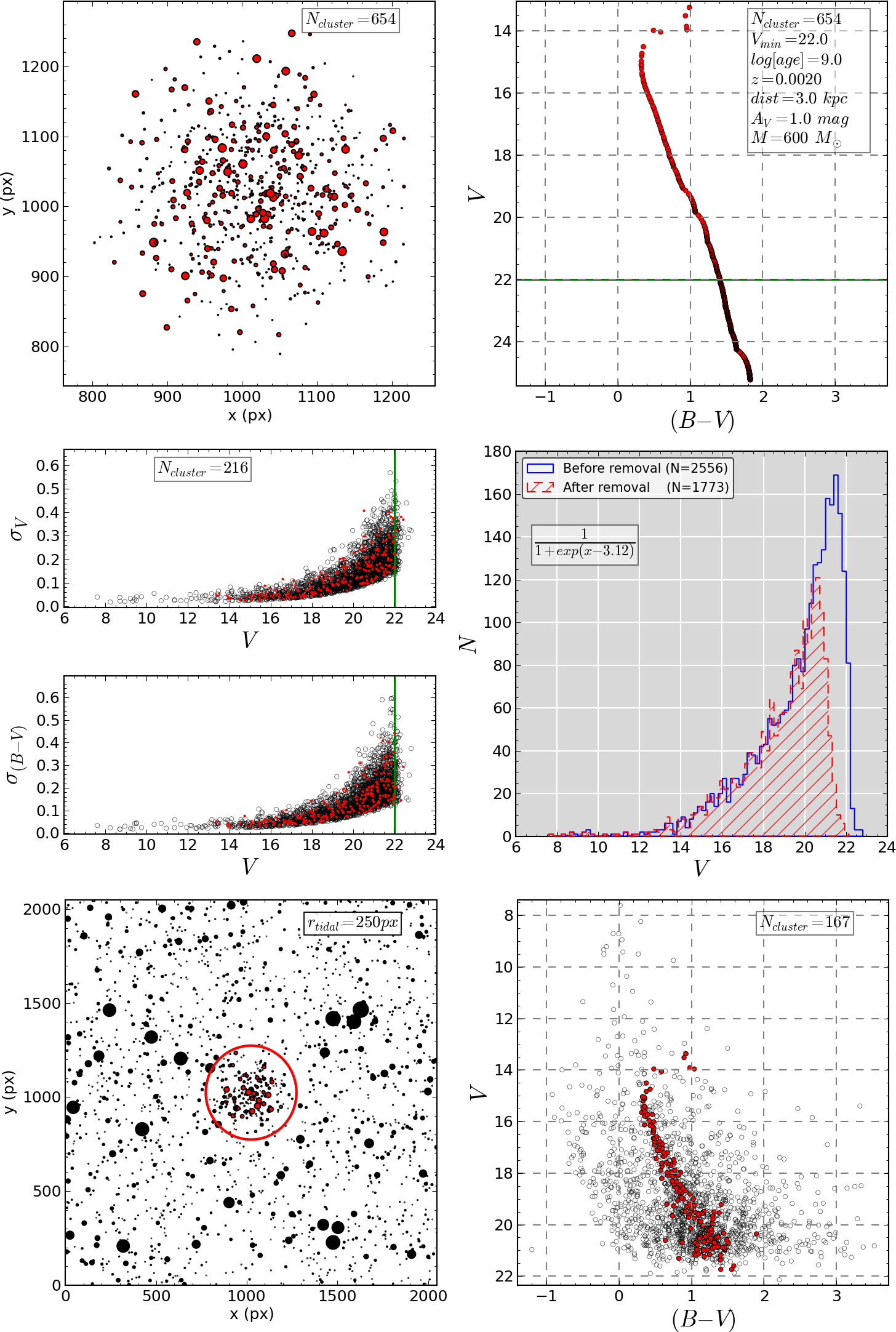}
\caption{\textit{Top}: Spatial distribution (left) of a SOC according to the
parameters labeled in the respective CMD (right), wherein the green line
indicates the magnitude limit adopted in the validation. Red symbols refer to
cluster stars. \textit{Middle}: Photometric errors (left) and the completeness
function (right) adopted. \textit{Bottom}: resulting cluster star spatial
distribution (left) with a circle representing the tidal radius and the
corresponding CMD (right).
\label{fig:err-comp-mass}}
\end{center}
\end{figure}

\begin{figure*}[ht]
\begin{center}
% One-column
% \includegraphics[width=1.2\columnwidth]{socs-examples.png}
% Two-columns
\includegraphics[width=2.\columnwidth]{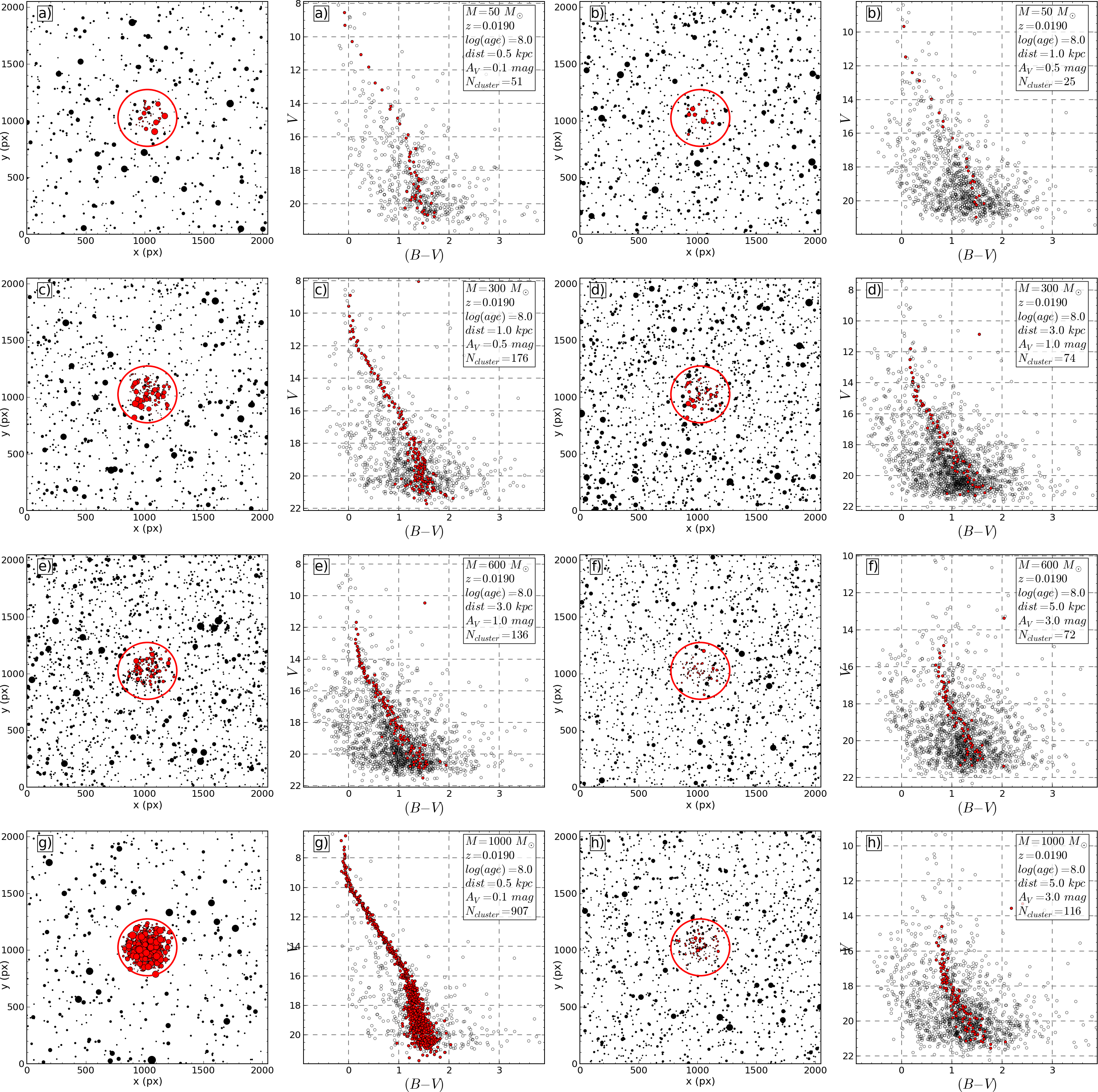}
\caption{Examples of SOCs produced as described in Sect. \ref{sec:validation}
with fixed metallicity and age and varying total mass, distance and visual
absorption values.
Symbols as in Fig. \ref{fig:err-comp-mass} (bottom panels).
\label{fig:socs-examples}}
\end{center}
\end{figure*}

Fig. \ref{fig:err-comp-mass} depicts an example for a $600\,M_{\odot}$ SOC.
The upper panels show the initial spatial stellar distribution (left) and CMD
(right), respectively; the middle panels illustrate the behavior of the error
distribution and incompleteness functions, while the bottom panels show the
resulting cluster star field and the respective CMD. Likewise,
Fig \ref{fig:socs-examples} shows eight examples of SOCs generated with
different masses and field-star contamination.

\subsection{\texttt{ASteCA} test with synthetic clusters}
\label{sec:resul-socs}

We applied \texttt{ASteCA} for the 432 SOCs in automatic mode, which means that
no initial values were given to the code with the exception of the ranges where
the GA should look for the optimal cluster parameters, shown in Table
\ref{tab:ga-range}.
The relations connecting $V$, $(B-V)$, visual absorption and distance are
of the form:

\begin{equation}
\begin{aligned}
V &= M_V -5 + 5 \log(d) + A_v \\
A_V &= R_V E_{B-V} \\
(B-V) &= (B-V)_0 + E_{B-V}
\end{aligned}
\label{eq:ext-rel}
\end{equation}

\noindent where $M_V$, $(B-V)_0$ are the absolute magnitude and intrinsic
color taken from the theoretical isochrone, $d$ is the distance in parsecs,
$A_V$ is the visual absorption and the extinction parameter is fixed
to $R_V=3.1$.
In its current version \texttt{MASSCLEAN} uses the \cite{Marigo_2008} \&
\cite{Girardi_2010} Padova isochrones, which adopt a solar metallicity of
$z_{\odot}{=}0.019$.

The results obtained by \texttt{ASteCA} are combined with the true
values used to generate the SOCs in the sense \textit{true value} minus
\textit{\texttt{ASteCA} value}:

\begin{equation}
\Delta param = param_{true} - param_{asteca}
\label{eq:delta-param}
\end{equation}

\noindent for the radius and the four cluster parameters, and the
resulting deltas compared against the CI (defined in Sect. \ref{sec:cont-ind})
assigned to the SOCs.
We choose to use the CI not only because it is simple to obtain
and useful to asses the field star contamination, but also because it can be
easily calculated even for observed clusters, i.e., there is no requirement to
know in advanced the true members of a cluster and the field stars within its
defined region to obtain its CI.
This independence from a priori unknown factors, in contrast for example with
the similar \textit{contamination rate} parameter defined in KMM14, means we
can use the CI to extrapolate, with caution, the limitations and strengths of
\texttt{ASteCA} gathered using SOCs, to observed OCs.
In some of the plots below the natural logarithm of the CI is used instead,
to provide a clearer graphical representation of the results.

The SOCs generated via the \texttt{MASSCLEAN} package act as substitutes for
observed OCs whose parameter values are fully known, and must not be confused
with the synthetic clusters generated internally by \texttt{ASteCA} as part
of the best model fitting method (Sect. \ref{sec:bf-method}).
In what follows the former will always be referred to as SOCs and the
latter either as ``synthetic clusters'' or ``models''.

\begin{table}[t]
\centering
\caption{Ranges used by the GA algorithm when analyzing \texttt{MASSCLEAN}
SOCs. The last column shows the number of values used for each
parameter for a total of $\sim3.4{\times}10^6$ possible models.}
\label{tab:ga-range}
\begin{tabular}{lcccc}
\hline
\hline\\[-1.85ex]
 Parameter & Min value & Max value & Step & N\\
\hline\\[-1.85ex]
$Metallicity\,(z)$ & 0.0005 & 0.03 & 0.0005 & 60\\
$log(age)$ & 6.6 & 10.1 & 0.05 & 70\\
$E_{B-V}$ & 0.0 & 1.5 & 0.05 & 30\\
$Distance\,modulus$ & 8.5 & 14. & 0.2 & 27\\
\verb1  1 $d\,(kpc)$ & 0.5 & 6.3 & & \\
\hline
\end{tabular}
\end{table}

\subsubsection{Structure parameters}
\label{sec:mass-center-rad}

As a first step we analyze the center and radius determination functions, which 
are based on spatial information exclusively.
To generate the final SOCs the original \texttt{MASSCLEAN} clusters have a
portion of their stars removed via the maximum magnitude cut-off
and incompleteness functions, as shown in Fig. \ref{fig:err-comp-mass}.
The original center and radius values used to
create them will clearly be affected by both processes, so an intrinsic
scatter around these true values is expected independent from the behavior of
\texttt{ASteCA}.

\begin{figure}[t!]
\begin{center}
\includegraphics[width=\columnwidth]{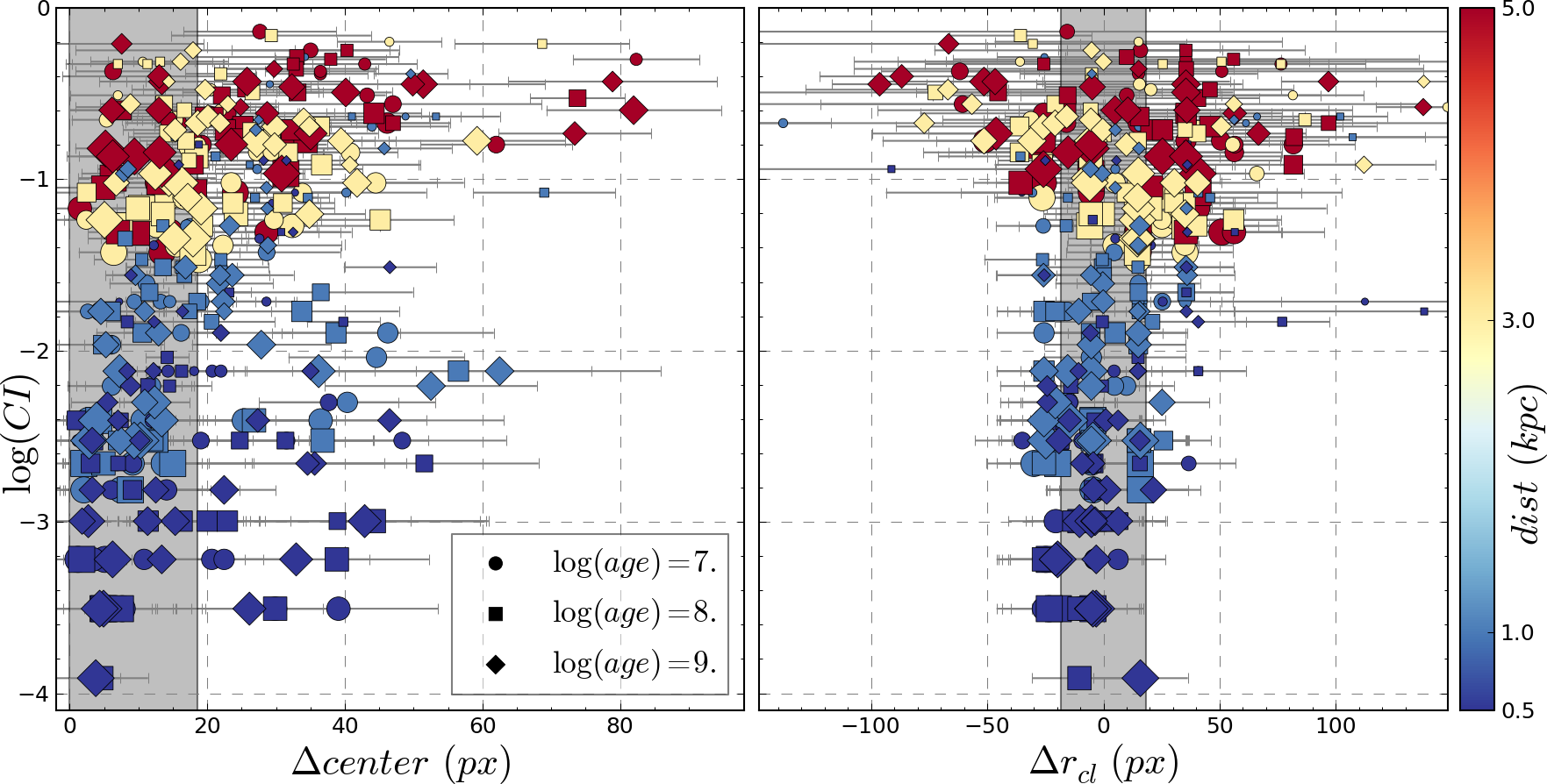}
\caption{\textit{Left}: Logarithm of the CI vs difference in center assigned
by \texttt{ASteCA} with true center. Sizes vary according to the initial masses
(larger mean larger initial mass) and colors according to the distance
(see colorbar to the right) \textit{Right}: Logarithm of the CI vs radius
difference (\texttt{ASteCA} minus true value) for each SOC. In both plots
shaded regions mark the range where $50\%$ of all SOCs are positioned and
estimated errors are shown as gray lines.
\label{fig:mass-cent-rad}}
\end{center}
\end{figure}

Fig. \ref{fig:mass-cent-rad} shows to the left the distribution of the
distances to the true center $(1024, 1024)\,px$ of the central coordinates found
by the code $(x_o, y_o)\,px$ as:

\begin{equation}
\Delta center = \sqrt{(x_o - 1024)^2 + (y_o - 1024)^2}
\end{equation}

\noindent for all SOCs located closer than $90\,px$ away from the true center,
that is almost $90\%$ (386) out of the 432 SOCs processed.
The remaining 46 SOCs that were given center values farther away have on
average less than 15 members and CI values larger than 0.7, thus
the difficulty in finding their true center.
Almost half of the sample was positioned less than $20\,px$ away from the
true center (shaded region in the figure), which represents $1\%$ of the
frame's dimension in either coordinate.

To the right of Fig. \ref{fig:mass-cent-rad} the delta difference for the
radius is shown, as defined in Eq. \ref{eq:delta-param}, only for the
sub-sample of 386 SOCs whose central coordinates
were positioned closer than $90\,px$ from the true center.
In this case $50\%$ of the sub-sample presented 
$r_{cl}$ radius values less than $19\,px$ away from the true value used to 
generate the SOCs ($r_{tidal}=250\,px$), with $90\%$ of the sub-sample found to 
have values less than $57\,px$ away. Considering the frame has an area of 
$2048\times2048\,px$, these results are quite good.

As stated in Sect. \ref{sec:rad-determ}, fitting a 3-parameters King profile 
is not always possible, especially for clusters that show a low density 
contrast with their surrounding fields. About $76\%$ of the above mentioned 
sub-sample of SOCs could have their 3P King profile calculated of which only 2 
converged to values $r_{tidal}<300\,px$ with approximately $\sim80\%$ returning 
values $r_{tidal}>400\,px$. This overestimation of the tidal radius is due to 
the high dependence of the fitting process with the shape of the RDP, 
particularly for SOCs with low members counts. Internal tests showed that 
modifying the bin width of the 2D histogram used to obtain the RDP (see Sect. 
\ref{sec:rad-determ}) by even $1\,px$ can cause the 3P King profile to 
converge to a significantly different $r_{tidal}$ value. Care should be 
exercised thus when utilizing or interpreting the structure of an OC based on 
this parameter.

A clear correlation can be seen in Fig. \ref{fig:mass-cent-rad} where the
dispersion in both distance to the true center and departure from the true
cluster radius increase for larger CI values.

\subsubsection{Probabilities and members count}
\label{sec:mass-probs-memb}

\begin{figure}[t!]
\begin{center}
\includegraphics[width=\columnwidth]{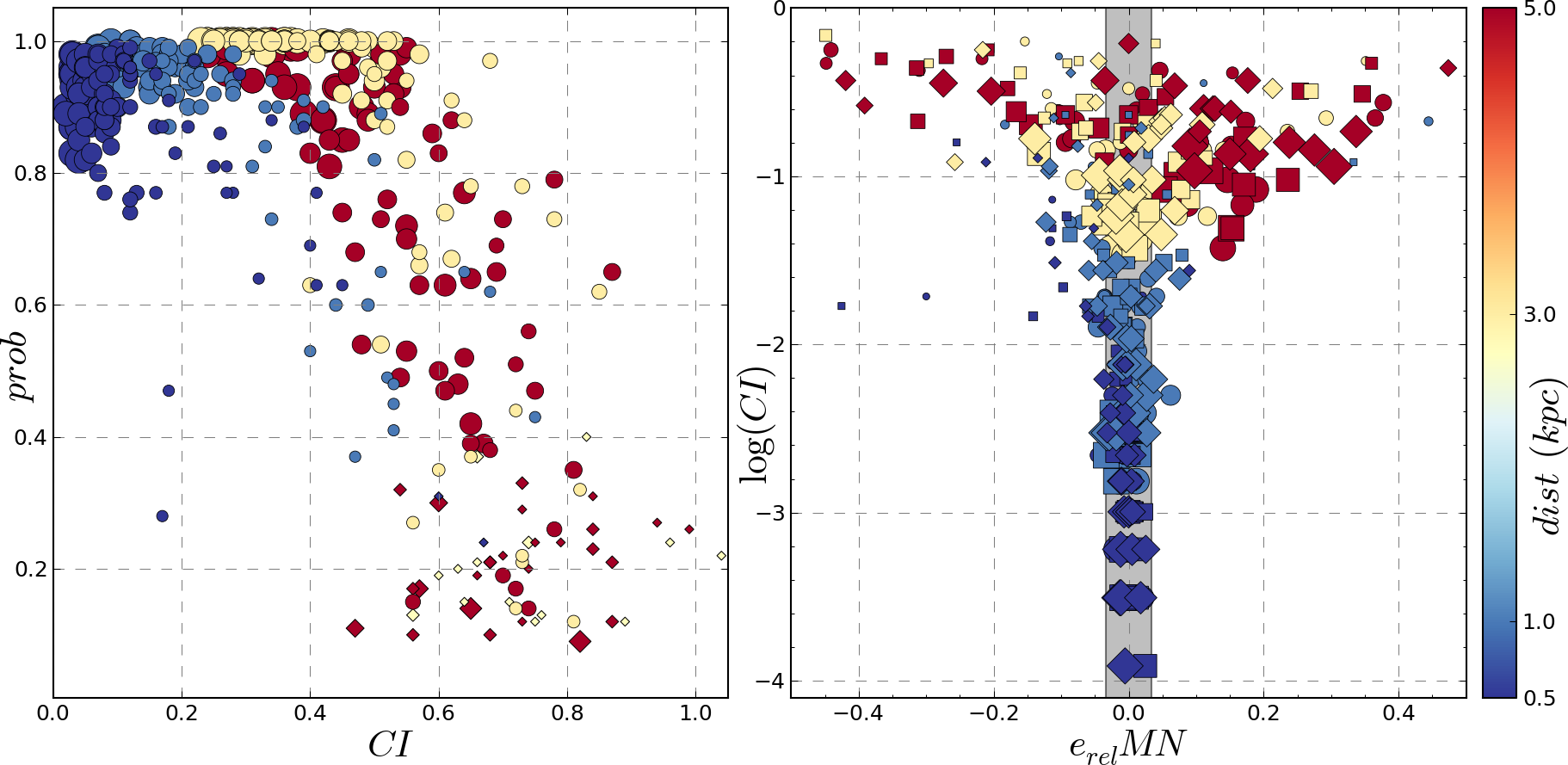}
\caption{\textit{Left}: distribution of real cluster probabilities vs CI.
Sizes vary according to the initial masses (bigger size means larger initial
mass) and colors according to the distance (see colorbar to the right). Circles
represent values for SOCs and diamonds values obtained for field regions.
\textit{Right}: Logarithm of the CI vs relative error for the true number of
members and the one predicted by \texttt{ASteCA} for each SOC. Point markers
are associated with ages as shown in the legend. The shaded region marks the
range where $50\%$ of all SOCs are positioned.
\label{fig:mass-pval-memb}}
\end{center}
\end{figure}

The probability of being a real star cluster rather than an artificial grouping
of field stars is assessed by the function described in Sect. \ref{sec:pvalue}. 
For each SOC whose central coordinates were found within a $90\,px$ radius
of the true central coordinates ($90\%$ of the SOCs, as stated in Sect.
\ref{sec:mass-center-rad}) their probability value can be seen to the left
of Fig. \ref{fig:mass-pval-memb} as filled circles.
Diamonds represent those 46 instances were the center was detected far
away from its true position meaning the comparison made to obtain
the p-values, from which the probability is derived, was done contrasting
mostly (and sometimes entirely) a field region with other field regions.
Expectedly, the probabilities found in these cases are very low having an
average value of $\sim0.2$.
A few low mass distant SOCs with an average number of members of 33 can also
be found in this region of low probability. This is unavoidable since high
field-stars contamination means that effectively separating true star
clustering from random overdensities is not a simple task, especially for
those systems with a very low number of members.
Nevertheless, the large majority of SOCs are assigned
high probability values which points to the function being reliable even for
clusters with high CIs and relatively few true members. In what follows these
46 ``clusters'' that were positioned far away from the true center, and are thus
almost entirely composed of field stars, are dropped from the analysis.

To the right of Fig. \ref{fig:mass-pval-memb} the relative error for the number
of members is shown following the relation:

\begin{equation}
e_{rel}\,MN = \frac{mn_{asteca} - mn_{true}}{mn_{true}}
\end{equation}

\noindent where $mn_{true}$ and $mn_{asteca}$ are the true number of members
and the number of members calculated by \texttt{ASteCA}. Half of the SOCs had
their number of members estimated within a $3\%$ relative error (shaded region)
and over $80\%$ of them within a $20\%$ relative error.
The error can be clearly seen to increase primarily with the CI but also with
age given that older SOCs usually have fewer true members making them more
susceptible to field-stars contamination.

\subsubsection{Decontamination algorithm}
\label{sec:decontam-algor}

To study the effectiveness of the decontamination algorithm (DA), described in
Sect. \ref{sec:field-star-cont}, in assigning membership probabilities (MP) to
true cluster members, we define two \textit{member index} (MI) relations as:

\begin{equation}
MI_1 = n_m /N_{cl} \;\;\;|\, p_m > 0.9
\label{eq:mi-1}
\end{equation}

\begin{equation}
MI_2 = \frac{\sum\limits^{n_m} p_m - \sum\limits^{n_f} p_f}{N_{cl}}
\label{eq:mi-2}
\end{equation}

\noindent where $n_m$ and $n_f$ are the number of true cluster members and of
field stars bound to be present within the cluster region (i.e.: inside the
boundary defined by the cluster radius), $p_m$ and $p_f$ are the MPs assigned
to each of them by the DA and $N_{cl}$ is the total number of true cluster
members.
The index in Eq. \ref{eq:mi-1}, $MI_1$, can be thought of as an equivalent to 
the True Positive Rate ($TPR$) defined in KMM14 as
\textit{``the ratio between the number of real cluster members in the high
probability subset and its total size''}, where the high probability subset
contains only true members with assigned probabilities over $90\%$ ($TPR_{90}$).
The maximum value for this index is 1, 
attained when all true members are recovered with $MP>90\%$.
The index $MI_2$ defined in Eq. \ref{eq:mi-2} rewards high probabilities given 
to true members but also punishes high probabilities given to field stars. The 
optimal value is 1, achieved when all true cluster members are identified as 
such with MP values of 1, and no field star is assigned a probability higher 
than 0. The index $MI_2$ dips below zero when the added probabilities of field 
stars is larger than that of true members. In this case the DA can be said to 
have ``failed'' since it assigned a greater combined MP to non-member (field) 
stars than to true cluster members.

\begin{figure}[t!]
\begin{center}
\includegraphics[width=\columnwidth]{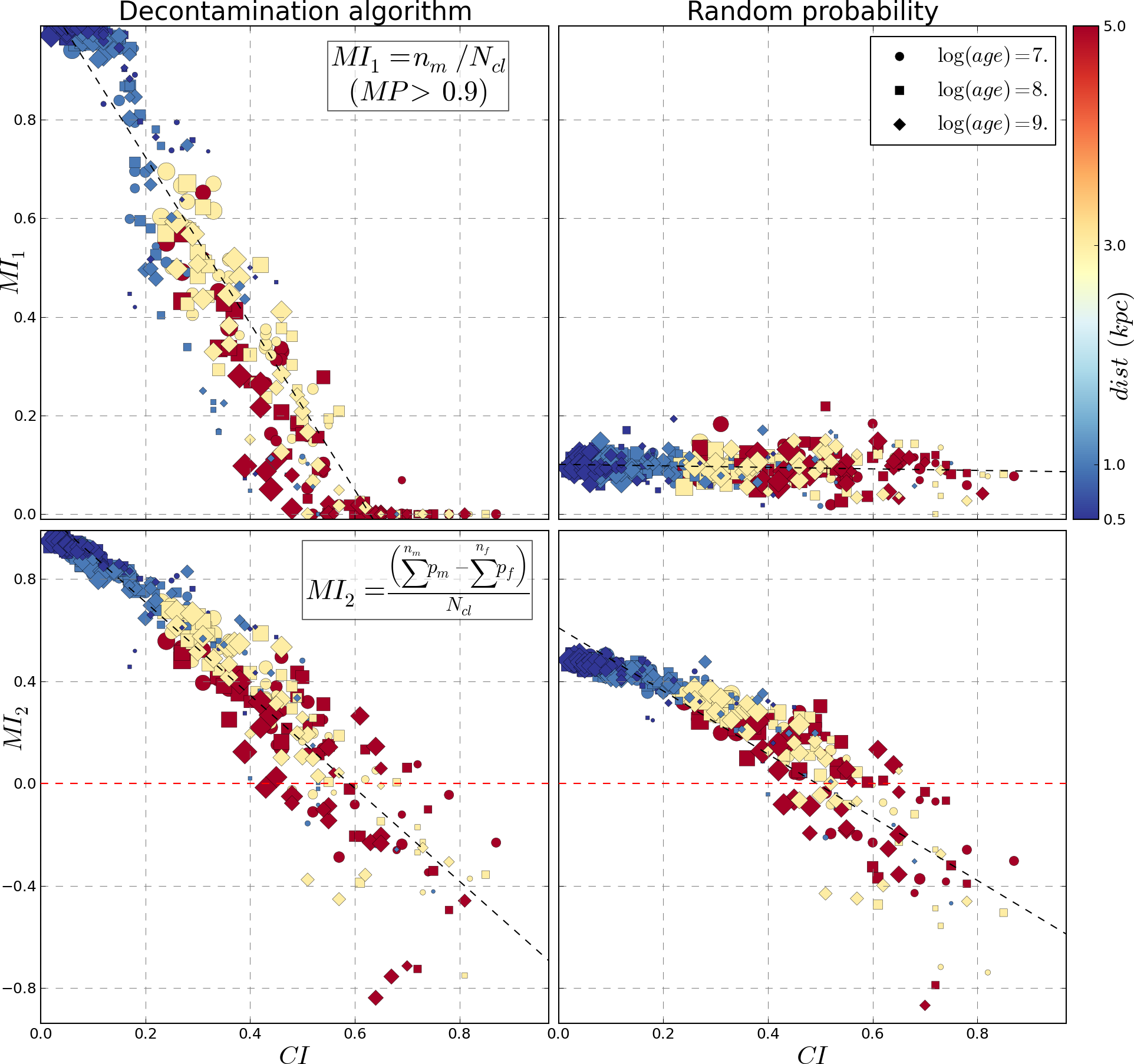}
\caption{\textit{Top left}: Member index defined in Eq. \ref{eq:mi-1} vs
contamination index. \textit{Top right}: same MI vs CI plot but using the
membership probabilities given by the random assignment algorithm.
\textit{Bottom left}: Member index defined in Eq. \ref{eq:mi-2} vs
contamination index. \textit{Bottom right}: same MI vs CI plot but using the
membership probabilities given by the random assignment algorithm. Linear
fits in each plot shown as dashed black lines. Sizes, shapes and colors as
in Fig. \ref{fig:mass-cent-rad}.
\label{fig_mass_decont_algor}}
\end{center}
\end{figure}

To provide some context for the results we compare both MIs obtained by the DA 
for each SOC with those returned by an algorithm of random probabilities 
assignation. The latter randomly assigns MPs uniformly distributed between $[0, 
1]$ to all stars within the cluster region of a SOC.
As can be seen in Fig. \ref{fig_mass_decont_algor} the DA behaves very good up 
to a CI value of 0.4 where it shows a value of $\overline{MI_1}\approx0.78$ 
(that is: an average $\sim78\%$ of true members being recovered with MPs higher 
than $90\%$) and $\overline{MI_2}\approx0.77$ which means field stars do not 
play an important role thus far. Between the range $CI=(0.4,0.6)$ the MIs 
obtained for the DA begin looking increasingly similar to those obtained for 
the random probability assignation algorithm. This is noticeable especially for 
$MI_2$ where the increased presence of field stars starts having a larger 
influence in its value. Beyond a CI of approximately 0.6 the DA breaks down as 
it apparently no longer represents an improvement over assigning MPs randomly.

\subsubsection{Cluster parameters}
\label{sec:cluster-parameters}

The results for all cluster parameters are presented here for the entire set
of SOCs, with the exception of those low-mass and highly contaminated systems
that were rejected because the center finding function assigned their position
far away from the real cluster center and are thus mainly a grouping of field
stars (46 as stated in Sect. \ref{sec:mass-center-rad}).

At this point it is important to remember that the accuracy of the
results obtained for the cluster parameters depends not only on the 
correct working of the methods written within the code,
but also on the intrinsic limitations of the photometric system selected
and the type and number of filters chosen (see \citealp{Anders_2004} and
\citealp{deGrijs_2005} for a discussion applied to the recovery of cluster
parameters based on fitting observed spectral energy distributions).
In our case, as stated at the beginning of Sect. \ref{sec:validation}, we
opted to generate the SOCs for the validation process using only
the $VB$ bands of the Johnson system provided by \texttt{MASSCLEAN} to
construct simple $V$ vs $(B-V)$ CMDs. This not only means we are relying on a
very reduced space of ``observed'' data (two-dimensional), but also
that the resolution power of the analysis is necessarily limited by our
selection of filters.
The presence of a third band, particularly one below or encompassing the
Balmer jump, would provide a CMD packed with more photometric information and
quite possibly help reduce uncertainties in general.
The $U$ and $B$ filters of the Johnson system for example can be combined to
generate the $(U-B)$ index, known to be sensitive to metal abundance, while
bands located toward the infrared part of the spectrum are less affected
by interstellar extinction.
It is also worth stressing that the isochrone matching method is an inherently
stochastic process. Even if the SOCs were generated without
errors and incompleteness perturbations as isochrones populated via a given
IMF, the randomness involved in producing the synthetic clusters used to
match the best model, as explained in Sect. \ref{sec:bf-method},
would introduce an inevitable degree of inaccuracy in the final
cluster parameters.

\begin{figure}[t!]
\begin{center}
\includegraphics[width=\columnwidth]{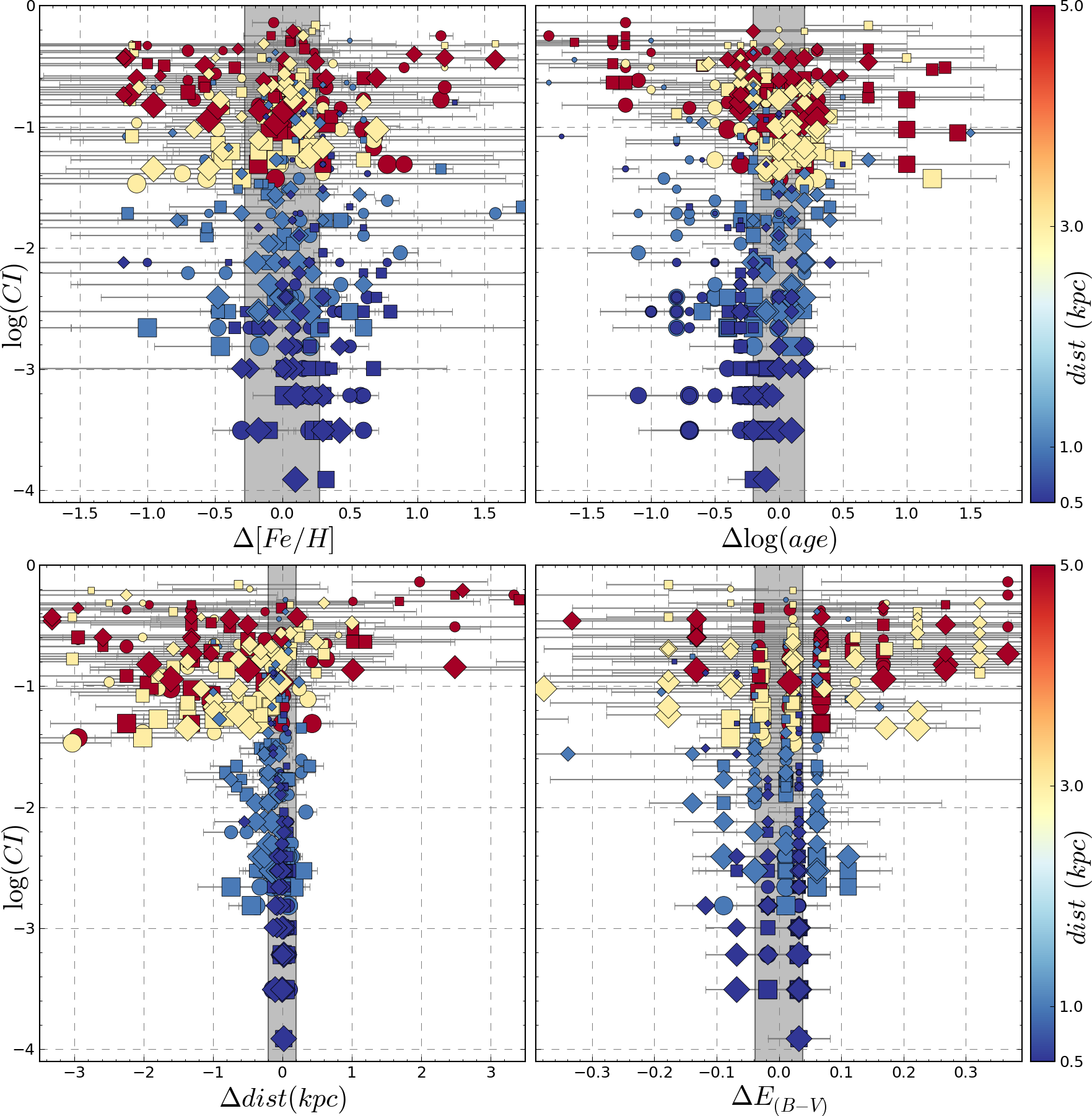}
\caption{\textit{Top left}: Metallicity differences in the sense true value
minus \texttt{ASteCA} estimate vs $\log(CI)$ with errors assigned by the code
as gray horizontal lines. \textit{Top right}: Idem for $\log(age)$.
\textit{Bottom left}: Idem for distance in kpc. \textit{Bottom right}: Idem
for $E_{B-V}$ extinction. Shaded regions mark the ranges where $50\%$ of
all SOCs are positioned. Sizes, shapes and colors as in
Fig. \ref{fig:mass-cent-rad}.
\label{fig:mass-params}}
\end{center}
\end{figure}

The plots in Fig. \ref{fig:mass-params} show the dependence of the differences
between the true value used to generate the SOCs and those estimated by
\texttt{ASteCA} ($\Delta param$) with the CI, for the four cluster
parameters.
The dispersion in the delta values tends to increase with the CI as do the
errors with which these parameters are estimated by the code (horizontal lines).

The metallicity is converted from $z$ to the more usual $[Fe/H]$ applying the 
standard relation $[Fe/H] = \log(z/z_{\odot})$ where $z_{\odot}=0.019$, and 
contains about $25\%$ of the sample below the largely acceptable error limit of 
$\pm0.1\,dex$ with $50\%$ showing errors under $\sim0.28\,dex$ (shaded 
region in top left plot of Fig. \ref{fig:mass-params}). There are no 
noticeable biases in the metallicities assigned by \texttt{ASteCA} although the 
dispersion increases rapidly even for SOCs with low CI values.

\begin{figure}[t!]
\begin{center}
\includegraphics[width=\columnwidth]{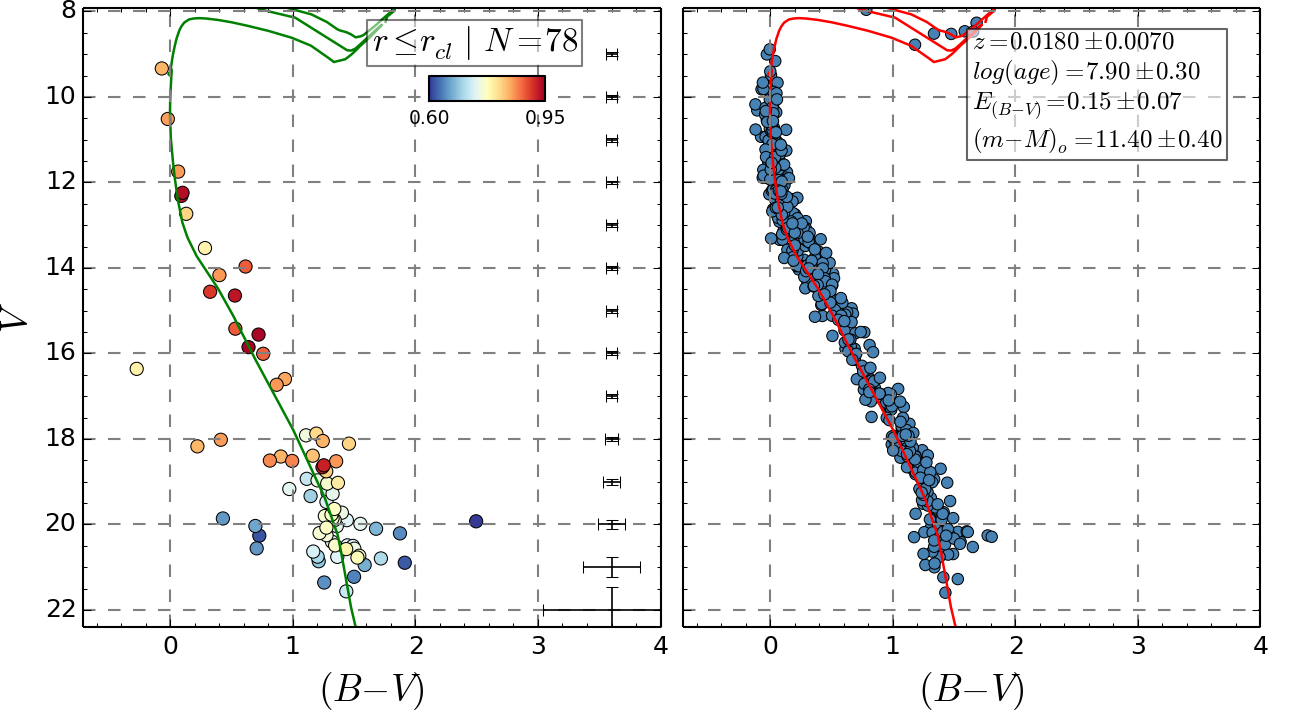}
\caption{\textit{Left}: CMD of cluster region for a \texttt{MASSCLEAN} young
SOC with MPs for each star shown according to the color bar at the top and
the best fit isochrone found shown in green. True age value is $\log(age)=7.0$.
\textit{Right}: best fit
synthetic cluster found along with the theoretical isochrone used to generate
it in red, cluster parameters and uncertainties shown in the top right box.
\label{fig:soc-young-age}}
\end{center}
\end{figure}

The top right plot in Fig. \ref{fig:mass-params} shows half of the
sample located within $\pm0.2$ of their true $\log(age)$ values (shaded region).
Many young SOCs with low CIs are assigned higher ages by \texttt{ASteCA} than
those they were created with. This effect arises from the difficulty in
determining the location of the TO point for these young clusters that have no
evolved stars in their sequences. An example is shown in Fig.
\ref{fig:soc-young-age} for a SOC of $\log(age)=7.0$ where the age is
recovered with a substantial error ($0.9$) even though the isochrone displays
a very good fit.

Taking the subset of 132 analyzed young SOCs with $\log(age)=7$, we find
that only $17\%$ of them (23) had assigned ages that differ more than $\Delta
\log (age) {>}1$ with their true age values.
This is the result of bad isochrone assignations, owing primarily to the
combination of high field star contamination ($CI{\simeq}0.5$, on average) and
low total masses ($\sim260\,M_{\odot}$, on average) of these clusters.
The remaining $83\%$ of young clusters presented on average $\Delta \log(age)
{\simeq}0.36$, distributed as follows: almost half (62) were given ages
deviating less than $\log(age){\simeq}0.3$ from their true values, a third (43)
showed $\Delta \log(age){\leq}0.2$ and a fourth (33) presented
$\Delta \log(age){\leq}0.1$, or less than $26\%$ relative error for the age
expressed in years, which is quite reasonable.
These results contrast with those obtained for the 254 analyzed older clusters
($\log(age)=[8,9]$) which show that $80\%$ of them have their true age values
recovered within $\Delta \log(age){\leq} 0.3$, with an average deviation
of  $\Delta \log(age){\simeq} 0.16$.

Both the distance and extinction parameters, Fig. \ref{fig:mass-params} bottom 
row, are recovered with a much higher accuracy, especially for SOCs with lower 
CI values. In the case of the distance half the sample is within a 
$\pm0.2\,kpc$ range from the true values (shaded region) and $\sim77\%$ of the
sample within $\pm1\,kpc$.
The dispersion in the $E_{B-V}$ color excess positions half of the 
sample below $\pm0.04\,mag$ (shaded region) and almost $90\%$
below $\sim0.2\,mag$. 
There seems to be a correlation in the portion of SOCs with high CI, 
where clusters appear to be located simultaneously at larger distances and with
lower extinction than their true values. Upon closer inspection we see that
this is not the case, as shown by the positive correlation coefficient
found between these two parameters.

The full correlation matrix (covariance matrix normalized by the standard
deviations) for the deltas of all cluster parameters can be seen in Table
\ref{tab:corr-matr}.
\begin{table}[tb]
\centering
\caption{Correlation matrix for the deltas defined for each cluster
parameter.}
\label{tab:corr-matr}
\begin{tabular}{lcccc}
\hline
\hline\\[-1.85ex]
$\Delta param$ & $\Delta[Fe/H]$ & $\Delta \log(age)$ & $\Delta dist$ & $\Delta E_{B-V}$ \\
\hline\\[-1.85ex]
$\Delta[Fe/H]$ &  1. & -0.13 & 0.37  & -0.21 \\
$\Delta \log(age)$ &  -- & 1.    & -0.38 & -0.48 \\
$\Delta dist$ &  -- & --    & 1.    & 0.40 \\
$\Delta E_{B-V}$ &  -- & --    & --    & 1. \\
\hline
\end{tabular}
\end{table}
The departures from the true distance and color excess values
($\Delta dist$ \& $\Delta E_{B-V}$) have no negative correlation but in
fact a small positive one (0.4), meaning that as one increases so does the other.
It is worth noting the small negative correlation value found between
metallicity and age, which points at a successful lifting of the
age-metallicity degeneracy problem by the method.
The well-known age-extinction degeneracy, whereby a young cluster affected by
substantial reddening can be fitted by an old isochrone with a small amount
of reddening, stands out as the highest correlation value
\citep[HVG08,][]{Meulenaer_2013}.
The positive correlation between metallicity values and the distance
has also been previously mentioned in the literature
\citep[e.g.:][]{Hasegawa_2008}.

\subsubsection{Limitations and caveats}
\label{sec:limit-caveats}

\begin{figure}[t!]
\begin{center}
\includegraphics[width=\columnwidth]{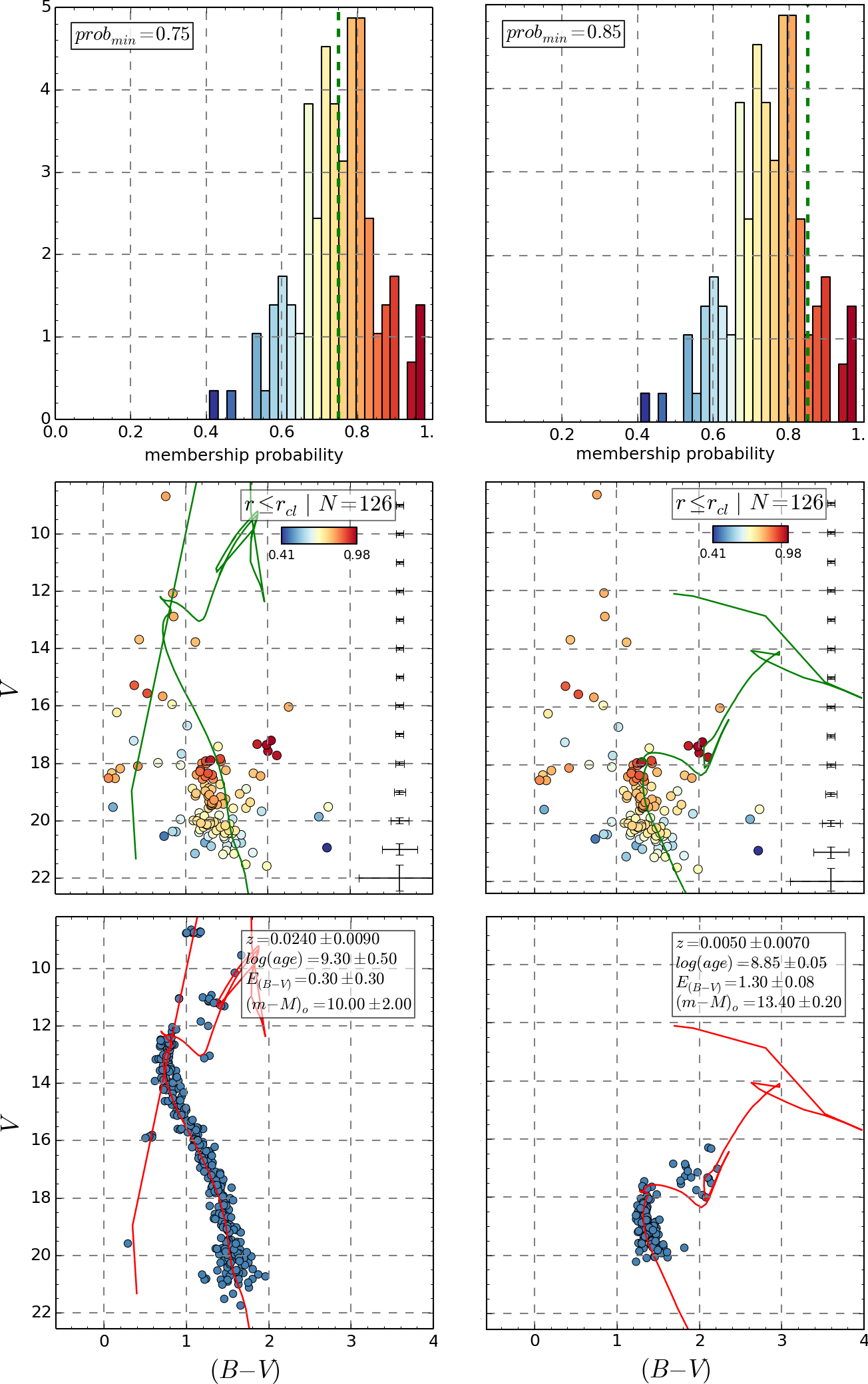}
\caption{\textit{Left column}: top, MPs distribution for an old SOC where the
cut is done for a value of $prob_{min}=0.75$; center, CMD of cluster region
with MPs for each star shown according to the color bar at the top and the
best fit isochrone found shown in green; bottom, best fit synthetic cluster
found along with the theoretical isochrone used to generate it in red, cluster
parameters and uncertainties shown in the top right box. \textit{Right column}:
top, idem above with probability cut now done at $prob_{min}=0.85$; center,
idem above with new best fit isochrone; bottom, idem above showing new
synthetic cluster and improved cluster parameters.
\label{fig:reduc-memb}}
\end{center}
\end{figure}

An important source responsible for inaccuracies when recovering the SOCs 
cluster parameters, mainly for those in the high CI range, comes as a 
consequence of the way the theoretical isochrones are employed in the synthetic 
cluster fitting process. In HVG08 the authors limit the isochrones to below the 
helium flash to avoid issues with non-linear variations in certain regions of 
the CMD. Unlike this work we chose to avoid ad-hoc cuts in the set of 
theoretical isochrones and instead use their entire lengths to generate the 
synthetic clusters, which means using also their evolved parts. Fig. 
\ref{fig:reduc-memb} shows how this affects the way synthetic clusters are 
fitted to obtain the optimal cluster parameters, for a SOC of $\log(age)=9$ 
located at $5\,kpc$ with a color excess of $E_{B-V}\simeq0.97\,mag$.
The left column of the figure shows at the top the distribution of MPs found
by the decontamination algorithm with the probability cut used by the GA made
at $prob_{min}=0.75$. In the middle left portion the CMD of the cluster region
is shown along with the best matched theoretical isochrone found using this
MP minimum value and the synthetic cluster generated from it at the bottom.
As can be seen the highly evolved parts of the isochrone are being populated
which means these few stars will also be matched with those from the SOC with
MPs above the mentioned limit, thus forcing the match toward very unreliable
cluster parameters.
In such a case, a simple solution is to increase the minimum MP value 
used by the genetic algorithm to obtain the best observed-synthetic cluster 
fit, as shown in the right column of Fig. \ref{fig:reduc-memb}.
Here we were able to change the unreliable cluster parameters obtained for
the SOC using a minimum MP value of $prob_{min}=0.75$, to a very good fit
with small errors by simply increasing said value to $prob_{min}=0.85$.
This slight adjustment restricts the SOC stars used in the fitting process
to those in the top $15\%$ MP range, which translates to a much more reasonable
isochrone matching and improved cluster parameters as seen in the middle and
bottom plots of the right column.
Were this process to be replicated in the full SOCs sample, we would surely see 
an overall improvement in the determination of the cluster parameters.\\

\begin{table}[tb]
\centering
\caption{Number of poorly populated ($N_{memb}\leq50$) SOCs that had their
parameters recovered with relative errors ($e_r$) below the given thresholds;
percentage of total sample (82) is shown in parenthesis. The sample also
represents the most heavily contaminated SOCs (CI>0.5).}
\label{tab:rel-errors}
\begin{tabular}{lccc}
\hline
\hline\\[-1.85ex]
 Parameter & $e_r<100\%$ & $e_r<50\%$ & $e_r < 25\%$ \\
\hline\\[-1.85ex]
$Metallicity\,(z)$ &  49 (60\%) & 37 (45\%) & 21 (26\%) \\
$Age\, (yr)$       &  54 (66\%) & 27 (33\%) & 13 (16\%) \\
$Distance\, (kpc)$ &  79 (96\%) & 59 (72\%) & 49 (25\%) \\
$E_{B-V}$          &  76 (93\%) & 53 (65\%) & 42 (51\%) \\
\hline
\end{tabular}
\end{table}

Clusters with a scarce number of stars and high field star
contamination are particularly difficult to analyze for obvious reasons.
The subset of our sample of SOCs containing the most poorly populated clusters
coincides with the highest contaminated SOCs analyzed: 82 clusters averaging 
less than 50 true members each with $CI{>}0.5$ (ie: more field stars than true
members within the cluster region) .
Table \ref{tab:rel-errors} shows the relative errors (defined as
[True value - \texttt{ASteCA} value] / True value) obtained in this subset
for each parameter. While distance and extinction display the best match with
the true values, we see that the age is the most affected parameter when
analyzing clusters with very few members.
Still, we find that over half of this subset shows absolute age errors below
$\log(age){=}0.3$, which is largely acceptable.
The 40 clusters with the largest age relative errors ($e_r>90\%$) show no
definitive tendency towards any particular age, being distributed as 16, 15
and 9 clusters for $\log(age)$ values of 7, 8 and 9 respectively.

\begin{table*}[htb]
\centering
\caption{Summary of validation results. Parameter values in the top row are
those used to generate the \texttt{MASSCLEAN} SOCs. The rest of the rows show
the mean and standard deviation of the values assigned by \texttt{ASteCA} to
the set of SOCs with that parameter value located in the CI range shown in
the first column.}
\label{tab:mass-met-age}
\begin{tabular}{c|cccc|ccc}
\hline
\hline\\[-1.85ex]
CI & \multicolumn{4}{c|}{Metallicity ($z\times1e02$)} & \multicolumn{3}{c}{log(age)} \\
 & 0.2 & 0.8 & 1.9  & \multicolumn{1}{c|}{3.0} & 7.0 & 8.0 & 9.0 \\
\hline\\[-1.85ex]
$<0.1$       & $0.4\pm0.4$ & $0.6\pm0.5$ & $1.2\pm0.6$ & $2.2\pm0.7$ & $7.6\pm0.3$ & $8.2\pm0.2$ & $9.0\pm0.1$\\
$[0.1, 0.2)$ & $0.8\pm0.9$ & $1.2\pm0.9$ & $1.4\pm0.8$ & $1.9\pm0.9$ & $7.5\pm0.4$ & $8.2\pm0.2$ & $9.0\pm0.2$\\
$[0.2, 0.3)$ & $0.6\pm0.7$ & $1.3\pm0.9$ & $2.1\pm0.3$ & $1.6\pm1.0$ & $7.2\pm0.4$ & $7.8\pm0.4$ & $8.9\pm0.2$\\
$[0.3, 0.4)$ & $0.9\pm1.0$ & $1.5\pm0.9$ & $1.3\pm0.7$ & $1.9\pm0.6$ & $7.3\pm0.5$ & $7.9\pm0.5$ & $8.8\pm0.4$\\
$[0.4, 0.5)$ & $1.1\pm1.1$ & $1.3\pm0.9$ & $2.0\pm0.8$ & $2.2\pm0.8$ & $7.2\pm0.4$ & $7.9\pm0.4$ & $9.1\pm0.2$\\
$[0.5, 0.6)$ & $1.3\pm0.9$ & $1.9\pm1.0$ & $1.8\pm0.8$ & $2.2\pm0.7$ & $7.6\pm0.9$ & $8.2\pm0.7$ & $9.0\pm0.3$\\
$[0.6, 0.7)$ & $2.2\pm0.7$ & $1.2\pm1.0$ & $1.2\pm0.9$ & $1.5\pm1.2$ & $7.7\pm0.6$ & $8.4\pm1.0$ & $9.0\pm0.4$\\
$[0.7, 0.8)$ & $2.2\pm0.7$ & $1.2\pm0.9$ & $1.6\pm1.1$ & $1.8\pm0.9$ & $8.9\pm0.5$ & $8.7\pm0.7$ & $8.3\pm1.0$\\
$\geq0.8$    & $2.4\pm0.0$ & $2.8\pm0.2$ & $2.2\pm0.1$ & $2.2\pm0.4$ & $8.5\pm1.0$ & $8.3\pm0.8$ & $9.2\pm0.0$\\
\hline
\end{tabular}
\end{table*}
\begin{table*}[htb]
\centering
\caption{Same as Table \ref{tab:mass-met-age} for visual absorption and distance.}
\label{tab:mass-dist-ext}
\begin{tabular}{c|cccc|cccc}
\hline
\hline\\[-1.85ex]
CI & \multicolumn{4}{c|}{$A_V$ (mag)} & \multicolumn{4}{c}{Dist (kpc)}\\
 & 0.1 & 0.5 & 1.0 & \multicolumn{1}{c|}{3.0} & 0.5 & 1.0 & 3.0 & 5.0\\
\hline\\[-1.85ex]
$<0.1$       & $0.1\pm0.1$ & $0.4\pm0.2$ & -           & -           & $0.5\pm0.1$ & $1.1\pm0.2$ & -           & -\\
$[0.1, 0.2)$ & $0.0\pm0.1$ & $0.5\pm0.2$ & -           & -           & $0.5\pm0.2$ & $1.2\pm0.3$ & -           & -\\
$[0.2, 0.3)$ & $0.2\pm0.2$ & $0.7\pm0.4$ & $1.0\pm0.3$ & $2.9\pm0.1$ & $0.6\pm0.1$ & $1.2\pm0.3$ & $4.2\pm0.7$ & $6.2\pm1.3$\\
$[0.3, 0.4)$ & $0.0\pm0.0$ & $0.4\pm0.3$ & $1.0\pm0.4$ & $2.8\pm0.1$ & $0.6\pm0.1$ & $1.2\pm0.5$ & $3.8\pm0.7$ & $5.9\pm0.7$\\
$[0.4, 0.5)$ & $0.4\pm0.1$ & $0.5\pm0.2$ & $0.9\pm0.4$ & $2.3\pm0.9$ & $0.6\pm0.0$ & $1.4\pm0.5$ & $3.7\pm0.9$ & $5.4\pm2.0$\\
$[0.5, 0.6)$ & -           & $0.5\pm0.2$ & $0.7\pm0.5$ & $2.5\pm0.8$ & -           & $1.3\pm0.4$ & $3.2\pm0.7$ & $5.6\pm1.8$\\
$[0.6, 0.7)$ & -           & $0.2\pm0.1$ & $0.6\pm0.5$ & $1.9\pm1.3$ & -           & $1.0\pm0.0$ & $3.4\pm1.4$ & $5.2\pm2.7$\\
$[0.7, 0.8)$ & -           & $0.5\pm0.0$ & $0.9\pm0.8$ & $1.5\pm1.1$ & -           & $1.0\pm0.0$ & $4.2\pm2.1$ & $3.8\pm2.2$\\
$\geq0.8$    & -           & -           & $1.2\pm0.3$ & $1.1\pm0.5$ & -           & -           & $4.3\pm1.0$ & $2.6\pm1.0$\\
\hline
\end{tabular}
\end{table*}

The results obtained in Sect. \ref{sec:cluster-parameters} are summarized in
Table \ref{tab:mass-met-age} for the metallicity and age, and Table
\ref{tab:mass-dist-ext} for the distance and visual absorption.
To create these tables we take all SOCs with the same originating value for
each parameter, group them into a given CI range and calculate the mean and
standard deviations of the values for that parameter estimated by
\texttt{ASteCA}.
Ideally, the means would equal the value used to create the SOCs with a null
standard deviation; in reality the values show a dispersion in the mean around
the true value that grows with the CI as does its standard deviation.

Both distance and extinction/absorption show very reasonable values below a
CI of 0.7. Beyond that value, ie: when the number of field stars
present in the cluster region is \textit{more than 2.3 times} the number of
expected cluster members,\footnote{This comes from the fact that the CI
can be written as: $CI=n_{fl}/(n_{fl}+n_{cl})$ where $n_{fl}$
and $n_{cl}$ are the number of field and cluster member stars expected
within the cluster region, respectively (see Eq. \ref{eq:cont-index}).
If $CI=0.7$, we get from the previous relation: $n_{fl}=2.3\, n_{cl}.$}
the accuracy diminishes as the value of these parameters increases.
The metallicity, as expected, displays the largest scatter
of the means and higher proportional standard deviation values than the rest
of the parameters. Without any extra information available, it is unnecessary
and possibly counter-productive to allow the code to search for metallicities
in such a large parameter space as we did here (60 values, see Table
\ref{tab:ga-range}). Unless a photometric system suitable for dealing with
metal abundances or a specific metallicity-sensitive color is used, it
is recommended to limit the metallicity values in the parameter space to just a
handful, enough to cover the desired range without forcing too much resolution.
Another approach could involve techniques specifically designed to
obtain accurate metallicities, like the ``differential grid'' method by
\cite{Pohnl_2010}.\footnote{This iterative semi-automatic procedure is
based on main-sequence stars and requires reliable initial values for the
cluster's age, distance and reddening along with a clean sequence of true
members, which is not always possible to obtain.}
% \footnote{Applied to photometric observations made
% with Johnson's $BV$ filters and extended in \cite{Netopil_2013} to make
% use of photometric data taken from other systems.}

As can be seen in Table \ref{tab:mass-met-age}, the code shows a tendency to
assign somewhat larger ages for younger clusters; this is due to the difficulty
in correctly locating the turn off point even when the SOC is heavily populated,
as was mentioned in the previous section.
Since the majority of young SOCs with high $\Delta \log(age)$ values had their
ages \textit{overestimated}, age values returned by the code for young clusters
with non-evolved sequences should be considered maximum estimates, especially
when high field contamination is present.

Although these results should be used with caution when dealing with real OCs,
which are usually more complicated than SOCs both in their spatial and
photometric structures, they are certainly of use in assigning levels of
confidence in the automatic analysis performed by \texttt{ASteCA}. This same
analysis could eventually be replicated for another photometric system and
CMDs if the internal accuracy for such particular case is required.

\section{Results on observed clusters}
\label{sec:oc-results}

To demonstrate the code's versatility and test how well it handles real
clusters, we applied it on 20 OCs observed in three different systems:
9 of them are spread throughout the third quadrant (3Q) of the Milky Way above
and below the plane and were observed with the $CT_1$ filters of the
Washington photometric system \citep{Canterna_1976}, 10 were observed with
Johnson's $UBV$ photometry\footnote{We made use of the
WEBDA database (accessible at \url{http://www.univie.ac.at/webda/}) to retrieve
the photometry for these clusters.} and are scattered throughout the Galaxy.
The remaining one is the template NGC 6705 (M11) cluster located in the first
quadrant, with photometry taken from its 2MASS $JH$ bands
\citep{Skrutskie_2006}.

In Tables \ref{tab:real-ocs} \& \ref{tab:real-ocs-cont} we show the names,
coordinates, photometric bands used in each study and parameter values for
the complete set of real OCs, both present in recent literature and those
found in this work.
On occasions the cluster parameters present in the reference articles
are not given as a single value but rather as two or even a range of possible
values; in these cases an average is calculated to allow a comparison with
the unique solutions returned by the code.
In the case of clusters with several studies available we restricted
those listed in the tables to the three most recent ones with at least two
parameters determined.

\begin{table*}[h!]
\centering
\caption[caption]{Observed OCs parameters. Literature values are shown in the
firsts rows for each cluster, \texttt{ASteCA} values are shown in the
two last rows for a manually fixed radius ($r_{cl,m}$) and an automatically
assigned radius $r_{cl,a}$ respectively, see Figs. \ref{fig:real-ocs-CT1},
\ref{fig:real-ocs-M11}, \ref{fig:real-ocs-BV1} and \ref{fig:real-ocs-BV2}.
Our values are rounded following the convention of one significant figure in
the error.\\
Abbreviations: (pg) photographic photometry; (pe) photoelectric photometry;
(vr) variable reddening.
% Articles marked with a $\dagger$ symbol represent those with equivalent
% photometry as the one used by \texttt{ASteCA}.
}
\label{tab:real-ocs}
\begin{tabular}{lccccccc}
\hline
\hline\\[-1.85ex]
Name   & ep (J2000) & \textit{Ref.} & Bands & \multicolumn{1}{c}{[Fe/H]} & log(age) & $E_{B-V}$ & Distance (kpc) \\
\hline\\[-1.85ex]
\object{Berkeley 7} & $\alpha=28.55º$ & [1]        & $UBVRI$ & $0.0\pm-$    & $7.1\pm0.1$ & $0.74\pm0.05$ & $2.6\pm0.1$\\
                    & $\delta=62.37º$ & [2]        & $UBV$   & $-$          & $6.6\pm-$   & $0.8\pm-$     & $2.57\pm-$\\
                    &                 & $r_{cl,m}$ & $BV$    & $-0.6\pm0.7$ & $7.3\pm0.2$ & $0.90\pm0.07$ & $1.8\pm0.3$\\
                    &                 & $r_{cl,a}$ & $BV$    & $-0.7\pm0.6$ & $7.0\pm0.2$ & $0.90\pm0.09$ & $1.7\pm0.2$\\[0.2ex]
\cline{3-8}\\[-1.85ex]
\object{Bochum 11} & $\alpha=161.81º$ & [3]        & $UBV$ (pe) & $-$          & $-$         & $0.54\pm0.13$ & $3.6\pm-$\\
                   & $\delta=-60.08º$ & [4]        & $UBV$      & $-$          & $6.2\pm0.4$ & $0.588\pm-$   & $3.47\pm-$\\
                   &                  & [5]        & $UBVRI$    & $0.0\pm-$    & $6.6\pm-$   & $0.58\pm0.05$ & $3.5\pm-$\\
                   &                  & $r_{cl,m}$ & $UV$       & $-0.1\pm0.3$ & $7.0\pm0.2$ & $0.45\pm0.06$ & $1.2\pm0.3$\\
                   &                  & $r_{cl,a}$ & $UV$       & $0.1\pm0.2$  & $7.3\pm0.2$ & $0.7\pm0.3$   & $1.2\pm0.2$\\[0.2ex]
\cline{3-8}\\[-1.85ex]
\object{Czernik 26} & $\alpha=97.74º$ & [6]             &  $BVI$  & $-0.4\pm-$    & $9\pm-$        & $0.38\pm-$    & $8.9\pm-$\\
					& $\delta=-4.18º$ & [7]             &  $CT_1$ & $0.0\pm0.2$   & $9.11\pm0.05$  & $0.05\pm0.05$ & $6.7\pm1.4$\\
                    &                 & $r_{cl,m}$      &  $CT_1$ & $0.14\pm0.08$ & $8.85\pm0.06$  & $0.30\pm0.06$ & $9.1\pm0.8$\\
                    &                 & $r_{cl,a}$      &  $CT_1$ & $-0.1\pm0.3$  & $10\pm1$       & $0.1\pm0.2$   & $6.\pm2.$ \\[0.2ex]
\cline{3-8}\\[-1.85ex]
\object{Czernik 30} & $\alpha=112.83º$ & [6]             & $BVI$  & $-0.4\pm-$    & $9.4\pm-$     & $0.25\pm-$    & $7.1\pm-$\\
					& $\delta=-9.97º$  & [7]             & $CT_1$ & $-0.4\pm0.2$  & $9.39\pm0.05$ & $0.26\pm0.02$ & $6.2\pm0.8$\\
                    &                  & $r_{cl,m}$      & $CT_1$ & $0.07\pm0.09$ & $9.5\pm0.3$   & $0.1\pm0.1$   & $8.\pm1.$\\
                    &                  & $r_{cl,a}$      & $CT_1$ & $-0.3\pm0.4$  & $8.9\pm0.2$   & $0.5\pm0.1$   & $8.\pm1.$\\[0.2ex]
\cline{3-8}\\[-1.85ex]
\object{Haffner 11} & $\alpha=113.85º$ & [8]             & $JHK_s$ & $0.0\pm-$    & $8.95\pm0.07$ & $0.36\pm0.03$ & $5.2\pm0.2$\\
                    & $\delta=-27.72º$ & [9]             & $UBVI$  & $-$          & $8.9\pm-$     & $0.32\pm0.05$ & $6.0\pm-$\\
                    &                  & [7]             & $CT_1$  & $-0.4\pm0.2$ & $8.69\pm0.09$ & $0.57\pm0.05$ & $6.1\pm1.1$\\
                    &                  & $r_{cl,m}$      & $CT_1$  & $0.05\pm0.1$ & $8.8\pm0.1$   & $0.55\pm0.06$ & $6.6\pm0.9$\\
                    &                  & $r_{cl,a}$      & $CT_1$  &$-0.1\pm0.2$  & $8.9\pm0.4$   & $0.5\pm0.1$   & $5.0\pm0.7$\\[0.2ex]
\cline{3-8}\\[-1.85ex]
\object{Haffner 19} & $\alpha=118.2º$  & [10]       & $UBVRI$     & $-$          & $6.65\pm0.05$ & $0.38\pm0.02$ & $6.4\pm0.65$\\
                    & $\delta=-26.28º$ & [11]       & $UBV(RI)_c$ & $-$          & $6.3\pm0.3$   & $0.42\pm0.01$ & $5.2\pm0.4$\\
                    &                  & [12]       & $UBV(RI)_c$ & $-$          & $6.8\pm-$     & $0.44\pm0.03$ & $5.1\pm0.2$\\
                    &                  & $r_{cl,m}$ & $BV$        & $-0.9\pm0.4$ & $7.1\pm0.2$   & $0.3\pm0.1$   & $3.0\pm0.8$\\
                    &                  & $r_{cl,a}$ & $BV$        & $-1.2\pm0.9$ & $7.0\pm0.1$   & $0.4\pm0.2$   & $3.2\pm0.9$\\[0.2ex]
\cline{3-8}\\[-1.85ex]
\object{NGC 133} & $\alpha=7.829º$ & [13]       & $UBVI$ & $-$         & $7\pm-$     & $0.6\pm0.1$ & $0.63\pm0.15$\\
                 & $\delta=63.35º$ & $r_{cl,m}$ & $BV$   & $0.3\pm0.1$ & $9.0\pm0.4$ & $0.0\pm0.2$ & $0.6\pm0.1$\\
                 &                 & $r_{cl,a}$ & $BV$   & $-1.5\pm1.$ & $7.5\pm0.7$ & $0.7\pm0.3$ & $0.9\pm0.3$\\[0.2ex]
\cline{3-8}\\[-1.85ex]
\object{NGC 2236} & $\alpha=97.41º$ & [14]            & $UBV$ (pg) & $-$           & $7.9\pm-$     & $0.76\pm-$ (vr) & $3.7\pm0.1$\\
                  & $\delta=6.83º$  & [15]            & $UBVRI$    & $0.0\pm-$     & $8.7\pm0.1$   & $0.56\pm0.05$   & $2.84\pm-$\\
                  &                 & [16]            & $CT_1$     & $-0.3\pm0.2$  & $8.78\pm0.04$ & $0.55\pm0.05$   & $2.5\pm0.5$\\
                  &                 & $r_{cl,m}$      & $CT_1$     & $-0.1\pm0.2$  & $8.75\pm0.07$ & $0.55\pm0.05$   & $2.8\pm0.3$\\
                  &                 & $r_{cl,a}$      & $CT_1$     & $-0.1\pm0.1$  & $8.70\pm0.08$ & $0.55\pm0.05$   & $2.9\pm0.1$\\[0.2ex]
\cline{3-8}\\[-1.85ex]
\object{NGC 2264} & $\alpha=100.24º$ & [17]       & $X{-}ray;VI_c$ & $-$          & $6.4\pm-$   & $0.08\pm-$      & $0.8\pm-$\\
                  & $\delta=9.895º$  & [18]       & $UBV$          & $-$          & $6.7\pm-$   & $0.075\pm0.003$ & $0.77\pm0.01$\\
                  &                  & [19]       & $BV$           & $0.0\pm-$    & $6.81\pm-$  & $0.04\pm-$      & $0.66\pm-$\\
                  &                  & $r_{cl,m}$ & $BV$           & $-0.6\pm0.2$ & $6.7\pm0.2$ & $0.10\pm0.05$   & $0.58\pm0.05$\\
                  &                  & $r_{cl,a}$ & $BV$           & $-1.5\pm0.2$ & $6.8\pm0.2$ & $0.0\pm0.07$    & $0.44\pm0.06$\\[0.2ex]
\cline{3-8}\\[-1.85ex]
\object{NGC 2324} & $\alpha=106.03º$ & [20]             & $UBV$ (pe) & $0.0\pm-$    & $8.9\pm-$     & $0.02\pm-$    & $3.8\pm-$\\
                  & $\delta=1.05º$   & [21]             & $UBVI$     & $-0.32\pm-$  & $8.8\pm-$     & $0.17\pm0.12$ & $4.2\pm0.2$\\
                  &                  & [22]             & $CT_1$     & $-0.3\pm0.1$ & $8.65\pm0.06$ & $0.20\pm0.05$ & $3.9\pm0.3$\\
                  &                  & $r_{cl,m}$       & $CT_1$     & $0.05\pm0.1$ & $8.85\pm0.06$ & $0.05\pm0.05$ & $4.2\pm0.2$\\
                  &                  & $r_{cl,a}$       & $CT_1$     & $0.02\pm0.1$ & $8.8\pm0.1$   & $0.10\pm0.07$ & $4.4\pm0.4$\\[0.2ex]
\hline\\[0.2ex]
\multicolumn{8}{p{.85\textwidth}}{
\textit{References}:
[1] \cite{Lata_2014}; [2] \cite{Phelps_Janes_1994}; [3] \cite{Moffat_1975};
[4] \cite{Fitzgerald_1987}; [5] \cite{Patat_2001}; [6] \cite{Hasegawa_2008};
[7] \cite{Piatti_2009};
[8] \cite{Bica_2005}; [9] \cite{Carraro_2013}; [10] \cite{Vazquez_2010};
[11] \cite{Moreno_2002}; [12] \cite{Munari_1996}; [13] \cite{Carraro_2002_2};
[14] \cite{Babu_1991} ; [15] \cite{Lata_2014}; [16] \cite{Claria_2007};
[17] \cite{Dahm_2007}; [18] \cite{Turner_2012}; [19] \cite{Kharchenko_8-2005};
[20] \cite{Mermilliod_2001}; [21] \cite{Kyeong_2001}; [22] \cite{Piatti_5-2004}
}
\end{tabular}
\end{table*}

\begin{table*}[h!]
\centering
\caption[caption]{Continuation of Table \ref{tab:real-ocs}.}
\label{tab:real-ocs-cont}
\begin{tabular}{lccccccc}
\hline
\hline\\[-1.85ex]
Name   & ep (J2000) & \textit{Ref.} & Bands & \multicolumn{1}{c}{[Fe/H]} & log(age) & $E_{B-V}$ & Distance (kpc) \\
\hline\\[-1.85ex]
\cline{3-8}\\[-1.85ex]
\object{NGC 2421} & $\alpha=114.05º$ & [23]       & $byH_{\alpha}$ & $-$          & $7.4\pm-$   & $0.47\pm-$    & $2.18\pm-$\\
                  & $\delta=-20.61º$ & [24]       & $UBVR_cI_c$    & $-0.45\pm-$  & $7.9\pm0.1$ & $0.42\pm0.05$ & $2.2\pm0.2$\\
                  &                  & [19]       & $BV$           & $0.0\pm-$    & $7.41\pm-$  & $0.45\pm-$    & $2.18\pm-$\\
                  &                  & $r_{cl,m}$ & $BV$           & $-1.2\pm0.4$ & $7.8\pm0.4$ & $0.45\pm0.05$ & $1.8\pm0.2$\\
                  &                  & $r_{cl,a}$ & $BV$           & $-1.2\pm0.4$ & $8.0\pm0.2$ & $0.45\pm0.05$ & $1.9\pm0.2$\\[0.2ex]
\cline{3-8}\\[-1.85ex]
\object{NGC 2627}    & $\alpha=129.31º$ & [25]             & $UBV$  & $-$          & $9.25\pm^{+0.1}_{-0.05}$   & $0.04\pm^{+0.1}_{-0.02}$ & $1.8\pm0.2$\\
                     & $\delta=-29.96º$ & [26]             & $CT_1$ & $-0.1\pm0.1$ & $9.09\pm0.07$              & $0.12\pm0.07$            & $1.9\pm0.4$\\
                     &                  & $r_{cl,m}$       & $CT_1$ & $-0.1\pm0.2$ & $8.9\pm0.8$                & $0.3\pm0.2$              & $3.\pm1.$\\
                     &                  & $r_{cl,a}$       & $CT_1$ & $-0.3\pm0.3$ & $9.2\pm0.2$                & $0.20\pm0.06$            & $1.9\pm0.4$\\[0.2ex]

\cline{3-8}\\[-1.85ex]
\object{NGC 6231} & $\alpha=253.54º$ & [23]       & $byH_{\alpha}$   & $-$          & $6.9\pm-$   & $0.44\pm-$    & $1.24\pm-$\\
                  & $\delta=-41.83º$ & [27]       & $UBVIH_{\alpha}$ & $-$          & $6.7\pm0.2$ & $0.47\pm-$    & $1.58\pm-$\\
                  &                  & [19]       & $BV$             & $0.0\pm-$    & $6.81\pm-$  & $0.44\pm-$    & $1.25\pm-$\\
                  &                  & $r_{cl,m}$ & $BV$             & $-0.9\pm0.7$ & $6.9\pm0.3$ & $0.40\pm0.05$ & $0.91\pm0.08$\\
                  &                  & $r_{cl,a}$ & $BV$             & $-1.2\pm0.9$ & $7.0\pm0.2$ & $0.40\pm0.05$ & $0.91\pm0.04$\\[0.2ex]
\cline{3-8}\\[-1.85ex]
\object{NGC 6705}         & $\alpha=282.77º$ & [28]       & $uvby\beta$ & $-0.06\pm0.59$ & $8.39\pm-$    & $0.45\pm-$    & $1.82\pm0.03$\\
\multicolumn{1}{c}{(M11)} & $\delta=-6.27º$  & [29]       & $BVIri$     & $0.10\pm0.06$  & $8.45\pm0.07$ & $0.40\pm0.03$ & $2.0\pm0.2$\\
                          &                  & [30]       & $JH$        & $0.0\pm-$      & $8.39\pm0.05$ & $0.42\pm0.03$ & $1.9\pm0.2$\\
                          &                  & $r_{cl,m}$ & $JH$        & $-0.1\pm 0.2$  & $8.60\pm0.07$ & $0.45\pm0.05$ & $1.45\pm0.07$\\
                          &                  & $r_{cl,a}$ & $JH$        & $0.1\pm0.1$    & $8.6\pm0.1$   & $0.30\pm0.08$ & $1.7\pm0.2$\\[0.2ex]
\cline{3-8}\\[-1.85ex]
\object{NGC 6383} & $\alpha=263.70º$ & [31]       & $uvby$                  & $-0.12\pm-$  & $6.6\pm-$     & $0.29\pm0.05$ & $1.7\pm0.3$\\
                  & $\delta=-32.57º$ & [32]       & $UBV(RI)_{c}H_{\alpha}$ & $-$          & $6.4\pm0.1$   & $0.32\pm0.02$ & $1.3\pm0.1$\\
                  &                  & [19]       & $BV$                    & $0.0\pm-$    & $6.71\pm-$    & $0.3\pm-$     & $0.985\pm-$\\
                  &                  & $r_{cl,m}$ & $BV$                    & $0.2\pm0.1$  & $6.7\pm0.6$   & $0.4\pm0.4$   & $1.3\pm0.5$\\
                  &                  & $r_{cl,a}$ & $BV$                    & $-0.1\pm0.4$ & $6.90\pm0.07$ & $0.20\pm0.05$ & $1.1\pm0.2$\\[0.2ex]
\cline{3-8}\\[-1.85ex]
\object{Ruprecht 1} & $\alpha=99.10º$  & [19]             & $BV$   & $0.0\pm-$     & $8.76\pm-$  & $0.15\pm-$    & $1.1\pm-$\\
                    & $\delta=-14.18º$ & [33]             & $CT_1$ & $-0.15\pm0.2$ & $8.3\pm0.2$ & $0.25\pm0.05$ & $1.7\pm0.3$\\
                    &                  & $r_{cl,m}$       & $CT_1$ & $0.1\pm0.2$   & $9.\pm1.$   & $0.0\pm0.4$   & $2.\pm1.$\\
                    &                  & $r_{cl,a}$       & $CT_1$ & $0.3\pm0.1$   & $8.9\pm0.8$ & $0.0\pm0.3$   & $1.6\pm0.7$\\[0.2ex]
\cline{3-8}\\[-1.85ex]
\object{Tombaugh 1} & $\alpha=105.12º$ & [34]             & $UBV$ (pe) & $-$            & $8.9\pm-$     & $0.27\pm0.01$ & $1.26\pm0.02$\\
                    & $\delta=-20.57º$ & [35]             & $VI$       & $0.02\pm-$     & $9\pm-$       & $0.40\pm0.05$ & $3.0\pm0.1$\\
                    &                  & [36]             & $CT_1$     & $-0.30\pm0.25$ & $9.11\pm0.05$ & $0.30\pm0.05$ & $2.2\pm0.3$\\
                    &                  & $r_{cl,m}$       & $CT_1$     & $-0.2\pm0.3$   & $8.6\pm0.2$   & $0.6\pm0.1$   & $3.3\pm0.5$\\
                    &                  & $r_{cl,a}$       & $CT_1$     & $-0.2\pm0.3$   & $9.0\pm0.1$   & $0.35\pm0.06$ & $2.6\pm0.2$\\[0.2ex]
\cline{3-8}\\[-1.85ex]
\object{Trumpler 1} & $\alpha=23.925º$ & [2]        & $UBV$   & $-$           & $7.43\pm-$    & $0.61\pm-$    & $2.63\pm-$\\
                    & $\delta=61.283º$ & [37]       & $UBVRI$ & $-$           & $7.6\pm0.1$   & $0.60\pm0.05$ & $2.6\pm0.1$\\
                    &                  & [38]       & $UBV$   & $0.13\pm0.13$ & $7.18\pm0.35$ & $0.65\pm0.06$ & $2.6\pm0.4$\\
                    &                  & $r_{cl,m}$ & $BV$    & $-0.3\pm0.4$  & $7.7\pm0.2$   & $0.60\pm0.05$ & $1.9\pm0.3$\\
                    &                  & $r_{cl,a}$ & $BV$    & $-0.2\pm0.3$  & $8.1\pm0.5$   & $0.55\pm0.08$ & $2.2\pm0.4$\\[0.2ex]
\cline{3-8}\\[-1.85ex]
\object{Trumpler 5} & $\alpha=99.18º$ & [39]             & $BVI$  & $0.0\pm-$      & $9.6\pm-$     & $0.58\pm-$ (vr) & $3.0\pm-$\\
                    & $\delta=9.43º$  & [40]             & $CT_1$ & $-0.30\pm0.15$ & $9.69\pm0.04$ & $0.60\pm0.08$   & $2.4\pm0.3$\\
                    &                 & $r_{cl,m}$       & $CT_1$ & $0.19\pm0.09$  & $9.70\pm0.06$ & $0.45\pm0.08$   & $2.8\pm0.3$\\
                    &                 & $r_{cl,a}$       & $CT_1$ & $-0.1\pm0.2$   & $9.5\pm0.2$   & $0.60\pm0.08$   & $2.9\pm0.1$\\[0.2ex]
\cline{3-8}\\[-1.85ex]
\object{Trumpler 14} & $\alpha=160.98º$ & [41]       & $JHK_s$   & $-$          & $6.2\pm0.2$   & $0.84\pm0.09$ & $2.6\pm0.3$\\
                     & $\delta=-59.55º$ & [42]       & $UBVI_c$  & $-$          & $6.3\pm0.2$   & $0.36\pm0.04$ & $2.9\pm0.3$\\
                     &                  & [19]       & $BV$      & $0.0\pm-$    & $6.67\pm-$    & $0.45\pm-$     & $2.75\pm-$\\
                     &                  & $r_{cl,m}$ & $BV$      & $-0.5\pm0.6$ & $7.1\pm0.1$   & $0.40\pm0.05$ & $1.3\pm0.2$\\
                     &                  & $r_{cl,a}$ & $BV$      & $-0.6\pm0.5$ & $7.05\pm0.08$ & $0.40\pm0.05$ & $1.3\pm0.2$\\[0.2ex]
\hline\\[0.2ex]
\multicolumn{8}{p{.85\textwidth}}{
\textit{References}:
[23] \cite{McSwain_2005}; [24] \cite{Yadav_2004}; [25] \cite{Ahumada_2005};
[26] \cite{Piatti_2003}; [27] \cite{Sung_2013}; [28] \cite{Beaver_2013};
[29] \cite{Cantat_2014}; [30] \cite{Santos_2005}; [31] \cite{Paunzen_2007};
[32] \cite{Rauw_2010}; [33] \cite{Piatti_2008}; [34] \cite{Turner_1983};
[35] \cite{Carraro_1995}; [36] \cite{Piatti_7-2004}; [37] \cite{Yadav_2002};
[38] \cite{Oliveira_2013}; [39] \cite{Kaluzny_1998}; [40] \cite{Piatti_4-2004};
[41] \cite{Ortolani_2008}; [42] \cite{Hur_2012}
% [23] McSwain et al 2005: http://adsabs.harvard.edu/cgi-bin/bib_query?2005ApJS%2E%2E161%2E%2E118M
% [24] Yadav & Sagar 2004: http://adsabs.harvard.edu/cgi-bin/bib_query?2004MNRAS%2E351%2E%2E667Y
% [25] Ahumada 2005: http://adsabs.harvard.edu/abs/2005AN....326....3A
% [26]
% [27] Sung et al 2013: http://adsabs.harvard.edu/cgi-bin/bib_query?2013AJ%2E%2E%2E%2E145%2E%2E%2E37S
% [28] Beaver et. al 2013; http://adsabs.harvard.edu/cgi-bin/bib_query?2013PASP%2E%2E125%2E1412B
% [29] Cantat-Gaudin et. al 2014: http://adsabs.harvard.edu/abs/2014arXiv1407.1510C
% [30] Santos et al. 2005: http://adsabs.harvard.edu/cgi-bin/bib_query?2005A%26A%2E%2E%2E442%2E%2E201S
% [31] Paunzen et al 2007: http://adsabs.harvard.edu/cgi-bin/bib_query?2007A%26A%2E%2E%2E462%2E%2E157P
% [32] Rauw et al 2010: http://adsabs.harvard.edu/cgi-bin/bib_query?2010A%26A%2E%2E%2E511A%2E%2E25R
% [34] Turner 1983: http://adsabs.harvard.edu/abs/1983JRASC..77...31T
% [35] Carraro & Patat 1995: http://adsabs.harvard.edu/abs/1995MNRAS.276..563C
% [36] 
% [37] Yadav & Sagar 2002: http://adsabs.harvard.edu/cgi-bin/bib_query?2002MNRAS%2E337%2E%2E133Y
% [38] Oliveira et al 2013: http://adsabs.harvard.edu/abs/2013A%26A...557A..14O
% [39] Kaluzny 1998: http://adsabs.harvard.edu/abs/1998A%26AS..133...25K
% [40]
% [41] Ortolani et al 2008: http://adsabs.harvard.edu/cgi-bin/bib_query?2008NewA%2E%2E%2E13%2E%2E508O
% [42] Hur et al 2012: http://adsabs.harvard.edu/cgi-bin/bib_query?2012AJ%2E%2E%2E%2E143%2E%2E%2E41H
}
\end{tabular}
\end{table*}

We divide the analysis in two parts. The cluster parameter values obtained by
\texttt{ASteCA} are compared separately with those taken from articles that
used the same photometric bands as the ones used by the code (when
available), and those that made use of different systems altogether.
%
% The works where the $CT_1$ Washington photometry was originally presented,
% in addition to the \cite{Santos_2005} $JH$ study of NGC 6705,
% are marked with a $\dagger$ symbol in Table \ref{tab:real-ocs}, representing
% those articles with equivalent photometry as the one used by \texttt{ASteCA}.
%
For brevity we will refer to the former set of articles as SP
(``same photometry'') and the set composed by the remaining articles as
DP (``different photometry'').
Furthermore, to stress the importance of a correct radius assignment in the
overall cluster analysis process, we applied the code on the full set twice:
first with a manually fixed radius value for each OC, $r_{cl,m}$, and after
that letting \texttt{ASteCA} find this value automatically,
$r_{cl,a}$, through the function introduced in Sect. \ref{sec:rad-determ}.
PARSEC v1.1 isochrones \citep{Bressan_2013} were employed by the
GA to obtain the best fit cluster parameters
% \footnote{This was the latest set of isochrones available in the CMD
% service when we performed the analysis. While writing this article the new
% PARSEC isochrones v1.2S were released \citep{Chen_2014}.}
which means the solar metallicity value used to convert the $z$ parameter
that \texttt{ASteCA} returns into $[Fe/H]$ was $z_{\odot}=0.0152$
\citep{Bressan_2012}.
The ranges for each cluster parameter where the GA searched for the best
fit model are given in Table \ref{tab:ga-mw-range}.
\begin{table}[tb]
\centering
\caption{Ranges used by the GA algorithm when analyzing the set of 9 Milky
Way OCs. The last column shows the number of values used for each
parameter for a total of $\sim9\times10^6$ possible models.}
\label{tab:ga-mw-range}
\begin{tabular}{lcccc}
\hline
\hline\\[-1.85ex]
 Parameter & Min value & Max value & Step & N\\
\hline\\[-1.85ex]
$Metallicity\,(z)$ & 0.0005 & 0.03 & 0.0005 & 60\\
$log(age)$ & 6.6 & 10.1 & 0.05 & 70\\
$E_{B-V}$ & 0.0 & 1.5 & 0.05 & 30\\
$Distance\,modulus$ & 8 & 15 & 0.1 & 70 \\
\verb1  1 $d\,(kpc)$ & 0.4 & 10 & & \\
\hline
\end{tabular}
\end{table}
The same general relations presented in Eq. \ref{eq:ext-rel} were used by the
GA for the $T_1$ vs $(C-T_1)$ and $J$ vs $(J-H)$ CMDs, with the ratios
$A_{T_1}/E_{B-V}{=}2.62$, $E_{C-T_1}/E_{B-V}{=}1.97$ and
$A_{J}/E_{B-V}{=}0.82$, $E_{J-H}/E_{B-V}{=}0.34$
taken from \cite{Geisler_1996} and \cite{Koornneef_1983} respectively.
% Geisler et. al 1996, pag 1539

\begin{figure}[h!]
\begin{center}
% One-column
% \includegraphics[width=0.8\columnwidth]{MW_ASteCA.png}
% Two-columns
\includegraphics[width=\columnwidth]{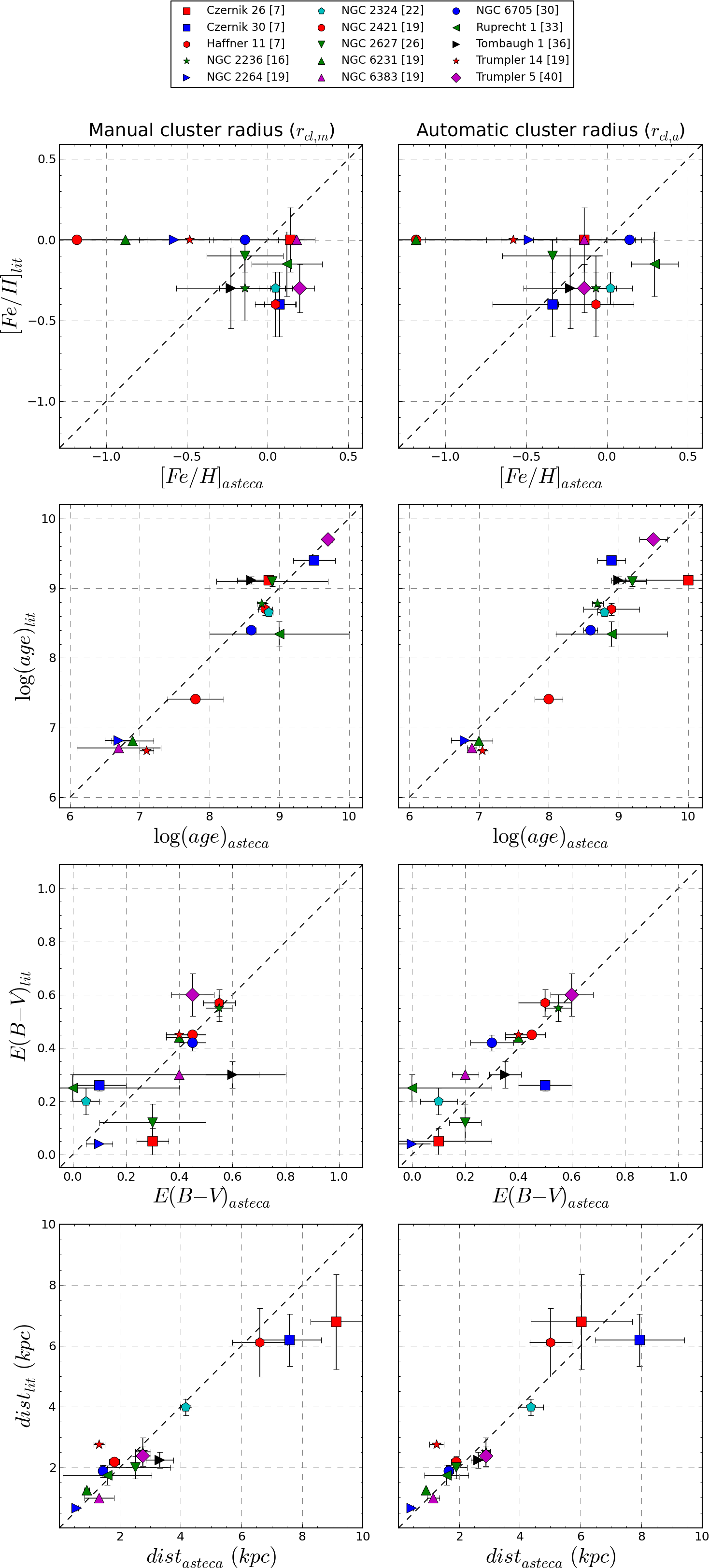}
\caption{Comparison of values found by \texttt{ASteCA} with those present
in the SP set (articles using the same photometric system).
\textit{Left column}: parameters obtained using a manually fixed
radius value for each OC vs literature values. \textit{Right column}: idem
but using  radius values automatically assigned by \texttt{ASteCA}. Identity
relation shown as a dashed black line.
\label{fig:MW-clusts}}
\end{center}
\end{figure}

\begin{figure}[h!]
\begin{center}
% One-column
% \includegraphics[width=0.8\columnwidth]{MW_ASteCA_lit2.png}
% Two-columns
\includegraphics[width=\columnwidth]{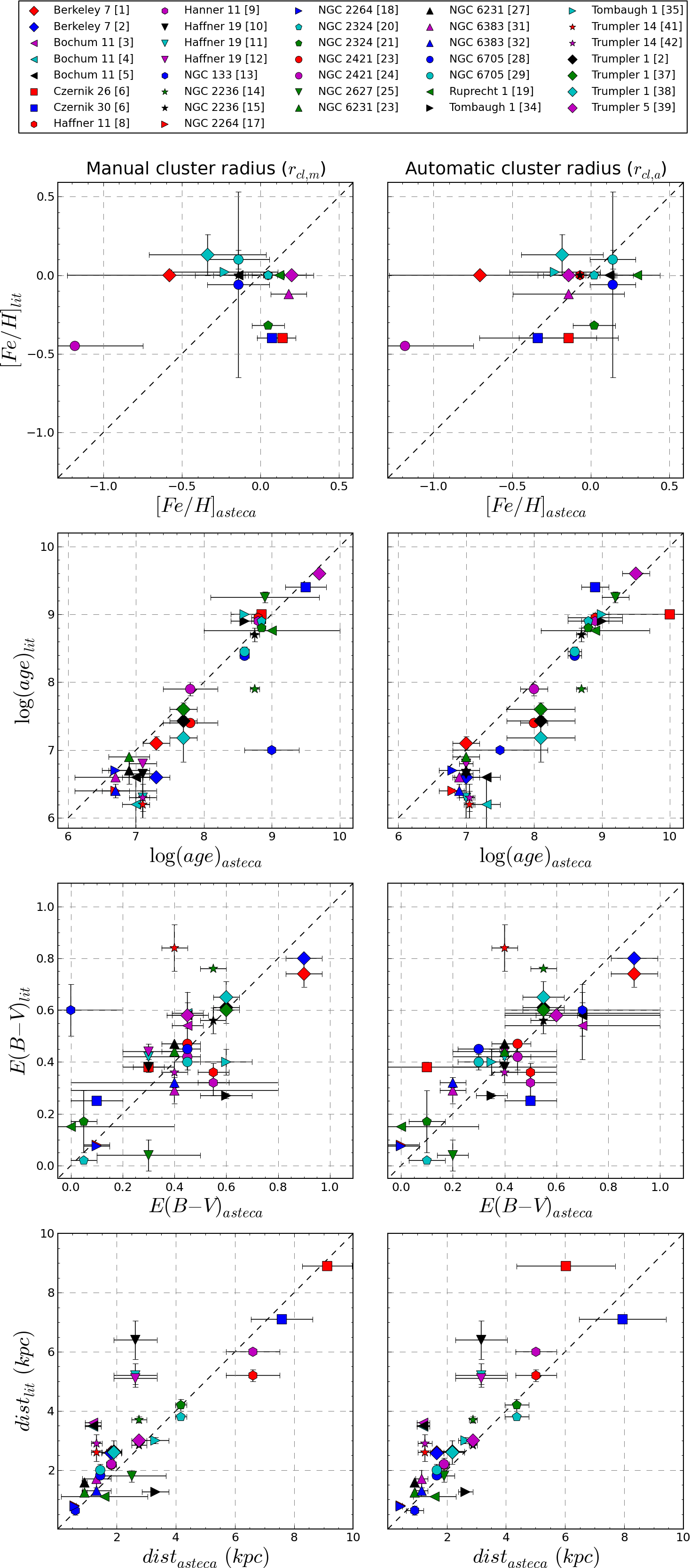}
\caption{Comparison of values found by \texttt{ASteCA} with those present
in the DP set (articles using different photometric systems).
\textit{Left column}: parameters obtained using a manually fixed
radius value for each OC vs literature values. \textit{Right column}: idem
but using  radius values automatically assigned by \texttt{ASteCA}. Identity
relation shown as a dashed black line.
\label{fig:MW-clusts-lit2}}
\end{center}
\end{figure}

In Fig. \ref{fig:MW-clusts} we show the cluster parameters for the OCs
determined by the code versus the SP values, while Fig.
\ref{fig:MW-clusts-lit2} is equivalent but showing a comparison with DP
results.
Left and right columns in both figures present the values for the parameters
found by \texttt{ASteCA} using the manual and automatic radii, respectively.
The figures \ref{fig:real-ocs-CT1}, \ref{fig:real-ocs-M11},
\ref{fig:real-ocs-BV1} and \ref{fig:real-ocs-BV2} in Appendix \ref{app:obs-ocs}
show for the 20 OCs analyzed their positional charts, observed CMD and CMD
of the best synthetic cluster match found by the code using both the manual
and automatic radii values.\\
%\footnote{In some cases the sequence of binaries
%lifted from the original isochrone can be distinguished in the synthetic
%cluster. Because the code creates its synthetic clusters using similar errors
%to those found in the observed photometry, if the latter are very small so will
%be }


The metal content for the manual radius analysis shows a slight tendency to
be over-estimated compared to those values assigned in the SP set
(Fig. \ref{fig:MW-clusts}) while the opposite seems to be true for the
DP articles (Fig \ref{fig:MW-clusts-lit2}).
In contrast, the automatic radius assignment exhibits a more balanced scatter
around the identity line in both cases.
The cluster that departs the most from the metallicity values assigned by
\texttt{AsteCA} is NGC 6231. According to \cite{Tadross_2003} it has a metal
content of $[Fe/H]{=}0.26$, while the code finds $[Fe/H]{=}-1$ on average.
NGC 2421 is given a particularly large negative
metallicity for the manual and automatic radius analysis. Ref. [19]
\citep{Kharchenko_8-2005} assigns by default metal content to the entire sample
of clusters they analyzed, which explains the discrepancy in the SP study, see
Fig. \ref{fig:MW-clusts} (the same happens with clusters NGC 6231, NGC 2264
and Trumpler 14 all showing negative metallicities that do not match the solar
metal content assigned by Ref. [19]). 
The DP source Ref. [24] \citep{Yadav_2004} on the
other hand, finds that NGC 2421 is metal deficient ($[Fe/H]{=}-0.45$), in
much closer agreement with the values for $[Fe/H]$ retrieved by the code, see
Fig. \ref{fig:MW-clusts-lit2}.
The young cluster Berkeley 7 is also given a low metallicity by the code,
$[Fe/H]{\simeq}-0.65$, which deviates significantly from the solar value
used in Ref. [1] \citep{Lata_2014}. In this work the authors employ a $z{=}0.02$
Girardi isochrone to derive the reddening, which is why we associate a
$[Fe/H]{=}0.0$ value to it.
\cite{Tadross_2001} and \cite{Tadross_2003} find for this cluster $[Fe/H]$
values of $-1.75$ and $-0.25$ respectively, while \cite{Paunzen_2010} give it an
unweighted average value of $-0.25$.
As with NGC 2421, we see that the code appears to be able to correctly
identify instances of metal deficient clusters.
Other clusters with negative metallicity values assigned by the code and 
found also in the literature are NGC 2264 ($[Fe/H]{=}-0.08$,
\citealp{Paunzen_2010}), Trumpler 1 ($[Fe/H]{=}-0.71$, \citealp{Tadross_2003})
and Trumpler 14 ($[Fe/H]{=}-0.03$, \citealp{Tadross_2003}).

Ruprecht 1 stands out in Fig. \ref{fig:MW-clusts} as it is given a
large positive metallicity when $r_{cl,a}$ is used,
larger than the one assigned in its SP study, Ref. [33] \citep{Piatti_2008}.
This reference mentions that the cluster ``might'' be of solar metallicity,
which is in better agreement with the values found by the code and
in the DP source Ref. [19].
A similar but more pronounced behavior can be seen for NGC 2324
in Fig \ref{fig:MW-clusts-lit2}, where it clearly separates itself from
the rest of the OCs to the right of the identity line in the case of the
DP article Ref. [21] \citep{Kyeong_2001} which reports $[Fe/H]{=}-0.32$.
Both $[Fe/H]$ values found by \texttt{ASteCA} are in good
agreement with the $[Fe/H]{=}0.0$ value given in the DP source
Ref. [20] \citep{Mermilliod_2001}, but not with the one found in the SP
article Ref. [22] ($[Fe/H]{=}-0.3\pm0.1$, \citealt{Piatti_5-2004}).
Both Czernik 26 and Czernik 30 are assigned low $[Fe/H]$ values by
Ref. [6] \citep{Hasegawa_2008} since they are \emph{assumed} to be of subsolar
metallicities given their galactocentric distances. These values show a good
match with the ones found by \texttt{ASteCA} using the automatic radius but
disagree with the above-solar values derived when the manual radius was
used (see Table \ref{tab:real-ocs}).

The uncertainties associated to the metallicities are quite large as expected,
not only the ones obtained by \texttt{ASteCA} but those present in the
literature as well.
We warn the reader that a single CMD analysis, as that performed by the
code in its current form, should not replace specific and metal sensitive
studies.
Metallicity values returned must be be considered, along with its uncertainties,
as probable ranges that should be further investigated with the right tools.\\

The age parameter is recovered very closely to the literature values for both
sets. In the case of older clusters ($\log(age){>}8$), two cases
stand out in the SP set: Ruprecht 1 and Czernik 26, see Fig. \ref{fig:MW-clusts}.
The former OC shows a larger age value, $\log(age){\sim}9.0$, than the one 
assigned in the literature ($\log(age){=}8.3$; Ref [33]) but with a substantial
uncertainty, almost $\log(age){\sim}1$, for both radii values.
This is a result of the cluster's low star density, which also makes it
the one with the lowest average probability of being a true OC
(${\sim}0.42$). A better match is found with the value
determined by the DP source Ref [19] of $\log(age){=}8.76$.
For Czernik 26, the automatic radius found is larger than the one fixed by eye
resulting in a lot more field stars being added to the cluster region; the
effect of this contamination is to disrupt the synthetic cluster fitting
process into selecting a less than optimal model with a low positioned
TO point (Fig. \ref{fig:real-ocs-CT1}, top row)and thus a larger age.
As explained in Sect. \ref{sec:limit-caveats} and shown in
Fig. \ref{fig:reduc-memb}, this issue can be addressed by increasing the
minimum MP value a star in the cluster region should have in order to affect
the best model fitting algorithm (ie: the $prob_{min}$ parameter).
This is also the only instance where we
can be completely confident that the code is producing a solution of lower
quality than that present in any reference article.

The $\log(age){=}7.9$ value given to NGC 2236 in Ref. [14] \citep{Babu_1991}
and shown in Fig. \ref{fig:MW-clusts-lit2} is much lower than the ones found
in the rest of the literature and those given by the code
($\log(age){\geq}8.7$, see Table \ref{tab:real-ocs}).
Taking into account that the $E_{B-V}$ extinction reported is also higher by
${\sim}0.2$ to all values assigned to this cluster, we can be relatively
certain that this is an example of the negative correlation effect
between the $\Delta {param}$ of $\log(age)$ and reddening established in
Sect. \ref{sec:cluster-parameters} (see Table \ref{tab:corr-matr}). We then
conclude that the age has been underestimated in this article.

For younger clusters we can mention three notable cases: NGC 2421,
NGC 133 and Trumpler 1.
NGC 2421 is assigned by the code an age of $\log(age){\simeq}7.9$
in agreement with Ref. [24] but slightly larger than the values given in
Ref. [19] (SP set) and Ref. [23] \citep[][DP set]{McSwain_2005}, where
$\log(age){\simeq}7.41$ and ${\simeq}7.4$ are given respectively.
The cluster NGC 133, see Fig. \ref{fig:real-ocs-BV1}, has a low member
count and a very high rate of field star contamination ($CI{\simeq}0.91$).
A single study has been performed on it, Ref [13] \citep{Carraro_2002_2}, where
an age of $\log(age){=}7$ is determined, very close to the value found by the
code using the automatic radius.
The manual radius used is larger and includes more contaminating field stars, in
particular one located around $[(B-V){=}0.7;\,V{=}10]$; the code recognizes
this star as an evolved member of the cluster thus resolving it as a much
older system of $\log(age){=}9$.
This is accompanied by a null reddening value, in contrast
with the high extinction assigned to this cluster by the reference,
$E_{B-V}{=}0.6$, and found by the code using the automatic radius,
$E_{B-V}{=}0.7$. Once again this is a clear consequence of the negative
age-reddening correlation.
The code finds Trumpler 1 to have an average age of $\log(age){\simeq}7.9$,
somewhat larger than the values given in the literature where the maximum is
$\log(age){=}7.6$ for Ref. [37] \citep{Yadav_2002}; this age is nonetheless
within the uncertainties associated to the code's values. As seen in Fig.
\ref{fig:real-ocs-BV2} the match of this cluster's upper sequence, containing
the majority of probable members, with the best isochrones found by
\texttt{ASteCA} is very good.

In the lower left corner of the age plots in Fig.
\ref{fig:MW-clusts-lit2} we can clearly appreciate the effect mentioned in
Sect. \ref{sec:limit-caveats} where very young clusters will tend to have
their ages overestimated by the code. Bochum 11, see Fig. \ref{fig:real-ocs-BV1},
is not only a good example of such behavior, but also a demonstration of the
code's ability to perform adequately even with extremely poorly populated
open clusters. Additionally, this cluster was analyzed using
a $V\,vs\,(U-V)$ CMD to show how the code can presently take advantage of the
$U$ filter.\\

Reddening values obtained via \texttt{ASteCA} applying manual radii appear to
be more dispersed when compared with those found in the literature for
either set, while the values that resulted from using the automatic radii
are more concentrated around the identity line for both sets of articles.
Particularly large values for this parameter derived when the
manual radius was used are shown in Figs. \ref{fig:MW-clusts} and
\ref{fig:MW-clusts-lit2} for the case of Tombaugh 1 and NGC 2627 for
the SP and DP set, and for Czernik 26 in the SP set.
This discrepancy is likely due to the lower age assigned to them in this
analysis, which forces the isochrone to be displaced in the CMD lower
(toward larger distance modulus) and to the right (increasing reddening) to
coincide with the cluster's TO.
The effect is also noticeable for the automatic radii analysis
in the case of Czernik 30 as seen in Fig. \ref{fig:MW-clusts} and Fig.
\ref{fig:real-ocs-CT1} (second row) and specially for NGC 133 in the
manual radius analysis for the DP set, Fig. \ref{fig:MW-clusts-lit2} left,
as mentioned previously.

Trumpler 14 displays a substantial difference of $0.44$ with the
large $E_{B-V}{=}0.84$ value assigned by Ref. [41] \citep{Ortolani_2008},
see Fig. \ref{fig:MW-clusts-lit2}, but a much closer match
to the remaining literature values in Ref. [19] and Ref. [42] \citep{Hur_2012},
which show a maximum deviation of ${\sim}0.5$ as seen in Table
\ref{tab:real-ocs-cont}.
Other studies determine for this cluster values of $E_{B-V}{=}0.56\pm0.13$
\citep{Vazquez_1996} and $0.57\pm0.12$ \citep[][where a mean value of
$R_v{=}4.16$ is also obtained]{Carraro_2004}.
In \cite{Yadav_2001} the authors derive a spatial variation of reddening in
the area of this cluster ranging from $E_{B-V}{=}0.44$ to $0.82$, which would
explain the discrepancies between the various values found in the literature.\\

Distances show on average an excellent agreement with literature values,
with two apparent exceptions.
The older cluster Czernik 26 is positioned ${\sim}2\,kpc$ farther away than
its SP Ref. [7] article \citep{Piatti_2009} when using the manual radius
but this large distance is in perfect coincidence with the DP value,
Ref. [6], of $8.9\,kpc$. The reason for this discrepancy is the low reddening
used in Ref. [7] ($E_{B-V}{=}0.05$) compared with the much higher value given
in Ref. [6] and found by the code with the manual radius ($0.38$ and $0.3$,
respectively), which results in a lower distance determination as the
positive distance-reddening correlation indicates (see Table
\ref{tab:corr-matr}).

The second exception is the young cluster Haffner 19 (see Fig.
\ref{fig:MW-clusts-lit2}) located at $d{\simeq}3.1\,kpc$,
between $1.9$ and $3.4\,kpc$ closer than DP
literature estimates ($d=6.4, 5.2, 5.1\,kpc$ in Ref. [10],
\citealp{Vazquez_2010}; [11], \citealp{Moreno_2002} and [12],
\citealp{Munari_1996}; respectively).
The age and reddening parameters obtained
by the code show very similar values to those present in the three DP sources
as seen in Table \ref{tab:real-ocs}, the only difference arises in the
assignment of the metal content. While \texttt{ASteCA} finds that this cluster
is markedly metal deficient ($[Fe/H]{\simeq}-1$), the three articles listed
as references carry out the analysis assuming solar metal content exclusively.
The positive metallicity-distance correlation effect discussed earlier would
indicate that this is the source of the lower distance derived by the code,
or equivalently, the larger distances estimated by the other studies.
Indeed, if \texttt{ASteCA} is run on this cluster constraining the
metallicity range around a solar value of $z_{\odot}{=}0.0152$, as opposed to
using the entire range shown in Table \ref{tab:ga-mw-range}, the resulting
distance is $d{\simeq}4.4\,kpc$, much more similar to that reported in the
references. If the old $z_{\odot}{=}0.019$ value is used, the distance obtained
is even larger: $d{\simeq}5.3\,kpc$.\\

Trumpler 5 is an interesting case not only because it is one of
the OCs with the highest average CI value ($CI{\simeq}0.79$), but also
because no field regions could be determined when the automatic radius
analysis was performed due to the large $r_{cl,a}$ value assigned.
Field regions need to have an equal area to that of the cluster region and in
this case, the frame was not big enough to fit even one.
With no field regions present, the DA could not be applied (field stars
are necessary to calculate the MPs, see Sect. \ref{sec:decont-algor-method})
so all stars within the cluster region were given equal probabilities of
being true members (see Fig. \ref{fig:real-ocs-M11}, bottom row, CMD to
the right).
The cluster parameters found by the GA are nevertheless very reasonable for
both the SP and the DP sets, which means the best fit method is robust
enough to handle highly contaminated OCs even when no MPs can be obtained.\\

\begin{figure}[tb]
\begin{center}
\includegraphics[width=\columnwidth]{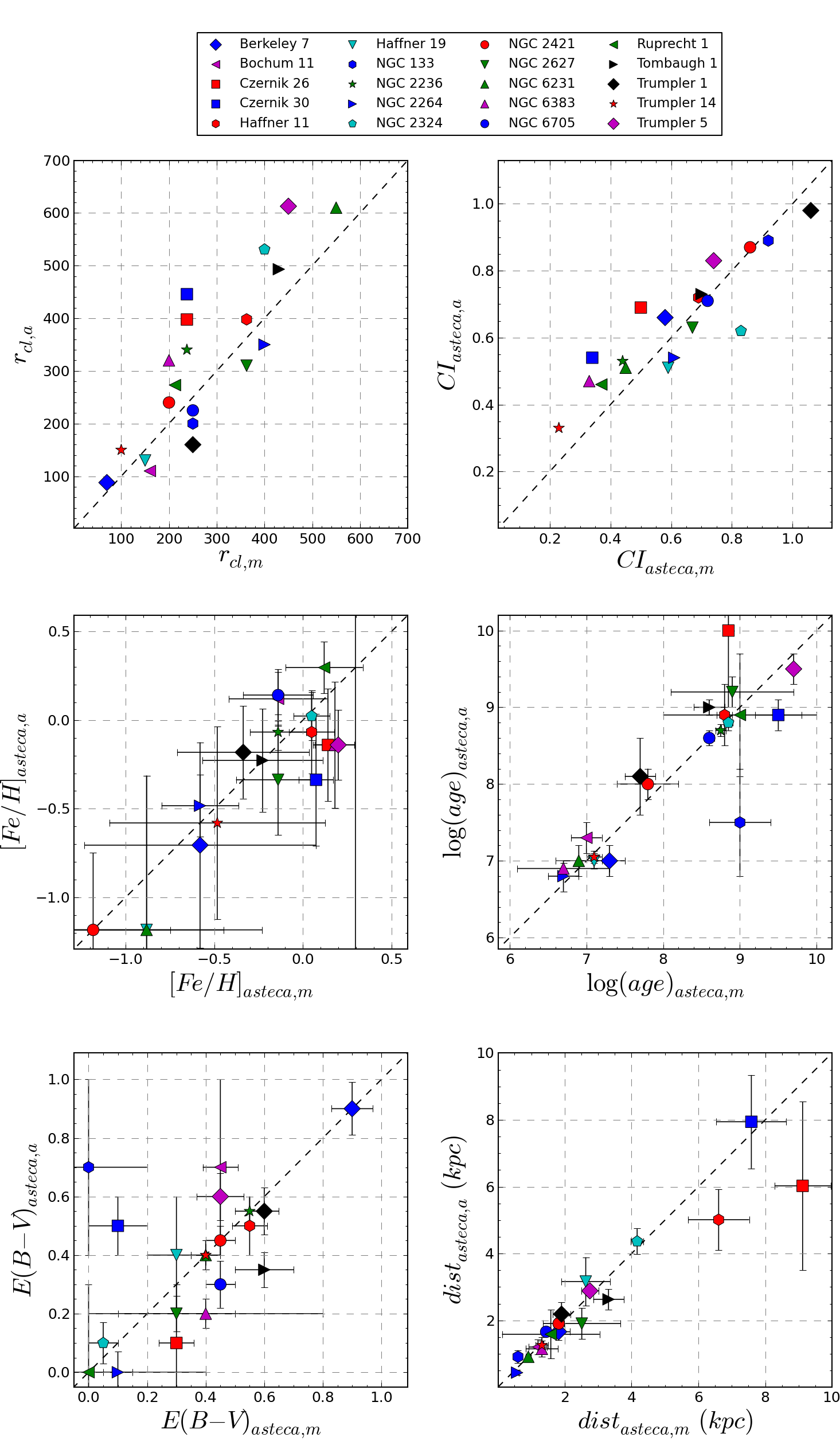}
\caption{\textit{Top}: manual vs automatic radius shown in the left diagram
and CIs obtained for the OCs with each radius to the right. For clusters whose
coordinates are given in degrees, their radii values were re-scaled for
plotting purposes.
\textit{Middle}: Metallicity and $\log(age)$ obtained by \texttt{ASteCA} using
each radius value. \textit{Bottom}: idem above for $E_{B-V}$ and distance.
\label{fig:MW-clusts-m-vs-a}}
\end{center}
\end{figure}

Finally, Fig \ref{fig:MW-clusts-m-vs-a} shows a direct comparison between the
cluster parameters obtained by \texttt{ASteCA} using the radii values assigned
manually and automatically.
The code by default attempts to include as many cluster members as possible
in the cluster region (i.e.: within the $r_{cl,a}$ boundary), so it will in
general return slightly larger radii than those assigned manually,
top left plot in Fig. \ref{fig:MW-clusts-m-vs-a}.
This results in higher CIs as more field stars are inevitably included in
the region. For the set of OCs analyzed the CI values are somewhat large,
up to ${\sim}1$ (top right plot in Fig. \ref{fig:MW-clusts-m-vs-a}),
due to the low overdensity of stars in the cluster regions (positional charts
for each OC can be seen in Appendix \ref{app:obs-ocs}).
The scatter around the identity line in all plots, albeit small in most cases
(with the exception of the incorrect age and reddening assignment for
Czernik 26 when the automatic radius was used and for NGC 133 when
the manual radius was used), points
to the importance of performing a careful analysis when determining the
radius of an OC, specially if little information (photometric and/or
kinematic) is available and the search space for the parameters is
large.
There is a delicate balance between a small radius that could potentially
leave out defining cluster members and a radius too large that introduces
substantial field star contamination, thus difficulting the process of
finding the optimal cluster parameters.
It is worth noting that these values are nonetheless no more scattered than
those that can be found in the literature by various sources, indicating
that the results returned by \texttt{ASteCA} using a minimal of photometric
information (i.e: only two bands) are \textit{at least} as good as those determined
by means of methods that require involvement by the researcher and make use
in general of more photometric bands, with the added advantages of
objectivity/reproducibility, statistical uncertainty estimation and full
automatization.

\section{Summary and Conclusions}
\label{sec:conclusions}

We presented \texttt{ASteCA}, a new set of open source tools dedicated
to the study of SCs able to handle large databases both objectively and
automatically.
Among others, the code includes functions to perform structure analysis,
LF curve and integrated color estimations statistically cleaned from field star
contamination, 
a Bayesian membership assignment algorithm and a synthetic cluster based best
isochrone matching method to simultaneously estimate a cluster's metal content
and age, along with its distance and reddening.
Its main objectives can be summarized as follows:

\begin{itemize}
\item provide an extensible open source template software for developing SC
analysis tools;
\item remove the necessity to implement the frequent and inferior by-eye
isochrone matching, replacing it with a powerful and easy to apply code;
\item facilitate the automatic processing of large databases of stars
with enough flexibility to encompass several scenarios, thus enabling
the compilation of homogeneous catalogs of SCs.
\end{itemize}

Exhaustive tests with artificial SCs generated via the \texttt{MASSCLEAN}
package have shown that \texttt{ASteCA} provides accurate parameters
for clusters suffering from low or moderate field star contamination, and
largely acceptable ones for highly contaminated objects.
The code introduces no biases or new correlations between the final cluster
parameter values it determines.
Being able to implement this internal validation makes it
possible to assess the limiting resolution \texttt{ASteCA} can achieve when
recovering parameters in any situation, provided enough SOCs can be
generated and analyzed.

We obtained fundamental parameters for 20 OCs with available $CT_1$,
$UBV$ and 2MASS photometry. The resulting values were compared with
studies using the same photometric system when available, as well as
a second set of studies done using many different photometric systems.
In both approaches the resulting age, distance and reddening estimates
showed very good agreement with those from the literature, the metallicity
showing a larger dispersion.
The radius assigned to an OC turned out to be an important
factor in estimating the cluster parameters, with small variations
in its value leading to rather diverse solutions.

The ability to provide an estimate for the metallicity and its uncertainty is
a crucial feature considering this is a seldom reliably
obtained parameter, much less one that can be found homogeneously
determined. In most studies its value is simply fixed as solar
\citep{Paunzen_2010}.
As stated in \cite{Oliveira_2013}, fewer than $10\%$ of the OCs in the
DAML02 catalog have their metal contents estimated in the literature, which
makes this an significant gap in the study of stellar evolution and the Galactic
abundance gradient, among other fields.
Although \texttt{ASteCA} is able to assign metallicities with an acceptable
level of accuracy for OCs with low field star contamination while
correctly accounting for the age-metallicity degeneracy issue, we find that
this parameter is expectedly the most difficult one to obtain and the
uncertainties attached to its values are by far the largest.
%
%It is therefore advised to restrict the number of metallicity values that the
%code is allowed to use when searching for the metal content of a stellar
%cluster, to avoid introducing noise by setting a high resolution that will
%end up being .
%
%Unless a photometric system suitable for dealing with
%metal abundances or a specific metallicity-sensitive color is used, it
%is recommended to limit the metallicity values in the parameter space to just a
%handful, enough to cover the desired range without forcing too much resolution.

In general, increasing the resolution of a parameter in the search space of the
best model fitting method will only lead to better results to the extent that
the available photometric information permits it.
It would be interesting then to investigate the increase in the attainable
accuracy for the cluster parameters, in particular for the estimated metal
abundance, when using larger spaces of observed data and/or different
photometric systems.
%Eventually the capability of processing kinematic data input could be added,
%which can substantially improve the membership determination.

Though the the code is applicable to a wide range of
situations, some limitations do apply. Clusters with very low member counts or
high field star contamination should be treated with caution since even
a single misinterpreted star can make a substantial difference, specially
when determining the age.
Very young clusters with no evolved stars are
particularly sensitive to contamination, which can induce the code to
assign larger age values by identifying bright field stars as spurious
members.
Regions affected by differential reddening also pose a great challenge,
as the code will by default assume a unique extinction value.
In all these cases it is advisable to err on the side of caution and treat
returned values as first order approximations. When more information is
made available, it should be used to either verify or dismiss the results.
Running the code more than once with different ranges for the input parameters
is a good idea.\\

\texttt{ASteCA} is meant to be considered a first step in the collaborative
aim toward an objective automatization and standardization in the study of
OCs.
It is written entirely in Python\footnote{\url{https://www.python.org/}}
(with one optional routine making use of the \texttt{R} statistical software
package, see Sect. \ref{sec:pvalue}) and works on 2.7.x versions up to
the latest 2.7.8 release, with 3.x support on the roadmap.

\texttt{ASteCA} is released under a general public license
(GPL v3\footnote{\url{https://www.gnu.org/copyleft/gpl.html}}) and can be
downloaded from it's official site.\footnote{\texttt{ASteCA}: \url
{http://http://asteca.github.io/}}
The code joins then a growing base of recent open source
astronomy/astrophysics software which includes the
set of \texttt{MASSCLEAN} tools,
\texttt{AMUSE}\footnote{Astrophysical Multipurpose Software
Environment: \url{http://www.amusecode.org/}},
AstroML \citep{astroML}\footnote{Machine Learning and Data Mining for
Astronomy: \url{http://www.astroml.org/}} and
Astropy \citep{astropy_2013}\footnote{Community-developed core Python
package for Astronomy: \url{http://www.astropy.org/}}.

\begin{acknowledgements}
G. I. Perren is very grateful to Drs. Girardi L., Popescu B., Moitinho A.,
Hernandez X., Kharchenko N., Dias W., Bonatto C. and Duong T. for the valuable
input provided at various stages during the making of this article.\\
Authors are very much indebted with Dr E.V. Glushkova for the
helpful comments and constructive suggestions that contributed to greatly
improve the manuscript.\\

The authors express their gratitude to the Universidad de La Plata and
Universidad de C\'ordoba.
G. I. Perren and R. A. V\'azquez acknowledge the financial support from the
CONICET PIP1359.
This work was partially supported by the Argentinian institutions CONICET and
Agencia Nacional de Promoci\'on Cient\'ifica y Tecnol\'ogica (ANPCyT).\\

This publication makes use of data products from the Two Micron All Sky Survey,
which is a joint project of the University of Massachusetts and the Infrared
Processing and Analysis Center/California Institute of Technology, funded by
the National Aeronautics and Space Administration and the National Science
Foundation.
This research has made use of the WEBDA database, operated at the Department 
of Theoretical Physics and Astrophysics of the Masaryk University.
\end{acknowledgements}

\bibliography{biblio}

\clearpage
\begin{appendix}

\section{Observed OCs}
\label{app:obs-ocs}

Figs. \ref{fig:real-ocs-CT1}, \ref{fig:real-ocs-M11}, \ref{fig:real-ocs-BV1}
and \ref{fig:real-ocs-BV2} show for the set of
real OCs analyzed, one row per OC, the following  plots:
positional chart (first column), observed CMD with MPs coloring
generated using the manual radius value (second column), CMD of the best
synthetic cluster match found by the code for the cluster region determined
by the manual radius value (third column), equivalent CMDs but generated
using the automatic radius value found (fourth and fifth columns).

\begin{figure*}[tb]
\begin{center}
% One-column
% \includegraphics[width=1.2\columnwidth]{obs_cl_5_CT1.png}
% Two-columns
\includegraphics[width=2.\columnwidth]{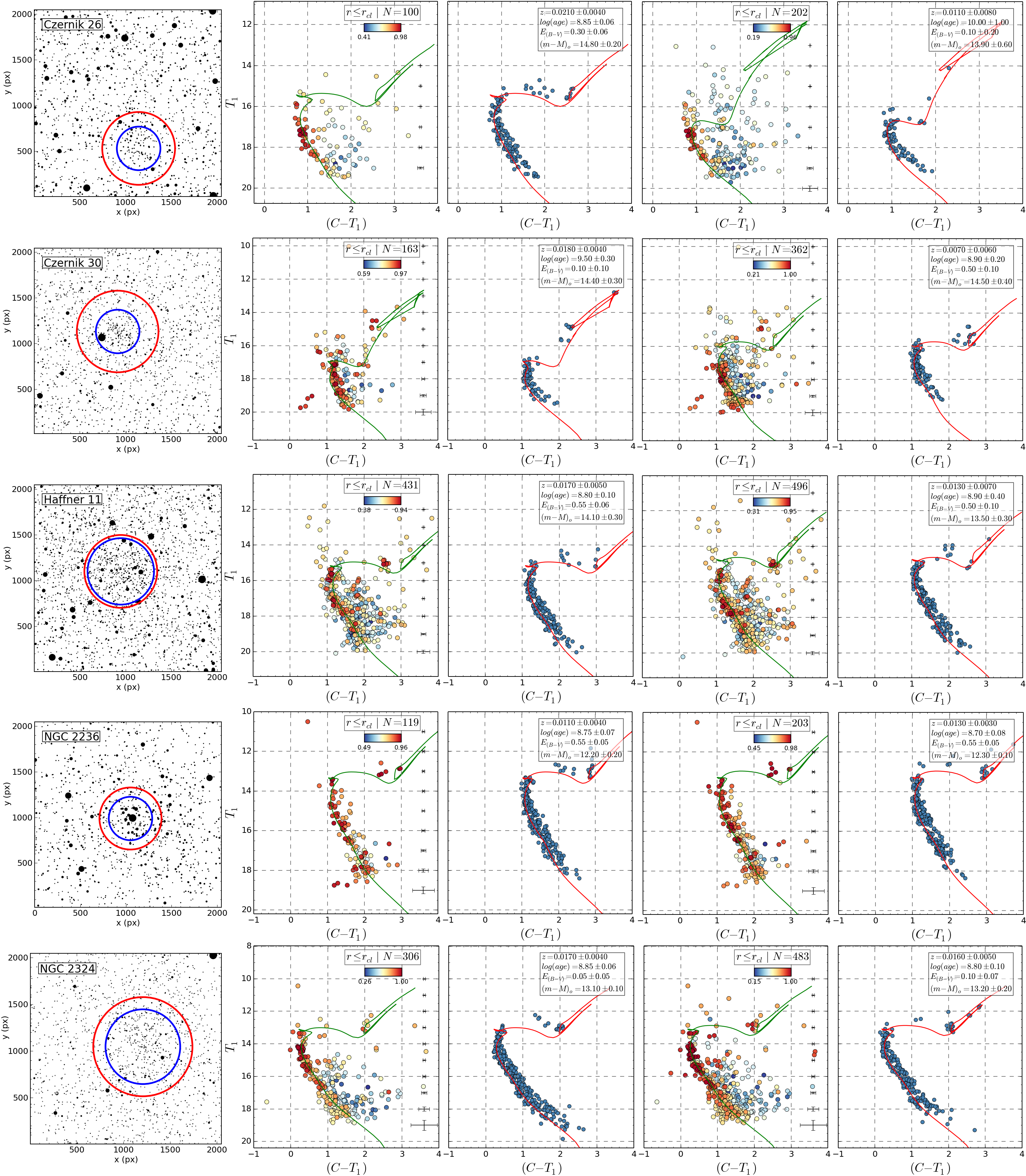}
\caption{Diagrams for observed OCs analyzed with \texttt{ASteCA} displayed in
rows for radii values assigned both manually ($r_{cl,m}$) and automatically
by the code ($r_{cl,a}$).
Leftmost plot is the star chart of the OC with $r_{cl,m}$ and $r_{cl,a}$ shown
as blue and red circles respectively.
Second and third plots are the observed cluster region CMD (colored according
to the MPs obtained by the DA) and best synthetic cluster found by the best
fit algorithm (see Sect. \ref{sec:bf-method}) respectively, using the
$r_{cl,m}$ radius.
Fourth and fifth plots are the same as the previous two but using the
$r_{cl,a}$ radius.
Cluster parameters and their uncertainties can be seen in the top right of the
best synthetic cluster CMDs and are summarized in Table \ref{tab:real-ocs}.}
\label{fig:real-ocs-CT1}
\end{center}
\end{figure*}

\begin{figure*}[tb]
\begin{center}
% One-column
% \includegraphics[width=1.2\columnwidth]{obs_cl_5_m11.png}
% Two-columns
\includegraphics[width=2.\columnwidth]{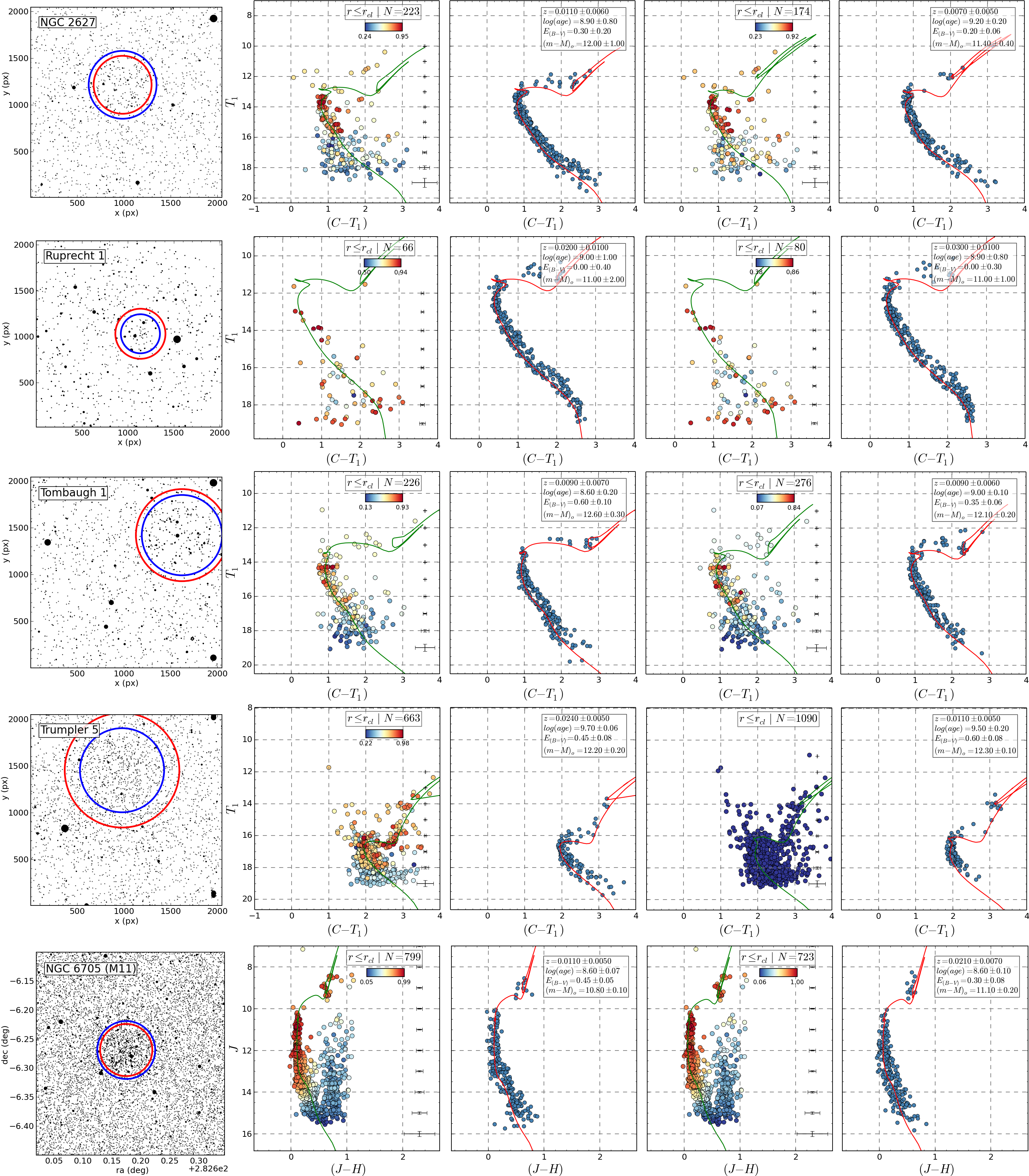}
\caption{Continuation of Fig. \ref{fig:real-ocs-CT1}.}
\label{fig:real-ocs-M11}
\end{center}
\end{figure*}

\begin{figure*}[tb]
\begin{center}
% One-column
% \includegraphics[width=1.2\columnwidth]{obs_cl_5_BV1.png}
% Two-columns
\includegraphics[width=2.\columnwidth]{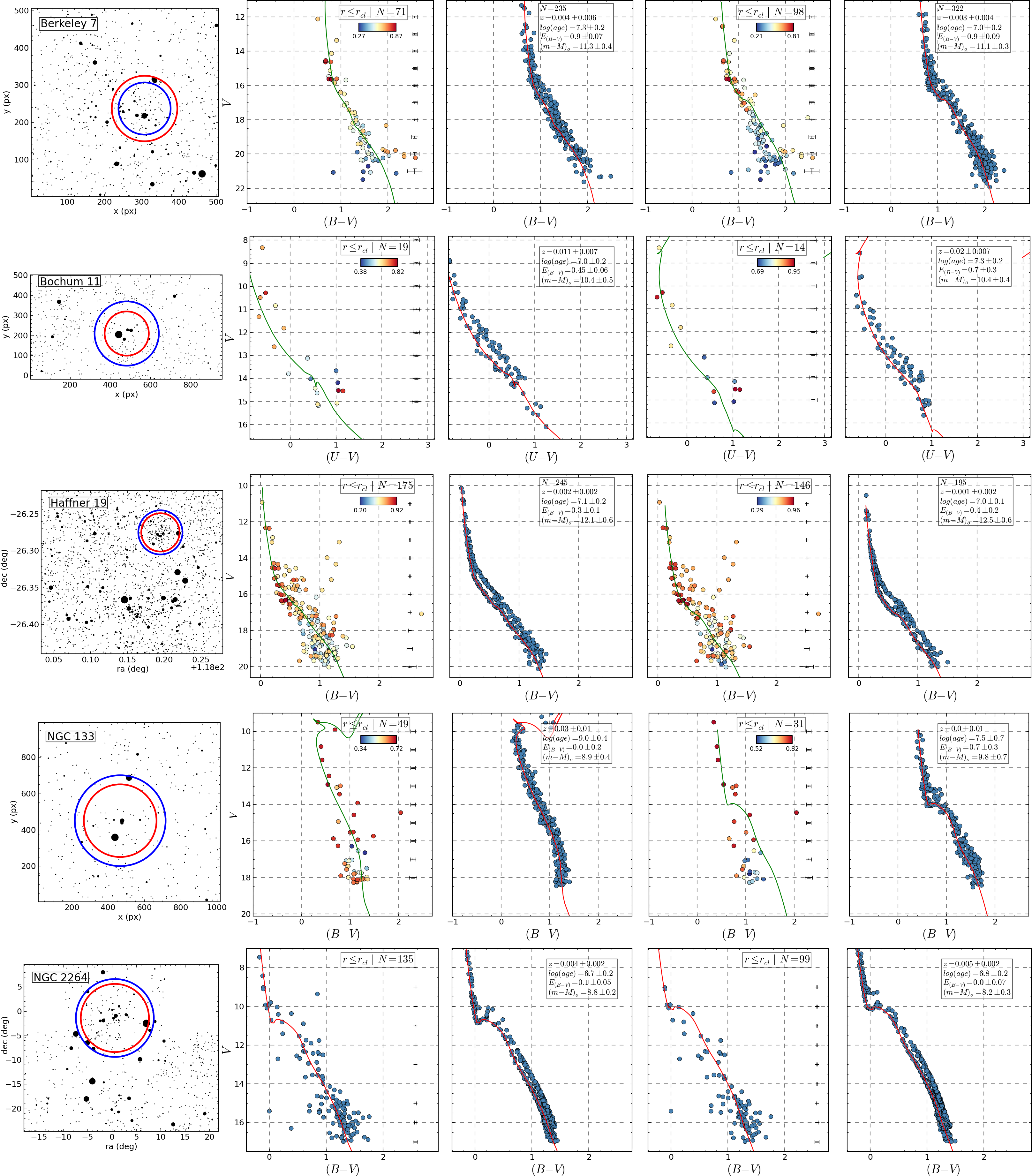}
\caption{Continuation of Fig. \ref{fig:real-ocs-CT1}.}
\label{fig:real-ocs-BV1}
\end{center}
\end{figure*}

\begin{figure*}[tb]
\begin{center}
% One-column
% \includegraphics[width=1.2\columnwidth]{obs_cl_5_BV2.png}
% Two-columns
\includegraphics[width=2.\columnwidth]{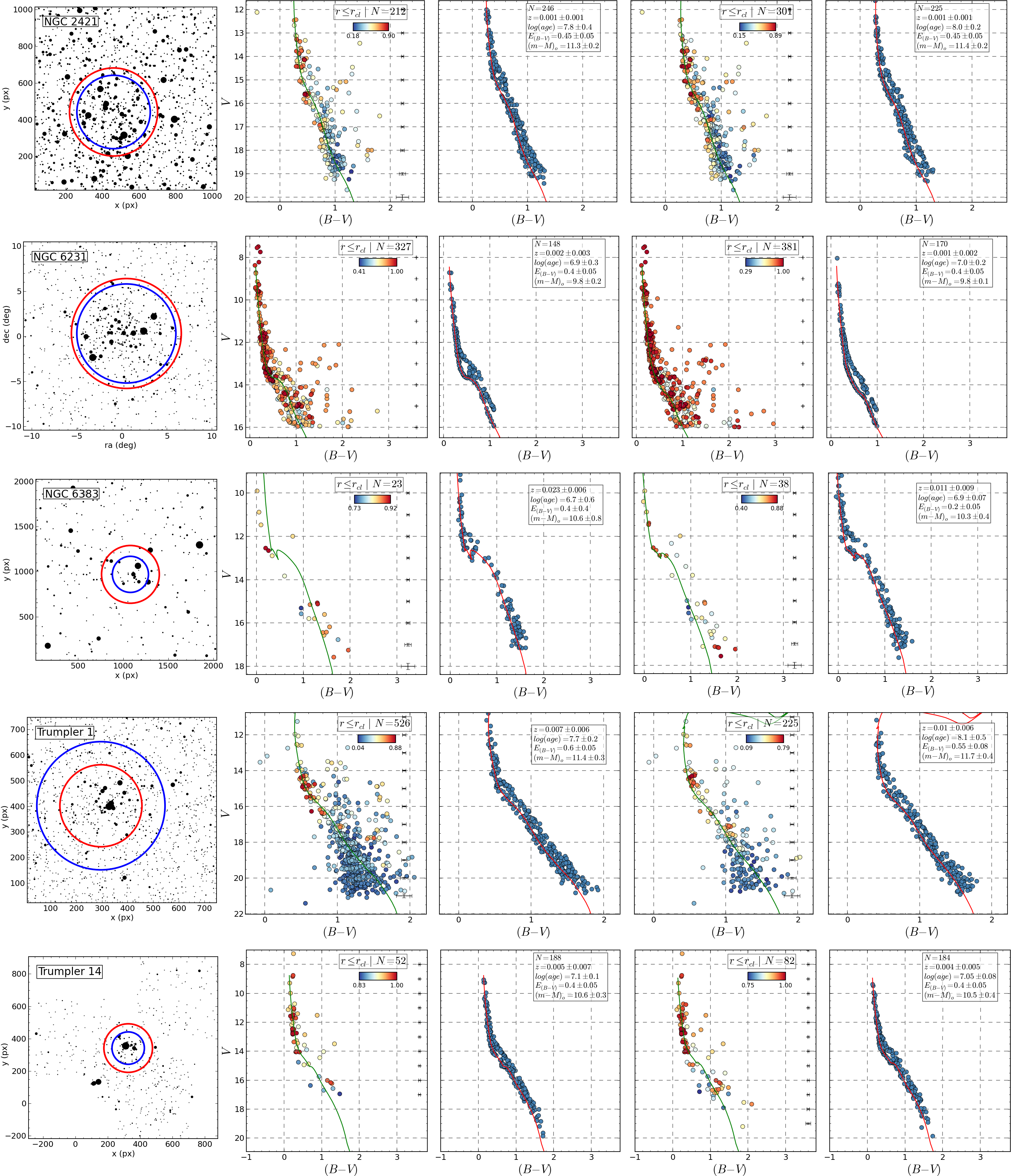}
\caption{Continuation of Fig. \ref{fig:real-ocs-CT1}.}
\label{fig:real-ocs-BV2}
\end{center}
\end{figure*}

\end{appendix}

\end{document}